\documentclass[12pt,a4paper]{report}
\usepackage[latin1]{inputenc}
\usepackage{cite}
\usepackage{hyperref}
\usepackage[usenames]{color}
\usepackage{amsmath}
\usepackage{amsthm}
\usepackage{amsfonts}
\usepackage{amssymb}
\usepackage{makeidx}
\usepackage{titlesec}
\usepackage{tocbibind}
\usepackage{graphicx}
\usepackage{setspace}
\usepackage{subfigure}
\usepackage{float}
\usepackage{lipsum}
\usepackage{fancyhdr}
\definecolor{custom}{rgb}{0.0, 0.0, 0.5}
\hypersetup{
     colorlinks   = true,
     citecolor    = custom
}

\numberwithin{thm}{chapter}
\titleformat{\chapter}[display]
  {\bfseries\Huge}
  {\filcenter\Huge\chaptertitlename~\thechapter}
  {3ex}
  {\titlerule[1.5pt]\vspace{1.5ex}\filcenter}
  [\vspace{1ex}{\titlerule[1.5pt]} ]
\newcommand{\beq}{\begin{eqnarray}}
\newcommand{\eeq}{\end{eqnarray}}

\newcommand{\la}{\lambda}

\newcommand{\s}{\sigma}
\newcommand{\qq}{(\s_{y}\otimes\s_{y})}
\newcommand{\uu}{\qq\rho^{*}\qq}
\newcommand{\tr}{\textrm{Tr}}
\makeindex
\begin{document}

\begin{center}

\vspace{0.5cm}
\begin{LARGE}
Corrected Copy\\
\textbf{Study on Entanglement And Its Utility In Information Processing}\\
\end{LARGE}
\vspace{5cm}

\begin{center}
{\large Thesis submitted for the degree of\\
Doctor of Philosophy\:(SCIENCE)\\
In\\
\textbf{APPLIED MATHEMATICS}\\
}

\end{center}

\vspace{3.5cm}

\begin{center}
by\\
{\large\textsc{SOVIK ROY}}\\
\vspace{1.28cm}
{\large\textbf{Department of Applied Mathematics}\\
\textbf{University of Calcutta}\\
2016}
\end{center}

\end{center}
\newpage
\thispagestyle{empty}
\begin{flushleft}
{\large \textsl{Dedicated to my father 
\textbf{ Late Asit Baran Roy}}}
\end{flushleft}
\newpage
\pagenumbering{roman}
\newpage
\thispagestyle{empty}
\chapter*{Acknowledgements}
\addcontentsline{toc}{chapter}{Acknowledgements}
\markboth{Acknowledgements}{}
No one deserves more thanks than my respected supervisor Prof. Archan S. Majumdar for the success of this work. The chance to watch him in action has shaped my scientific way of thought. He has been a valued teacher. His integral view on research has made a deep impact on me.\\\\
A few people have resolutely influenced how I think about quantum information science. The enjoyable thought provoking discussions with them refined my perspectives on quantum computation and quantum information. In this regard I am deeply indebted to Dr. Satyabrata Adhikari (Department of Mathematics, Birla Institute of Technology, Mesra) and Dr. Biplab Ghosh (Department of Physics, Vivekananda College for Women, Kolkata). Dr. Adhikari always provided me with useful insights and enriching suggestions whereas Dr. Ghosh has helped me in many ways to complete this project.\\\\
My heartiest thanks goes to many of my associates, a special one is always reserved for Dr. Nirman Ganguly (Department of Mathematics, Heritage Institute of Technology, Kolkata).\\\\
My sincere thanks goes to Dr. Juthika Sen Gupta, (Head of the Department of Mathematics) and Dr. Rina Paladhi, (Director) of Techno India, Salt Lake, for giving me opportunity in continuing my research project and also I would like to thank my colleagues Dr. Abhik Sur and Dr. Kaustubh Dutta (of Department of Mathematics, Techno India, Salt Lake) and $`$Techno India Management' for their support as when required.\\\\
I thank the authorities at S.N Bose National Centre for Basic Sciences for offering me with a prospect to work in the institution for my thesis project.\\\\
I also acknowledge my friend Late Soham Dutta. His sudden demise left me with a deep sense of hollow in my heart.\\\\ 
My family has been always a great source of support and encouragement throughout my life and during my last five years of work there has been no exception. I, therefore, feel an immense pleasure in acknowledging my Mother Mrs. Swapna Roy and my wife Mrs. Poulami Roy for their constant cooperation and mental support during these days.\\\\
Last but not the least, my lovely daughter Ms. Tridhara Roy has played a very precious role too. Her innocent smile has always purged all my tiredness after day's hard work and helped me start afresh with new enthusiasm.\\\\

\hspace{3.9in}\textsc{Sovik Roy}
\\\\ 
\newpage
\thispagestyle{empty}
\chapter*{Abstract}
\addcontentsline{toc}{chapter}{Abstract}
\markboth{Abstract}{}
Quantum mechanics has many counter-intuitive consequences which contradict the common intuitions that are based upon the theory of classical physics. In quantum mechanics one can prepare two particles in such a way that the correlation between them cannot be explained classically. These correlated states, known as entangled states, are often used in quantum information processing tasks like quantum teleportation, quantum dense coding, quantum secret sharing etc. Entanglement, the basic feature of quantum information theory, can be studied from various perspectives, its quantification, characterization and its detection. The present context is mainly based upon its applications. Another important aspect of quantum information theory is the study of mixed entangled states and how can these states be effectively used in quantum information protocols remain as the fundamental concern. Here, the efficacies of maximally and that of non-maximally entangled mixed states as teleportation channels have been studied. A new class of non-maximally entangled mixed states have been proposed also. Their advantages as quantum teleportation channels over existing non-maximally entangled mixed states have been verified. The mixed states can also be generated using quantum cloning machines. We have also studied how one can utilize the mixed entangled states obtained as output from a state independent quantum cloning machine in teleportation and dense coding. Mixed states can also be used in secret sharing. A new protocol of secret sharing has been devised where four parties are involved. One of them is honest by nature and the other is dishonest. The dishonest one has two associates who are also involved in the protocol. It has been shown how the honest party, by using quantum cloning machine, can prevent the dishonest party from communicating the secret message to his accomplices. A relation has been established among entanglement of the initially prepared state, entanglement of the mixed state received by the recipients of the dishonest one after honest party applies the cloning scheme and the cloning parameters of the cloning machine. The usefulness of tri-partite and four-partite entangled states in controlled dense coding has been discussed. The thesis concludes with the summary and possible courses of future works.
\newpage
\tableofcontents
\newpage 
\listoffigures
\newpage
\pagenumbering{arabic}
\chapter{Foreward}
\label{ch:fwd}
\textit{$``$ I cannot seriously believe in it(quantum theory) because the theory cannot be reconciled with the idea that physics should represent a reality in time and space, free from spooky actions at a distance (\textit{spukhafte Fernwirkungen})"}
\begin{flushright}
- Albert Einstein\\
(Letter to Max Born (3 Mar 1947). In Born-Einstein Letters (1971), 158).
\end{flushright}
\vspace{2.8cm}
\noindent Information processing is an evolving science. In the beginning of 21st century much of information processing has evolved from its classical to quantum counterparts and it will reach a new era in the coming years with the advent of quantum computers. At the heart of all these, there lies a feature, known as Entanglement, one of the most remarkable and enthralling facets of quantum mechanics. This idea goes back a long way - all the way back to the year 1935 when Albert Einstein, Boris Podolsky and Nathan Rosen (commonly known as EPR), published a philosophically sounding paper titled -$``$ \textsf{Can the quantum mechanical description of reality be considered complete?"} \cite{epr1935}. This paper, written by the quantum misanthropist Einstein, was actually designed to show that quantum theory was an incomplete idea.\\\\
Schrodinger however commenting on the EPR paper, coined the term Entanglement which is the exact translation of the German word \textit{Verschrankung} \cite{sch1935}. Entanglement is a very puzzling phenomenon and it is very difficult to provide an intuitive and simple explanation of this subtle phenomenon which has no classical analogy. Richard Feynman rightly said, $``$\textit{A description of the world in which an object can apparently be in more than one place at the same time, in which a particle can penetrate a barrier without breaking it, in which widely separated particles can cooperate in an almost psychic fashion, is bound to be both thrilling and bemusing}" \cite{feynman1977}.\\\\
In 1964, Bell accepted the EPR conclusion as a working premise and formalized the EPR deterministic world idea in terms of Local Hidden Variable Model (LHVM) \cite{bell1964}. He showed entanglement as that $`$characteristic trait' of quantum mechanics which made impossible to simulate the quantum correlations within any classical formalism. He formulated certain inequalities, now a days famously known as \textit{Bell-inequalities}. Bell-inequalities enlightened us with the fact that entanglement was  fundamentally a new resource in the world that goes essentially ahead of classical resources. Bell even showed that some entangled states violate these inequalities. Aspect \textit{et. al} however were first to present a persuasive test of violation of the Bell-inequalities in laboratories \cite{aspect1981,aspect1982}.\\\\
Greenberger, Horne and Zeilinger (GHZ) went beyond Bell-inequalities by showing that \textit{entanglement} of more than two particles leads to a contradiction with LHVM for non-statistical predictions of quantum formalism \cite{ghz1989}.\\\\
Apart from being central to the investigations of the fundamentals of quantum mechanics, since its inception, quantum entanglement or simply entanglement, is also being used in various quantum information processing protocols. These protocols include quantum teleportation \cite{bennett1993,bouwmeester1997}, superdense coding or dense coding \cite{bennett1992,mattle1996}, quantum secret sharing \cite{hillery1999,cleve1999}, quantum cryptography \cite{bennett1984,ekert1991,bennett1992(2)} and many more. The basic idea behind all such protocols is to realize fast and secured communication. Here the entanglement plays a vital role to accomplish the desired goal.\\\\
One of the important aspects of quantum information processing are the states used by the parties involved, which may be either pure or mixed\footnote{Pure and mixed states have been defined in chapter 2.}. But in practical scenario, it is very difficult to prepare a pure state due to the \textit{de-coherence}\footnote{De-coherence is synonymous to $`$Noise'.} effects of nature. This means that if a state is used as information processing channel, external noises (such as those created by $`$eavesdropping' or due to some other physical factors) may affect the state, thus making it a mixed state. Interactions of the system with the environment cannot be avoided.  So the above quantum information protocols, such as dense coding, teleportation etc. have been studied for mixed states too. The mixed state dense coding have been studied in \cite{barenco1994,bose1999}. Likewise teleportation with mixed states have been discussed in \cite{verstraete2003,albeverio2003}. Secret sharing tasks have been pulled off for mixed states too in \cite{gottesman2000,adhikari2010}.\\\\
Another perspective of quantum information theory is \textit{Quantum Cloning} \cite{scarani2005}. Encoding of information in quantum systems is a very important standpoint. The process of encoding of information, not in the individual constituents but rather to the state of the system as a whole  is basically known as \textit{quantum cloning}. More technically, the information being encoded in the state $\vert\psi\rangle$ of quantum systems as well as the process of the state's replication is known as \textit{quantum cloning}. But, just like Heisenberg's uncertainty principle \cite{heisenberg1925}, \textit{no - cloning theorem} defines an intrinsic impossibility as it prohibits perfect copying \cite{wootters1982}. Yet imperfect clones of quantum state can be produced \cite{buzek1996}. After this there was no looking back. Many quantum copying machines, like Gisin-Massar optimal cloning machine \cite{gisin1997}, Buzek-Hillery universal optimal cloning machine \cite{buzek1998}, Optimal universal state dependent copying machine \cite{bru1998}, probabilistic cloning machine \cite{duan1998}, arbitrary \textit{d} dimensional cloning machine \cite{zanardi1998,fan2001}, sequential cloning machine \cite{delgado2007,dang2008}, hybrid cloning machine \cite{adhikari2007} and many more, had been proposed thereafter. The cloned outputs obtained from cloning machines, after tracing out the ancilla states, however can also be used in information processing tasks as has been shown in \cite{adhikari2008}.\\\\
In all the above cases, entanglement always lies as the central figure which allows one to manipulate with it for achieving the desired objectives of diverse information processing tasks.
The dissertation is thus mainly concerned with some further investigations on \textbf{\textit{entanglement}}, and its relevance has been tested in different other set-ups of information etiquette.
\section*{Plan of the thesis:}
The thesis is organised as follows.\\
\begin{itemize}
\item \emph{\textbf{Chapter $2$}} deals with mathematical preliminaries and physical pre-requisites necessary for the development of the contents of this dissertation.
We also briefly describe some of the known protocols in quantum information science, like, teleportation, dense coding, controlled dense coding, secret sharing and cloning. These protocols remain central to the investigations throughout the advancement of this thesis.\\
\item \emph{\textbf{Chapter $3$}} discusses various classes of non-maximally entangled mixed states. Their entanglement properties, utility of such states in teleportation, Bell-inequality violation, and their advantages over other non-maximally entangled mixed states are studied thereafter. The chapter also discusses the mixed states of maximally entangled types from the view point of quantum teleportation. The usefulness of such states in teleportation, their behaviour corresponding to Bell violation, and entanglement properties have been emphasized also.\\
\item \emph{\textbf{Chapter $4$}} talks about the usefulness of an entangled two qutrit output as a resource obtained from a universal quantum cloning machine in information processing tasks such as teleportation and dense coding. Both the optimal and non - optimal forms of the output have been considered. The optimal fidelities of teleportation and capacities of dense coding protocols of these $3\otimes 3$ dimensional output states have been examined.\\
\item \emph{\textbf{Chapter $5$}} highlights the scheme known as controlled dense coding. Different tripartite and quadripartite pure entangled states have been put in order for this purpose. The central part of this chapter is to look into the effectiveness of these different kinds of multi-partite states in controlled dense coding. Controlled dense coding has also been analysed for qutrit system here.\\
\item \emph{\textbf{Chapter $6$}} introduces a secret sharing protocol. The success probability of such a protocol has been shown to be controlled by an honest party, who  using a quantum cloning circuit, will try to prevent a dishonest one in leaking some secret information. The interdependence of this success probability and cloning parameters has been scrutinized. A relation among the concurrence of initially prepared state, entanglement of the mixed state received by the receivers after the cloning scheme and the cloning parameters of cloning machine have been shown. \\
\item Lastly in \emph{\textbf{Chapter $7$}} we provide a summary of the results obtained in this thesis and some comments on possible future directions of research on these topics while in \emph{\textbf{Chapter $8$}} a brief appendix on quantum gates is presented.
\end{itemize}
\chapter{Mathematical and Physical Pre-requisites}
\label{ch:mpp}
\textit{$``$Mathematical science in my opinion an indivisible whole, an organism whose vitality is conditioned upon the connection of its parts."}
\begin{flushright}
- David Hilbert
\end{flushright}
\textit{$``$Not only is the Universe stranger than we think, it is stranger than we can think."}
\begin{flushright}
- Werner Heisenberg
\end{flushright}
\vskip1cm
\section{\textbf{Introduction:}}
The chapter deals with several mathematical and physical concepts for proper understanding of theory of entanglement. We begin by providing certain mathematical definitions, which are crucial for this purpose \cite{nielsenbook,mcmahonbook,mintertreview2005,horodeckireview,plenioreview,adhikarithesis}.
\section{\textbf{State Space and State Vector:}}
A complex vector space with inner product (also known as Hilbert space) is  associated with an isolated physical system. Such a space is called \textit{state space} of the system. A system is completely described by a unit vector in the system's state space, known as the \textit{state vector}. \textit{Quantum bit} or simply \textit{qubit} on which quantum computation and quantum information are built upon, has a two dimensional state space. If $\vert 0\rangle$ and $\vert 1\rangle$ form an orthonormal basis\footnote{If $V$ is an inner product space, then a subset of $V$ is an orthonormal basis for $V$ if it is an ordered basis that is orthonormal.} for the state space, then an arbitrary state vector in the state space is written as\\
\begin{eqnarray}
\vert \psi_{qubit}\rangle = a\:\vert 0\rangle + b\:\vert 1\rangle.
\label{qubit}
\end{eqnarray}\\
Here, $ \vert a\vert^{2} + \vert b\vert^{2}= 1$, a condition that is equivalent to the fact that $\vert \psi\rangle$ is a unit vector as $\langle \psi\vert \psi\rangle = 1$ and is known as \textit{normalization condition}.\\\\
Similar to this, for a higher dimensional system, say $n \otimes n$ system, the idea of qubit is extended to \textit{qunit}. The state vector of  qunit is described as \\
\begin{eqnarray}
\vert \psi_{qunit}\rangle = \sum^{n}_{i=1}\alpha_{i}\:\vert \psi_{i}\rangle,
\label{qunit}
\end{eqnarray}\\
where $\sum^{n}_{i=1}\vert \alpha_{i}\vert^{2}=1$ and $\lbrace\vert \psi_{i}\rangle \rbrace$ forms an orthonormal set.\\\\
The evolution of a closed quantum system is described by a \textit{unitary transformation} \footnote{An operator $U$ is said to be unitary if $U^{\dagger}\:U=I$, where $I$ is the identity operator and $U^{\dagger}$ is the hermitian conjugate of $U$. For example, Pauli spin operators are unitary operators.}. This implies that if a system is in a state $\vert \psi\rangle$ at a certain time $t_{1}$, then it will evolve to another state $\vert \psi^{/}\rangle$ at time $t_{2}$ by an unitary operator $U=U(t_{1},t_{2})$. The evolution of state is mathematically expressed as\\
\begin{eqnarray}
\vert \psi^{/}\rangle = U\:\vert \psi\rangle.
\label{evolve1}
\end{eqnarray}
\section{\textbf{Density Operator:}}
Apart from using state vectors for formulating quantum mechanics, an alternative but useful approach is often used for its formulation, known as the \textit{the density matrix} or \textit{the density operator} approach. This alternative way of formalization is mathematically sound and is also equivalent to the state vector approach. It provides much more expedient idiom for thinking about commonly encountered scenarios in quantum mechanics. The density operator language provides a convenient way for describing quantum systems whose states are not completely known.\\\\
More precisely, suppose a quantum system is in one of a number of possible states $\lbrace\vert \psi_{i}\rangle\rbrace$ with respective probabilities $p_{i}$ (for all $i$). Then $\lbrace p_{i},\vert \psi_{i}\rangle\rbrace$ is called an ensemble of pure states. The density operator is then defined as\\
\begin{eqnarray}
\rho = \sum_{i} p_{i}\:\vert \psi_{i}\rangle\langle \psi_{i}\vert
\label{densityop},
\end{eqnarray}\\
where, $\sum_{i}\vert \psi_{i}\rangle\langle \psi_{i}\vert = I$.
The evolution of density operator $\rho$ is given by the following.\\
\begin{eqnarray}
\rho= \sum_{i}\:p_{i}\vert \psi_{i}\rangle \langle \psi_{i}\vert \longrightarrow \sum_{i}\:p_{i}\:U\:\vert\psi_{i}\rangle\langle \psi_{i}\vert\:U^{\dagger}=\:U
\:\rho\:U^{\dagger}.
\label{evolve2}
\end{eqnarray}\\
If a vector state $\vert \psi\rangle$ of system is normalized, the state $\rho=\vert \psi\rangle\langle \psi\vert$ is pure and $\rho^{2}=\rho$. Otherwise, $\rho$ is in a \textit{mixed state}, the mixture of pure states.  For a pure state $\rho$ we have $Tr\:(\rho^{2})=1$, while a mixed state satisfies the inequality $Tr\:(\rho^{2})<1$.
\section{\textbf{Reduced Density Operator:}}
Perhaps the deepest application of the density operator is as a descriptive tool for sub-systems of a composite quantum system. Such a description is provided by the \textit{reduced density operator}.\\\\ For two physical systems $A$ and $B$, whose combined state is described by a density operator $\rho^{AB}$ acting on the Hilbert space $H_{A\:B}=H_{A}\otimes H_{B}$, the tensor product of $H_{A}$ and $H_{B}$, the reduced density operator for system $A$ (or $B$) is defined by\\
\begin{eqnarray}
\rho^{A}=Tr_{B}\:(\rho^{AB}),\nonumber\\
\mbox{or}\nonumber\\
\rho^{B}=Tr_{A}\:(\rho^{AB}).
\label{redensityop}
\end{eqnarray}  \\
$Tr_{B}$ (or $Tr_{A}$) is a map of operators known as the \textit{partial trace} over the Hilbert space $H_{B}$ (or $H_{A}$). The reduced density operators $\rho^{A}$ and $\rho^{B}$ can be explicitly written as\\
\begin{eqnarray}
\rho^{A} = \sum_{j=1}^{n_{B}}(I_{A}\otimes \langle \phi_{j}\vert)\rho^{AB}(I_{A}\otimes \vert \phi_{j}\rangle)\nonumber\\
\mbox{and}\nonumber\\
\rho^{B} = \sum_{j=1}^{n_{A}}(\langle \psi_{j}\vert\otimes I_{B})\rho^{AB}(\vert \psi_{j}\rangle\otimes I_{B} ),
\label{ptrace}
\end{eqnarray}\\
where $I_{A}$ and $I_{B}$ are the identity operators in $H_{A}$ and $H_{B}$ and $\vert \phi_{j}\rangle$ ($j=1,2,\cdots, n_{B}$) is an orthonormal basis in $H_{B}$. Similarly, $\vert \psi_{j}\rangle$ ($j=1,2,\cdots, n_{A}$) is an orthonormal basis in $H_{A}$.
\section{\textbf{Schmidt Decomposition and Purification:}}
Two additional tools which are of great importance in quantum information processing science are the \textit{Schmidt decomposition} and \textit{purification}. Let us suppose $\vert \psi\rangle$ is a pure state of a composite system (say, $AB$). Then there exist orthonormal vectors $\vert i_{A}\rangle$ for a system $A$ and orthonormal vectors $\vert i_{B}\rangle$ for a system $B$, such that $\vert \psi\rangle$  can be expressed as \\
\begin{eqnarray}
\vert \psi\rangle = \sum^{k}_{i}\lambda_{i} \:\vert i_{A}\rangle \:\vert i_{B}\rangle,
\label{schmidtform}
\end{eqnarray}
which is known as \textit{Schmidt} form and the non-negative real numbers $\lambda_{i}'s$ are known as \textit{Schmidt numbers} or \textit{Schmidt coefficients} satisfying the condition $\sum_{i}\lambda^{2}_{i}=1$. The number $k$ is called \textit{Schmidt rank}.\\\\
\textit{Purification} is another related technique for quantum computation and quantum information. If we are given a state $\rho^{A}$ for a quantum system $A$, then to purify the state, it is possible to introduce another system say, $R$ (known sometimes as \textit{fictitious} or \textit{reference} or \textit{dummy} or \textit{ancilla} system) which may not have a direct physical significance. The system $R$ has a Hilbert space $H_{R}$ unitarily equivalent to $H_{A}$. If $\vert i_{A}\rangle$ and $\vert i_{R}\rangle$ are the respective orthonormal bases for systems $A$ and $R$, then pure state $\vert AR\rangle \in H_{A}\otimes H_{R}$ can then be defined for the joint system $AR$ such that $\rho^{A}=Tr_{R}(\vert AR\rangle \langle AR\vert)$. $\vert AR\rangle$ is the purification of $\rho^{A}$. Purification, a purely mathematical procedure, allows one to associate pure states with mixed states.
\section{\textbf{Quantum Entanglement:}}
The more intrinsic quantum mechanical sense in which quantum states can embody vastly more information than its classical counterparts is due to the non-classical feature of \textit{quantum entanglement} or simply \textit{entanglement}. The phenomenon of entanglement features predominantly in most aspects of quantum information theory.\\\\
Let us consider a system consisting of two sub-systems where each sub-system is associated with a Hilbert space. Let $H_{A}$ and $H_{B}$ denote these two Hilbert spaces with respect to the two sub-systems $A$ and $B$ respectively. Let $\vert i\rangle_{A}$ and $\vert j\rangle_{B}$, ($i, j = 1,2,3,...$) represent two complete orthonormal bases for $H_{A}$ and $H_{B}$ respectively. The two sub-systems taken together is associated with the Hilbert space $H_{A} \otimes H_{B}$, spanned by the states $\vert i\rangle_{A} \otimes \vert j\rangle_{B}$ (or simply by $\vert i\rangle_{A}\vert j\rangle_{B}$ or by $\vert i j\rangle_{AB}$). Any linear combination of the basis states $\vert i\rangle_{A} \otimes \vert j\rangle_{B}$ is a state of the composite system $AB$. The pure state $\vert \psi\rangle_{AB}$ of the system can be written as\\
\begin{eqnarray}
\vert \psi \rangle_{AB} = \sum_{i,j}\:c_{ij}\:\vert i\rangle_{A} \otimes \vert j\rangle_{B},
\label{entanglement}
\end{eqnarray}\\
where $c_{ij}$'s are the complex coefficients satisfying the normalization condition $\sum_{i,j}\:\vert c_{ij}\vert^{2}=1$.\\\\
If $\vert \psi \rangle_{AB}$ factors into a normalized state $\vert \psi \rangle_{A}=\sum^{dim\:(H_{A})}_{i}\:c_{i}\:\vert i\rangle_{A}$ in $H_{A}$ and a normalized state $\vert \psi \rangle_{B}=\sum^{dim\:(H_{B})}_{j}\:c_{j}\:\vert j\rangle_{A}$ in $H_{B}$ , i.e. $\vert \psi \rangle_{AB}=\vert \psi\rangle_{A}\otimes \vert \psi \rangle_{B}$, then the state $\vert \psi \rangle_{AB}$ is called a \textit{separable state} or \textit{product state}.
If a state belonging to the Hilbert space $H_{A} \otimes H_{B}$ is not a product state, then such a state is called \textit{\textit{entangled state}}.\\\\
If $\vert \psi \rangle_{AB}$ represents a pure state of a composite system consisting of two Hilbert spaces $H_{A}$ and $H_{B}$ for the individual systems $A$ and $B$, then $\vert \psi\rangle_{AB}$ can always be written in the Schmidt form as \cite{nielsenbook}\\
\begin{eqnarray}
\vert \psi \rangle_{AB} = \sum^{k \leq min\: \lbrace dim\:H_{A}, dim\:H_{B}\rbrace}_{i}\:\sqrt{\lambda_{i}}\vert i\rangle_{A} \otimes \vert i \rangle_{B},
\label{purestateentanglement}
\end{eqnarray}\\
where $\vert i\rangle_{A}$  and $\vert i\rangle_{B}$ are two orthonormal bases of systems $A$ and $B$ respectively with the conditions $\lambda_{i}\geq 0$, $\sum \lambda_{i}=1$, $\lambda_{i}$'s being the Schmidt coefficients. If two or more Schmidt coefficients are non-zero, then the state $\vert \psi \rangle_{AB}$ is referred to as \textit{Pure entangled state}.\\\\
However, a quantum system may not always be in a pure state. In other words, it may not be possible to express it in the form given in eq. (\ref{entanglement}). In that case, the state may be observed as a mixture of states, which are not necessarily orthogonal to each other.  A mixed state of quantum systems consisting of various subsystems is supposed to represent entanglement if it is inseparable \cite{bennett1996,horodecki1998,kent1999}, i.e. cannot be written in the form \\
\begin{eqnarray}
\rho = \sum_{i} \:p_{i}\:(\rho^{A_{1}}_{i}\otimes \rho^{A_{2}}_{i} \otimes \cdots), \: p_{i} \geq 0, \:\:\sum_{i} p_{i}=1,
\label{mixedstateentanglement}
\end{eqnarray} \\
where $\rho^{A_{i}}_{i}$ , $i=1,2,...$ are states for the sub-systems $A_{i}$ $(i=1,2,...)$. Such an entangled state is known as \textit{Mixed entangled state}. \\\\
Entanglement has certain basic properties which can be categorized as follows \cite{plenioreview}:
\begin{itemize}
\item Separable states contain no entanglement.
\item All non-separable states allow some tasks to be achieved better than that by LOCC \index{LOCC}\footnote{In LOCC, LO stands for Local Operations and CC stands for Classical Communication.} alone, hence all non-separable states are entangled.
\item The entanglement of states does not increase under LOCC transformations.
\item Entanglement does not change under Local Unitary Operations.
\item There are maximally entangled states.\\\\
Any bipartite (i.e. two party) pure state which is local unitarily equivalent to \\
\begin{eqnarray}
\vert \psi^{+}_{n} \rangle = \frac{\vert 0,0 \rangle + \vert 1,1 \rangle + \cdots + \vert n-1,n-1 \rangle }{\sqrt{n}},
\label{mes}
\end{eqnarray}\\
is maximally entangled. Here $n$ is the dimension of the system involved.
\end{itemize}
Considering the two $`$two-state particles', the well known and often used two qubit maximally entangled pure states are Bell states, which are defined below \cite{bell1964,alberbook}.
\section{\textbf{Bell states and Bell-CHSH inequality:}}
Quantum superposition principle is another fundamental feature which lies at the heart of entanglement. In classical world various two-state systems are identified with one of their two possible states. As for example, a coin when flipped will always result in either of the two possible states as head $\vert h\rangle$ or tail $\vert t \rangle$, an atom can always be found in either ground $\vert g \rangle$ or excited $\vert e\rangle$ states or a photon can be found in either vertical $\vert V\rangle$ or horizontal $\vert H\rangle$ polarization states. Its quantum mechanical counterpart shows, a two-state quantum system to be found in any superposition of two possible basis states. A quantum mechanical coin is found in a state like $\frac{\vert h\rangle + \vert t\rangle}{\sqrt{2}}$, for an atom we have $\frac{\vert g\rangle + \vert e\rangle}{\sqrt{2}}$ or for a photon it is $\frac{\vert H\rangle + \vert V\rangle}{\sqrt{2}}$ and so on. As the superposition principle holds for more than one quantum system, two quantum particles can be in any superposition thereof, for example in the entangled state  $\frac{\vert hh\rangle + \vert tt\rangle}{\sqrt{2}}$ for two coins, $\frac{\vert gg\rangle + \vert ee\rangle}{\sqrt{2}}$ for two atoms or $\frac{\vert HH\rangle + \vert VV\rangle}{\sqrt{2}}$ for two photons. For two two-state particles, thus, a basis of four orthogonal maximally entangled states are defined as\\
\begin{eqnarray}
\vert \Psi^{+}\rangle = \frac{\vert 00\rangle + \vert 11\rangle}{\sqrt{2}},\nonumber\\
\vert \Psi^{-}\rangle = \frac{\vert 00\rangle - \vert 11\rangle}{\sqrt{2}},\nonumber\\
\vert \Phi^{+}\rangle = \frac{\vert 01\rangle + \vert 10\rangle}{\sqrt{2}},\nonumber\\
\vert \Phi^{-}\rangle = \frac{\vert 01\rangle - \vert 10\rangle}{\sqrt{2}}.
\label{bellstates}
\end{eqnarray}\\
The above states are known as \textit{Bell-states}.
It is important to notice here that one can still encode two bits of information, that is one has four different possibilities. But interestingly this encoding is done in such a way that none of the bits carries any well defined information on its own. All information is encoded into relational properties of the two qubits. In order to read out the information one has to have access to both qubits. The corresponding measurement is called \textit{Bell state measurement}. In eq. (\ref{bellstates}), the qubit obeys \textit{fermionic} symmetry in the case of $\vert \phi^{-}\rangle$ and \textit{bosonic} symmetry in case of the other three states. The states defined in (\ref{bellstates}) are termed as Bell-states since they maximally violate a Bell inequality \cite{bell1964}. This inequality was deduced in the context of so-called local realistic theories and gives a range of possible results for certain statistical tests on identically prepared pairs of particles. Bell showed that EPR claim implies an inequality that some quantum correlations do not satisfy. Five years later, Clauser, Horne, Shimony and Holt (CHSH) generalized Bell's inequality \cite{chsh1969}. The CHSH inequality, applied to photon polarization, allows a practical test of the EPR claim.\\\\
To understand the inequality, we consider a bipartite or two-party, (one is Alice and the other one is Bob, say) system. It is assumed that Alice can measure two quantities on her part. These can be labelled as $A$ and $A^{/}$. Bob can also measure two quantities on his part as well, labelled as $B$ and $B^{/}$. Here  $a_{i}$ and $b_{j}$ ($i,j=1,2$) are possible results of measurements of $A$ and $B$ respectively, which can take values $\pm 1$ each. The correlation between measurements $A$ and $B$ can be obtained as\\
\begin{eqnarray}
C(A, B)=\sum_{i,j}\:a_{i}\:b_{j}\:P(A,B;a_{i},b_{j}),
\label{cabchsh}
\end{eqnarray}\\
where $P(A,B;a,b)$ is the probability that measurements of $A$ and $B$ on photon pair yield the outcomes $a$ and $b$ respectively. Then the CHSH inequality is defined as\\
\begin{eqnarray}
-2\:\leq C(A,B) + C(A^{/},B) + C(A, B^{/}) - C(A^{/}, B^{/}) \:\leq 2.
\label{chshineq}
\end{eqnarray}\\
The CHSH inequality follows from the very assumption that local results exist, whether or not anyone measures them.
\section{\textbf{Cirelson's Bound:}}
Even though quantum correlations violate Bell's inequality, they satisfy weaker inequalities of similar types \cite{cirelson1980}. Let there be two spin-$\frac{1}{2}$ particles $A$ and  $B$. Let $A_{1}$, $A_{2}$, $B_{1}$ and $B_{2}$ can be interpreted as spin components along two different directions of these two particles. Then for a suitable choice of directions and for the choice of a density matrix $W$ if we define\\
\begin{eqnarray}
C_{11} + C_{12} + C_{21} - C_{22} = Tr [(A_{1}B_{1} + A_{1}B_{2} + A_{2}B_{1} - A_{2}B_{2})W]
\label{cirelson1},
\end{eqnarray}\\
then, it can be shown that\\
\begin{eqnarray}
C_{11} + C_{12} + C_{21} - C_{22} \leq 2\:\sqrt{2}.
\label{cirelson2}
\end{eqnarray}\\
The above inequality (\ref{cirelson2}) holds for arbitrary quantum observables $A_{1}$, $A_{2}$, $B_{1}$ and $B_{2}$ whereas $2\:\sqrt{2}$ is the greatest possible value for the particular linear combination of spin correlations.\\\\
Various types of mixed entangled states will now be discussed. Mixed entangled states can be categorized into two different classes. One such class is called \textit{Maximally Entangled Mixed States} and another is known as \textit{Non-maximally Entangled Mixed States}. One of the important mathematical entities which needs to be defined for discussing some of these classes of entangled mixed states, is called \textit{singlet fraction}.
\section{\textbf{Maximal singlet fraction:}} 
The concept of maximal singlet fraction of a state is defined as the maximal overlap of the state with a maximally entangled state. For the general state $\rho$, it is defined as \cite{horodecki601999,verstraete2002,bose2000}\\
\begin{eqnarray}
F (\rho) = max\:\langle \Psi\:|\: \rho \:|\:\Psi\rangle,
\label{singletfraction}
\end{eqnarray}\\
where, the maximum is taken over all maximally entangled states $\vert \Psi\rangle$. 
\section{\textbf{Maximally  and Non-maximally entangled Mixed States:}}
Those states that achieve the greatest possible entanglement for a given degree of mixedness are known as \textit{Maximally Entangled Mixed States} (in short MEMS) otherwise they are called \textit{Non-maximally entangled mixed states} (NMEMS) \cite{wei2003}. The forms of maximally entangled mixed states, however, may vary with the combination of entanglement and mixedness measures chosen for them. The notion of MEMS was actually pioneered by Ishizaka and Hiroshima \cite{ishizaka2000}. Below some forms of MEMS and NMEMS are displayed.
\subsection{\textbf{Ishizaka Hiroshima (IH) MEMS:}} The states proposed by Ishizaka \textit{et. al} are those obtained by applying any local unitary transformation to states of the following types.\\
\begin{eqnarray}
\vert S_{IH} \rangle= p_{1} \vert \Phi^{-} \rangle \langle \Phi^{-} \vert + p_{2} \vert 00 \rangle \langle 00 \vert + p_{3} \vert \Phi^{+} \rangle \langle \Phi^{+} \vert + p_{4} \vert 11 \rangle \langle 11 \vert,
\label{ishizakamems1}
\end{eqnarray}\\
where $\vert \Phi^{\pm} \rangle = \frac{\vert 01 \rangle \pm \vert 10 \rangle}{\sqrt{2}}$ are Bell states and $\vert 00 \rangle$ and $\vert 11 \rangle$ are product states orthogonal to $\vert \Phi^{\pm}\rangle$. Here $p_{i}$'s are the eigenvalues of $\vert S_{IH} \rangle$ in decreasing order $(p_{1} \geq p_{2} \geq p_{3} \geq p_{4})$, and $p_{1}+ p_{2}+ p_{3}+ p_{4} = 1$.\\\\
These include states such as \\
\begin{eqnarray}
\vert S^{/}_{IH} \rangle = p_{1} \vert \Psi^{-} \rangle \langle \Psi^{-} \vert + p_{2} \vert 01 \rangle \langle 01 \vert + p_{3} \vert \Psi^{+} \rangle \langle \Psi^{+} \vert + p_{4} \vert 10 \rangle \langle 10 \vert,
\label{ishizakamems2}
\end{eqnarray}\\
where $\vert \Psi^{\pm} \rangle = \frac{\vert 00 \rangle \pm \vert 11 \rangle}{\sqrt{2}}$ are also Bell states, and include those that are obtained by exchanging $\vert \Phi^{-} \rangle \leftrightarrow \vert \Phi^{+} \rangle$, $\vert 00 \rangle \leftrightarrow \vert 11 \rangle$ in eq. (\ref{ishizakamems1}) or $\vert \Psi^{-} \rangle \leftrightarrow \vert \Psi^{+} \rangle$, $\vert 01 \rangle \leftrightarrow \vert 10 \rangle$ in eq. (\ref{ishizakamems2}). For these states however the degree of entanglement cannot be increased further by any unitary operations. \textit{Werner state} is one such example \cite{werner401989}.
\subsection{\textbf{Werner state (as a special case of IH-class of MEMS):}}
Ishizaka and Hiroshima \cite{ishizaka2000} showed that the entanglement of formation \cite{wootters1998} (which will be discussed in the subsequent sections as a measure of entanglement) of the Werner state cannot be increased by any unitary transformation. Therefore, the Werner state \cite{werner401989} can be regarded as a maximally entangled mixed state. Werner state is a mixture of the maximally entangled state and the maximally mixed state. The state, however, can be expressed in various ways. One of the ways is to express it in terms of singlet fraction. The Werner state can thus be written in the form \\
\begin{eqnarray}
\rho_{werner} = \frac{1-F_{werner}}{3}\:I_{4} + \frac{4\:F_{werner}-1}{3}\:\vert \Phi^{-} \rangle \langle \Phi^{-}\vert
\label{wernerstate},
\end{eqnarray}\\
where, $\vert \Phi^{-} \rangle = \frac{\vert 01\rangle - \vert 10 \rangle}{\sqrt{2}}$ is the singlet state and $F_{werner}$ is the maximal singlet fraction corresponding to the Werner state. $I_{4}$ is the identity operator.
\subsection{\textbf{Munro \textit{et. al} class of MEMS:}}
Munro \textit{et. al} \cite{wei2003,munro2001} showed that there exist a class of states that have significantly greater degree of entanglement for a given mixedness than that of  Werner state. The analytical form of the MEMS class proposed by Munro \textit{et. al} is \\
\begin{eqnarray}
\rho_{mjwk} = \left(%
\begin{array}{cccc}
h(C)&0&0&\frac{C}{2}\\
0&1-2h(C)&0&0\\
0&0 &0&0\\
\frac{C}{2}&0&0&h(C)
\end{array}%
\right)
\label{munromems}, 
\end{eqnarray}\\
where\\
\beq
h(C)=\left\{\begin{array}{cccc} \frac{C}{2}
& & & C \geq \frac{2}{3}\\
\frac{1}{3} & & & C < \frac{2}{3} 
\end{array}
\right. \label{munrocon},
\eeq\\
with $C$ denoting the concurrence (which is another measure of entanglement and will be defined in the subsequent section) of $\rho_{mjwk}$ defined in eq. (\ref{munromems}).
\subsection{\textbf{Wei class of MEMS:}}
A much wider class of maximally entangled mixed states was proposed by Wei \textit{et. al} in \cite{wei2003}. The general form of two qubit density matrix comprising a mixture of the maximally entangled Bell state $\vert \Psi^{+}\rangle$ and mixed diagonal state is given by\\
\begin{eqnarray}
\rho_{wei} = \left(%
\begin{array}{cccc}
x+\frac{\gamma}{2}&0&0&\frac{\gamma}{2}\\
0& a& 0& 0\\
0&0 &b &0\\
\frac{\gamma}{2}&0&0& y+\frac{\gamma}{2}
\end{array}%
\right)
\label{weimems}, 
\end{eqnarray}\\
where, $a$, $b$, $x$, $y$ and $\gamma$ are non negative real parameters. The normalization condition gives $x+y+\gamma+a+b=1$.
\subsection{\textbf{The Werner derivative (a type of NMEMS):}}
Hiroshima and Ishizaka \cite{hiroshima2000} studied a particular class of mixed states called \textit{Werner derivative} states which are obtained by applying  non-local unitary operations on Werner state. The non-local unitary transformation $U$ on the Werner state (\ref{wernerstate}), (i.e. $\rho_{wd} = U\:\rho_{werner}\:U^{\dagger}$), transforms it in to  a new density matrix. The Werner derivative is described by the density operator\\
\begin{eqnarray}
\rho_{wd}=\frac{1-F_{werner}}{3}\:I_{4}\:+\:\frac{4\:F_{werner}-1}{3}\:\vert \psi\rangle\langle \psi\vert,
\label{wernerderivative1}
\end{eqnarray}\\
where $\vert \psi\rangle =U\:\vert \Psi^{-}\rangle = \sqrt{a}\:\vert 00\rangle + \sqrt{1-a}\:\vert 11\rangle$, with $\frac{1}{2}\:\leq\:a\:\leq\: 1$. It has been shown in \cite{hiroshima2000} that the state (\ref{wernerderivative1}) is entangled if and only if \\
\begin{eqnarray}
\frac{1}{2}\:\leq\:a\:<\:\frac{1}{2}\:\left\{ 1\:+\:\frac{\sqrt{3\:(4\:F_{werner}^{2}-1)}}{4\:F_{werner}-1\:}\:\right\}
\label{wenerderivative2}.
\end{eqnarray}
This further gives a restriction on $F_{werner}$ as $\frac{1}{2}\:<F_{werner}\:\leq 1$. It is also known that although it is generally possible to increase the entanglement of a single copy of a Werner derivative by LOCC, the maximal possible entanglement cannot exceed the entanglement of the original Werner state \cite{hiroshima2000}.\\\\
Subsequent discussions on the above classes of mixed entangled states from the view point of their non-locality structure as well as their efficacies as teleportation (one of the fascinating information processing tasks in quantum information theory) channels have been studied extensively in Chapter $3$. We now discuss some of the measures of pure and mixed state entanglement in the following sections.
\section{\textbf{Measure of pure state entanglement:}}
One of the measures of pure state entanglement is as follows:
\subsection{\textbf{Entropy of entanglement:}}
Let Alice ($A$) and Bob ($B$) share a pure entangled state $\vert \psi\rangle_{AB}$. Quantitatively, a pure state's entanglement is conveniently measured by its \textit{entropy of entanglement} as \cite{nielsenbook,adhikarithesis}\\
\begin{eqnarray}
E\:(\vert \psi_{AB}\rangle\langle \psi_{AB}\vert) = S(\rho_{A}) = S(\rho_{B}).
\label{entropypure}
\end{eqnarray}\\
Here $S(\rho)=\:-Tr\:(\rho\:\log_{2}\:\rho)$ is the von-Neumann entropy and $\rho_{A}=Tr_{A}\:(\vert \psi\rangle_{AB}\langle \psi \vert)$, $\rho_{B}=Tr_{B}\:(\vert \psi\rangle_{AB}\langle \psi \vert)$ denote the reduced density matrices obtained by tracing out the whole system's pure state density matrix $\vert \psi\rangle_{AB}\langle \psi \vert$ over Bob's and Alice's degrees of freedom respectively.
\section{\textbf{Measures of mixed state entanglement:}}
It is already known that entanglement cannot be created using LOCC operations. Mintert \textit{et. al} showed that the quantities that do not increase under LOCC operations \cite{mintertreview2005}, can be used to quantify entanglement. Any scalar valued function that satisfies this criterion is called \textit{entanglement monotone}.\\\\
Entanglement monotones that satisfy certain additional axioms are called \textit{entanglement measures} and is generally denoted by $E$. Such potential axioms can be listed below. For any mixed state $\varrho$,
\begin{itemize}
\item $E(\varrho)$ vanishes exactly for separable states.
\item The entanglement of several copies of a state adds up to $n$ times the entanglement of a single copy. Symbolically, $E(\varrho^{\otimes\:n})=n\:E(\varrho)$.
\item The entanglement of two states ($\varrho$ and $\varrho^{/}$) is not larger than the sum of the entanglement of both individual states. Symbolically, $E(\varrho\otimes\varrho^{/})\:\leq E(\varrho) + E(\varrho^{/})$.
\item Entanglement measure $E$ satisfies the convexity property, i.e. $E(\lambda\:\varrho + (1-\lambda)\:\varrho^{/})\:\leq\:\lambda\:E(\varrho) + (1-\lambda)\:E(\varrho^{/})$, where, $0\:\leq\lambda\:\leq\: 1$.
\end{itemize}
There are many measures of mixed state entanglement which satisfy the above properties and are often used. A few of such measurements are mentioned below \cite{adhikarithesis,wei2003}.
\subsection{\textbf{Entanglement of formation:}}
\textit{Entanglement of formation} \cite{wootters1998} quantifies the amount of entanglement necessary to create the entangled state. It is defined as\\
\begin{eqnarray}
E_{F}(\rho) = \min_{\lbrace p_{i},\vert\psi_{i}\rangle\rbrace}\:\sum_{i}\:p_{i}\:E(\vert \psi_{i}\rangle \langle \psi_{i}\vert)
\label{eef1},
\end{eqnarray}
where the minimum is taken over those probabilities $\lbrace p_{i}\rbrace$ and pure states $\lbrace \vert\psi_{i}\rangle\rbrace$ that, when taken together, reproduce the density matrix $\rho=\sum_{i}\:p_{i}\:\vert \psi_{i}\rangle\langle \psi_{i}\vert$. $E(\vert \psi_{i}\rangle \langle \psi_{i}\vert)$ is the entropy of entanglement.\\\\
For two qubit systems, $E_{F}$ can be expressed explicitly as \\
\begin{eqnarray}
E_{F}(\rho) = h\:\left\{\:\frac{1+\sqrt{1-C^{2}(\rho)}}{2}\:\right\},
\label{eef2}
\end{eqnarray}\\
where, $h(x)=\:-x\:\log_{2}\:x-(1-x)\:\log_{2}\:(1-x)$ is Shannon's entropy function and $C(\rho)$ is the \textit{concurrence} of the state $\rho$. The quantity $C^{2}(\rho)=\tau$ is sometimes called $`$concurrence squared' or $`$tangle' \cite{coffman2000}. The entanglement of formation $E_{F}$ is a strictly monotonic function of $\tau$, the maximum of $\tau$ corresponds to the maximum of $E_{F}$. Hence $`$tangle' can also be considered as direct measure of entanglement. For a maximally entangled pure state, $\tau=1$ while for an un-entangled state, $\tau=0$.
\subsection{\textbf{Concurrence:}}
\textit{Concurrence} is a non-negative real number. For a bipartite mixed state $\rho$ (for dimension   $2\otimes2$ or $2\otimes3$), it is defined in \cite{wootters1997,wootters1998}  as\\
\begin{eqnarray}
C(\rho)=\max\,(0, \sqrt\la_1-\sqrt\la_2-\sqrt\la_3-\sqrt\la_4),
\label{concurrence}
\end{eqnarray} \\
where the $\la_{i}$'s, ($i=1,2,3,4$), are the  eigenvalues of $\rho\,\tilde{\rho}$ in decreasing order. The spin flipped density matrix $\tilde{\rho}$ is defined as $\uu$ . Since $E_{F}$ is a monotonic function of $C$ and $C$ ranges from zero to one (i.e. for un-entangled to maximal entanglement), so the concurrence $C$ is also a measure of entanglement.
\subsection{\textbf{Entanglement cost:}}
An empirical measure associated with the entanglement of formation is the \textit{entanglement cost} \cite{bennett1996}, denoted by $E_{c}$. This is defined as follows\\
\begin{eqnarray}
E_{c}\:(\rho) = \lim_{n\rightarrow \infty}\:\frac{E_{F}(\rho^{\otimes\:n})}{n}.
\label{entanglementcost}
\end{eqnarray}\\
This is the asymptotic value of the average entanglement of formation. $E_{c}$ is generally difficult to calculate.
\subsection{\textbf{Relative entropy of entanglement:}}
The \textit{relative entropy of entanglement} \cite{audenaert2005,miranowicz2004,plenio1998} is based on distinguish-ability and geometrical distance. The idea is basically to compare a given quantum state $\sigma$ of a pair of particles with separable states. The \textit{relative entropy of entanglement} of a given state $\sigma$ is denoted by $E_{re}\:(\sigma)$ and is defined as\\
\begin{eqnarray}
E_{re}\:(\sigma) = \min_{\rho \in M}\: D(\sigma \parallel \rho).
\label{re}
\end{eqnarray}\\
Here $M$ denotes the set of all separable states and $D$ can be any function that describes a measure of separation between two density operators. A particular form of the function $D$ is the relative entropy which is defined as $S(\sigma \parallel \rho)=\:Tr\:(\sigma\:\ln\:\sigma -\sigma\:\ln\:\rho)$.
\subsection{\textbf{Negativity:}}
The concept of \textit{negativity} \cite{adhikarithesis} of a state is closely related to the well-known Peres-Horodecki criterion for the separability of a state \cite{horodecki1996,peres1996}.\\\\
Peres-Horodecki criterion states that a necessary and sufficient condition for the state $\rho$ of two spins to be inseparable is that at least one of the eigenvalues of the partially transposed operator, defined as $\rho^{T_{2}}_{m\mu\:,\:n\nu} = \rho_{m\nu\:,\:n\mu}$, is negative. This is equivalent to the condition that at least one of the following two determinants of eq. (\ref{peres1}) is negative.\\
\begin{eqnarray}
W_{3}=\left|%
\begin{array}{ccc}
\rho_{00\:,\:00}&\rho_{01\:,\:00}& \rho_{00\:,\:10}\\
\rho_{00\:,\:01}&\rho_{01\:,\:01}&\rho_{00\:,\:11}\\
\rho_{10\:,\:00}&\rho_{11\:,\:00}&\rho_{10\:,\:10}\\
\end{array}%
\right|~~~~ \mbox{and}~~~~
W_{4}=\left|%
\begin{array}{cccc}
\rho_{00\:,\:00}&\rho_{01\:,\:00}& \rho_{00\:,\:10}&\rho_{01\:,\:10}\\
\rho_{00\:,\:01}&\rho_{01\:,\:01}&\rho_{00\:,\:11}&\rho_{01\:,\:11}\\
\rho_{10\:,\:00}&\rho_{11\:,\:00}&\rho_{10\:,\:10}&\rho_{11\:,\:10}\\
\rho_{10\:,\:01}&\rho_{11\:,\:01}&\rho_{10\:,\:11}&\rho_{11\:,\:11}\\
\end{array}%
\right|, \nonumber\\\nonumber\\
\label{peres1}
\end{eqnarray}\\
and the determinant of eq. (\ref{peres2}) is non-negative.\\
\begin{eqnarray}
W_{2}=\left|%
\begin{array}{cc}
\rho_{00\:,\:00}&\rho_{01\:,\:00}\\
\rho_{00\:,\:01}&\rho_{01\:,\:01}\\
\end{array}%
\right|
\label{peres2}.
\end{eqnarray}\\
If a state is separable (i.e. not entangled), then the partial transpose of its density matrix is again a valid state i.e. it is positive semi-definite \footnote{A linear self adjoint map $\Lambda: \Re(H_{B})\rightarrow \Re(H_{C})$ is called positive semi-definite if for all $\rho \in \Re(H_{B})$, (where $H_{B}, \: H_{C}$ are the Hilbert spaces and $\Re(H_{i})_{i\:=B,\:C}$ are the set of linear operators acting on $H_{i}$), with $\rho\:\geq\: 0\Rightarrow \Lambda(\rho)\geq 0$. The map $\Lambda$ is Positive definite if $\Lambda(\rho)> 0$.}. It also turns out that  the partial transpose of a non-separable state may have one or more negative eigenvalues \cite{wei2003}.\\\\
The \textit{negativity} of a state, however, \cite{miranowicz2004,audenaert2001} indicates the extent to which a state violates the positive partial transpose (separability) criterion. The negativity of the state $\rho$ is defined as follows\\
\beq
N^{\rho} = 2\:\mbox{max}\:(0\:,\:-\lambda_{neg}),
\label{negativity1}
\eeq\\
where $\lambda_{neg}$ is the sum of the negative eigenvalues of $\rho^{T_{B}}$. In $2\otimes2$ systems, it can be shown that the partial transpose of the density matrix can have at most one negative eigenvalue. It was proved later that negativity is an entanglement monotone and hence a good measure of entanglement \cite{wei2003}. For mixed states, Eisert and Plenio \cite{eisert1999} conjectured that \textit{negativity never exceeds concurrence} and the conjecture was proved later by Audenaert \textit{et. al} in \cite{audenaertarxiv}\\\\
For higher dimensions, the negativity can be generalized as \cite{slee2003}\\
\beq
E_{N}=\frac{\parallel \rho^{T_{A}} \parallel - 1}{n-1},
\label{negativity2}
\eeq\\
where, $n$ is the smaller of the dimensions of the bipartite system.\\\\
Apart from all these, there are various other measures of entanglement like, geometric measure of entanglement, comb monotones and many more, all of which have been discussed in details in \cite{guhnereview,horodeckireview2009}. \\\\
The focus is now on some different types of quantities for studying state's mixedness. These mixedness measures have been used throughout this dissertation as and when required. 
\section{\textbf{Measures of mixedness of the states :}}
The two fundamental measures are \textit{von-Neumann entropy} and \textit{linear entropy} \cite{wei2003}. Although von-Neumann entropy has a natural significance stemming from its connection with statistical physics, linear entropy is comparatively easier to calculate. It is also a recognised fact that, the two measures show similar trend, if they are used for those density matrices that are completely mixed \footnote{The completely mixed state for one qubit is $\frac{I}{2}$ and for two qubits is $\frac{I\otimes I}{4}$.}.
\subsection{\textbf{von-Neumann entropy:}}
The standard measure of randomness of a statistical ensemble, described by a density operator, is von-Neumann entropy. If we consider a state described by a density matrix $\rho$ in the Hilbert space of dimension $n$, where $\lambda_{i}$'s are the eigenvalues of $\rho$, then the von-Neumann entropy is denoted by $S_{V}(\rho)$ and is defined by\\
\begin{eqnarray}
S_{V}(\rho) = -\: Tr(\rho\:\log\:\rho) = \:-\sum_{i}\:\lambda_{i}\:\log\:\lambda_{i},
\label{vonneumannentro}
\end{eqnarray}\\
where, $\log$ is taken to the base $n$. For pure states we have $S_{V}(\rho)=0$ whereas for completely mixed states we have $S_{V} (\rho)=1$. Computation of von-Neumann entropy requires the full knowledge of the eigenvalue spectrum of the state.
\subsection{\textbf{Linear entropy:}}
The measure of linear entropy is based on the purity of a state, $P = Tr\:(\rho^{2})$. $P$ ranges from $1$ (for a pure state) to $\frac{1}{n}$ (for a completely mixed state with dimension $n$). The linear entropy is denoted by $S_{L}$ and is defined as \\
\begin{eqnarray}
S_{L}(\rho) = \frac{n}{n-1}\left\{1-Tr(\rho^{2})\right\}.
\label{linearentropy1}
\end{eqnarray} \\
The lower limit of $S_{L}$ is  $0$ (for a pure state) and the upper limit is $1$ (for a maximally mixed state). For bi-partite systems, the linear entropy, can thus be explicitly expressed as\\
\begin{eqnarray}
S_{L}(\rho) = \frac{4}{3}\left\{1-Tr(\rho^{2})\right\}.
\label{linearentropy2}
\end{eqnarray}
It is to be noted in this context that, there are some intrinsic connections between entanglement and mixedness of the states. The states with such connections can be termed as \textit{frontier} states. These states are maximally entangled for a given value of mixedness or they are maximally mixed for a given value of entanglement \cite{wei2003}. Maximally entangled mixed states (MEMS) are of that type. Several MEMS have been already discussed in this chapter earlier.
\section{\textbf{Distance measurements for quantum states:}}
Another important concept to study quantum information science is about the closeness between two quantum states. Basically during the preparation of entangled states through copying (or quantum cloning) one often becomes interested about the distance between the original state and the copied states characteristically, i.e. how perfectly the state has been copied, since it is a well known fact that quantum states cannot be perfectly cloned \cite{wootters1982}. Below a few distance measurements are presented \cite{nielsenbook}.
\subsection{\textbf{Trace distance:}}
If there are two quantum states $\rho$ and $\sigma$, then the \textit{trace distance} between them is defined as \\
\begin{eqnarray}
D (\rho\:,\:\sigma) = \frac{1}{2}\:tr\:|\:\rho-\sigma\:|.
\label{tracedistance}
\end{eqnarray}\\
It is to be noted here that for an arbitrary operator $A$, we define $|\:A\:|=\sqrt{A^{\dagger}\:A}$, to be the positive square root of $A^{\dagger}\:A$\footnote{$A^{\dagger}$ is the Hermitian conjugate or adjoint of the matrix $A$ i.e. $A^{\dagger}=(A^{T})^{*}$.}.
\subsection{\textbf{Fidelity:}}
For two states, represented by density matrices, $\rho$ and $\sigma$, the fidelity is defined as\\
\begin{eqnarray}
F(\rho,\sigma) = tr\:\sqrt{\:\rho^{\frac{1}{2}}\:\sigma\:\rho^{\frac{1}{2}}}.
\label{fidelity1}
\end{eqnarray}\\
The fidelity between a pure state $\vert \psi\rangle$ and an arbitrary state $\rho$, is however defined as\\
\begin{eqnarray}
F(\:\vert \psi\rangle,\:\rho) = \:\tr\:\sqrt{\langle\:\psi\:|\:\rho\:|\:\psi\:\rangle}.
\label{fidelity2}
\end{eqnarray}\\
There are certain properties which are satisfied by fidelity and are summarized below.
\begin{itemize}
\item The fidelity is symmetric in its inputs, i.e. $F(\rho\:,\:\sigma)=F(\sigma\:,\:\rho)$.
\item  Generally, $0\:\leq\:F(\rho\:,\:\sigma)\:\leq\:1$. Now $F(\rho\:,\:\sigma)=0$ if and only if $\rho$ and $\sigma$ have support on orthogonal subspaces. Also, $F(\rho\:,\:\sigma)=1$ if and only if $\rho=\sigma$.
\item  $F(\rho\:,\:\sigma)=\max_{\vert \varphi\rangle}\:|\langle\:\psi\:|\:\varphi\:\rangle\:|$, where $\vert \psi\rangle$, is any fixed purification of $\rho$. The maximization is taken over all purifications of $\sigma$. The result is known as \textit{Uhlmann's theorem} \cite{uhlmann,joz1994}.
\item  Fidelity is strongly concave, which means $F\:(\sum_{i}\:p_{i}\:\rho_{i},\:\sum_{i}\:q_{i}\:\sigma_{i})\:\geq\:\sum_{i}\:\sqrt{p_{i}\:q_{i}}\:F\:(\rho_{i}\:,\:\sigma_{i}\:)$.
\item \textit{Entanglement fidelity}, $F\:(\rho\:,\:\varepsilon)$ is a measure of how well entanglement is preserved during a quantum mechanical process, starting with the state $\rho$ of a system $Q$, which is assumed to be entangled with another system $R$, and applying the quantum operation $\varepsilon$ to system $Q$.
\end{itemize}
\subsection{\textbf{Hilbert - Schmidt norm:}}
The Hilbert - Schmidt distance \cite{filip2002} between two quantum states $\rho$ and $\sigma$ is defined by\\
\begin{eqnarray}
D_{HS}\:(\rho\:,\:\sigma) = \parallel\:\rho-\sigma\:\parallel = Tr\:(\rho - \sigma)^{2}.
\label{hilbertschmidt}
\end{eqnarray}\\
Hilbert-Schmidt distance serves as a good measure of quantifying the distance between pure states as it is easier to calculate. It is conjectured that the Hilbert-Schmidt norm is a reasonable candidate of a distance to generate an entanglement measure \cite{plenioconjecture}.
\subsection{\textbf{Bures distance:}}
For two states $\rho$ and $\sigma$, Bures' distance is measured by the following formula\\
\begin{eqnarray}
d_{B}\:(\rho\:,\:\sigma) = \sqrt{2}\:\left\{\:1-\:Tr\:\sqrt{\rho^{\frac{1}{2}}\:\sigma\:\rho^{\frac{1}{2}}}\:\right\}^{\frac{1}{2}}.
\label{bures}
\end{eqnarray}\\
It is to be noted in this context that, if no a priori information about the \textit{in} state of the original system is available, then it is reasonable to require that all pure states should be copied equally well. This can be done by designing a quantum copier such that the distances between density operators of each system at the output $\hat{\rho}^{(out)}$ (where $j=a,b$) and the ideal density operator $\hat{\rho}^{(id)}$ which describes the \textit{in} state of the original mode are input state independent.\\\\
With respect to Bures' distance, one can say, then the quantum copier should be such that\\
\begin{eqnarray}
d_{B}(\hat{\rho_{j}}^{(id)},\hat{\rho_{j}}^{(out)}) = \mbox{constant},~~~~~~~~~ j= a,b.
\label{condition}
\end{eqnarray}\\
Let us now briefly discuss entanglement from another standpoint. So far we have talked about entanglement, its characteristics and its quantification. One another important aspect of entanglement is about its detection.
\section{\textbf{Entanglement witness:}}
\textit{Entanglement witness} is a type of mathematical tool \cite{guhnereview}, that constitutes a very general method to distinguish entangled states from the separable ones. Entanglement witness relies on Hahn-Banach theorem \cite{hahnbanach}. Entanglement witness or simply witness is  defined as follows.\\\\
An observable $W$ is called an witness, if
\begin{itemize}
\item $Tr\:(W\:\varrho_{s})\:\geq\:0$ for all separable $\varrho_{s}$.
\item $Tr\:(W\:\varrho_{e})\:<\:0$, for at least one entangled $\rho_{e}$.
\end{itemize}
Thus if one measures $Tr\:(W\:\varrho)\:<\:0$, one knows for sure that the state $\varrho$ is entangled. We call a state with $Tr\:(W\:\varrho)\:<\:0$ to be detected by $W$. It has been proved in \cite{guhnereview} that $``$\textit{for each entangled state $\varrho_{e}$ there exists an entanglement witness detecting it''}. Elaborate discussions on entanglement witness can be found in \cite{guhnereview,horodeckireview2009}.\\\\
We shall now talk about a few important information processing protocols below.\\\\
\section{\textbf{Information processing protocols}}
\textit{$``$All things physical are information theoretic in origin and this is a participatory universe...Observer participancy gives rise to information; and information gives rise to Physics."}
\begin{flushright}
- John Archibald Wheeler
\end{flushright}
\textit{$``$My greatest concern was what to call it. I thought of calling it information, but the word was overly used, so I decided to call it uncertainty. When I discussed it with John von-Neumann, he suggested me to call it entropy''.}
\begin{flushright}
- Claude Elwood Shannon
\end{flushright}
\vskip1cm
\subsection{\textbf{Introduction:}}
The problem of sending classical information by encoding the message into letters in alphabet,  or through speech, or via string of bits or by any other known classical means defines the domain of classical information theory. The communication channels through which the classical messages are en-route, operate in accordance with the classical laws of physics. Quantum information theory is however motivated by the study of communication channels which have a much wider domain of applications. Basically, the laws of physics and in particular, the laws of quantum mechanics limit one's ability to process information increasingly faster and cheaper using present day solid state technologies. The question was whether the strange world of quantum phenomena can be exploited in information theory in an effective way. The answer is YES ! It turns out that computer technology and communication theory using quantum effects have remarkable consequences. The practical sense of doing information theory through quantum effects means either encoding of information into the spin of an electron or encoding it into the polarization of a photon and sending these information from sender to the receiver through quantum channels. In due process, various interesting protocols like teleportation \cite{bennett1993}, dense coding\cite{bennett1992}, secret sharing\cite{hillery1999}, cryptography\cite{bennett1984}, cloning\cite{wootters1982,buzek1996} and many more emerged. Entanglement made things more interesting for physicists to work out with these procedures.\\\\
The two most important and contrasting methods in the arena of quantum information theory are teleportation and dense coding proposed by Charles H. Bennett and group \cite{bennett1993,bennett1992}. Quantum teleportation is the transmission of qubits by classical information, but contrarily dense coding is the transmission of classical bits by qubits. In fact, R. F. Werner proved that $``$\textit{under the condition of tightness\footnote{By tightness, Werner meant in \cite{werner2001} that the classical capacity of the quantum channel is exactly doubled by dense coding, and teleportation requires twice as much classical channel capacity as the quantum capacity of the channel set up by this scheme.} and with the maximally entangled states, quantum teleportation and dense coding have one-to-one correspondence, which means that any teleportation scheme works as a dense coding scheme and vice versa''}\cite{werner2001}.\\\\
These two protocols will remain the central point of investigations in chapters $3$ and $4$.
\subsection{\textbf{Teleportation:}}
Teleportation is one of the fundamental of all the other protocols designed in quantum information theoretic science \cite{bennett1993,danceofphoton,physicsinf}.
Let us suppose Alice and Bob are two persons where Alice has a particle in some quantum state that, in general, is unknown to her. She wants to communicate this particle to Bob in the sense that she actually wants Bob to receive a particle exactly in that state, in which her particle was intially in. Measuring the particle on Alice's side and telling Bob the result won't help as any sort of measurement by Alice on her part will change the original state of the particle which she actually wanted to communicate. So Alice and Bob together decide to generate for themselves auxiliary pairs of entangled particles. Alice gets say $A$ twin from each auxiliary pair and Bob gets the $B$ twin. The twin particles $A$ and $B$ are pairwise entangled, which means that if measured in the same way, the particles will exhibit the same result, that they will turn out to be identical. This entanglement connection between the two twin particles is the \textit{$``$spooky action at a distance''}, which Einstein did not believe in. This generated pair between Alice and Bob will however be the quantum channel shared by them. Bennett \textit{et. al} in their paper \cite{bennett1993} assumed this quantum channel shared by Alice and Bob to be EPR pair which are in turn can be any one of the four possible Bell states. Let $X$ be the original particle which Alice actually wants to send to Bob ($X$ is unknown to Alice). The unknown state, which is to be teleported, can be represented as $a\:\vert 0\rangle_{X} + b\:\vert 1\rangle_{X}$, with the normalization condition $|\:a\:|^{2}+|\:b\:|^{2}=1$. Alice now entangles $X$ with her twin particle $A$ of the EPR pair. Alice's entangling measurement is called Bell state measurement. The entangling here actually means that the unknown state $X$ loses its own individual properties. The state of the two particles $X$ and $A$ will turn out to be identical if they are measured. Neither the unknown particle $X$ nor Alice's twin particle $A$ from the EPR pair have any features of their own left, once they become entangled with each other. Now the question is how can Alice send Bob the quantum state of the unknown particle she possesses? The answer lies in the fact that an ancillary pair of entangled particles (EPR pair) will be shared as quantum channel by Alice and Bob. We consider the shared channel between Alice and Bob to be $\frac{\vert 01\rangle_{AB} - \vert 10\rangle_{AB}}{\sqrt{2}}$. The important characteristics of such an entangled state is as follows: if a measurement on one of the particles projects it on to a certain state, that can be any normalized linear superposition of $\vert 0\rangle$ and $\vert  1 \rangle$, the other state has to be in the orthogonal state. Thus the complete state of the three particles can be expressed as\\
\begin{eqnarray}
\vert \Lambda\rangle_{XAB} = \frac{1}{2}\:\lbrace \vert \phi^{-}\rangle_{XA}(-a\vert 0\rangle_{B}-b\vert 1\rangle_{B})+\vert \phi^{+}\rangle_{XA}(-a\vert 0\rangle_{B}+b\vert 1\rangle_{B})\nonumber\\
\vert \psi^{-}\rangle_{XA}(a\vert 1\rangle_{B}+b\vert 0\rangle_{B})+\vert \psi^{+}\rangle_{XA}(a\vert 1\rangle_{B}-b\vert 0\rangle_{B})\rbrace,
\label{completestate1}
\end{eqnarray}\\
where, $\vert \phi^{+}\rangle, \vert \phi^{-}\rangle, \vert \psi^{+}\rangle, \vert \psi^{-}\rangle$ are the Bell states as defined in eq. (\ref{bellstates}). After this Alice performs a Bell state measurement on particles $X$ and $A$ i.e. she projects her two particles on to one of the four Bell states. As a result of the measurement, Bob's particle will be found in a state which is directly related to the initial state. For example, if the result of Alice's Bell state measurement is $\vert \psi^{-}\rangle_{XA}$, then the particle $B$ in the hands of Bob is in the state $a\vert 1\rangle_{B}+b\vert 0\rangle_{B}$. All that Alice has to do now is to inform Bob about her measurement result and Bob consequently can perform appropriate unitary transformation on particle $B$ in order to obtain the initial state of particle $X$. Thus Bob's twin particle $B$ now ends up with the properties of the original particle $X$. We say that, all the features of $X$ have been teleported over to Bob. The procedure of teleportation is not fully quantum in the sense that at one point of time Alice does need to communicate with Bob classically to tell him about her measurement outcome. This classical communication may be a simple telephonic conversation with Bob. Thus to achieve an accurate teleportation in all cases, Alice needs to tell Bob about the outcome of her measurement classically. Bob, after knowing from Alice, applies the required rotation operator to transform the state of his particle $B$ into a replica of $X$. Alice, on the other hand, is left with particles $X$ and $A$ in one of the states, either $\frac{1}{\sqrt{2}}\:(\vert 01\rangle_{XA} \pm \vert 10\rangle_{XA})$ or $\frac{1}{\sqrt{2}}\:(\vert 00\rangle_{XA} \pm \vert 11\rangle_{XA})$, without any trace of the original state $X$. Teleportation is a kind of linear operation applied to the quantum state $X$. It not only works for pure entangled states, but can be worked out with mixed entangled states too \cite{verstraete2003,albeverio2003,senshi2001,leekim2000}.
\subsection{\textbf{Dense coding:}}
Dense coding involves two parties Alice and Bob, who are distant from one another. Alice wants to transmit some classical information to Bob. She is in possession of two classical bits say, $0$ and $1$ but is allowed to send only a single qubit to Bob. Classically, there are four possible polarisation combinations for a pair of such particles; viz. $00$, $01$, $10$, and $11$ and consequently two classical bits on each of these types. The scheme for quantum dense coding, theoretically proposed by Bennett and Wiesner \cite{bennett1992} utilises entanglement between two qubits, each of which individually has two orthogonal states, $\vert 0\rangle$ and $\vert 1\rangle$. In order to send a classical message to Bob, Alice uses quantum particles, all prepared in the same state by some source. Alice translates the bit values of the message by either leaving the state of the qubit unchanged or flipping it to the other orthogonal states and Bob, consequently, will observe the particle in one or the other state. That means that Alice can encode one bit of information in a single qubit. For Alice to avoid errors, the states arriving at Bob have to be distinguishable, which is only guaranteed when orthogonal states are used.  The particle which Alice gets from the source is entangled with another particle, which was sent directly to Bob. The two particles are in one of the four possible Bell states. Let it be $\vert \Phi^{-}\rangle$, say. The particle feature of Bell basis is that, if one of the two entangled particles is manipulated, then it suffices to transform to any other of the four Bell states. Alice, now can perform, any of the following four possible transformations. If she wishes to send the bit string $00$ to Bob, she does nothing at all to her qubit. If Alice wishes to send $01$, she applies the phase flip $Z$ to her qubit. For sending the bit string $10$ Alice needs to apply quantum NOT gate, $X$, to her qubit. Again if she wants to send $11$ to Bob, $iY$ gate is applied by Alice to her qubit\footnote{Different gates are discussed in Appendix}. In this way, Alice, interacting with only a single qubit, is able to transmit two bits of information to Bob. Of course, two qubits are involved in the protocol, but Alice never needs interaction with the second qubit. Classically, the task Alice accomplishes would have been impossible had she only transmitted a single classical bit. This scheme enhances the information capacity of the transmission channel to two bits compared to the classical maximum of one bit. While it is clear that this scheme enhances the information capacity of the transmission channel accessed by Bob to two bits, it is noticed that the channel carrying the other photon transmits $0$ bit of information, implying the fact that the total transmitted information does not exceed two bits.\\
\subsection{\textbf{Controlled dense coding:}} 
When we involve three parties in dense coding scheme, any one of them can act as controller who controls the situation, whether the remaining two parties may be able to share a maximally entangled channel between them and subsequently communicating bits of information. Such a protocol is known as Controlled Dense Coding. This was first proposed by Hao, Li and Guo in \cite{hao2001}, where initially the three parties shared a maximally tripartite entangled state among them. That formed the basis of further investigations which will be discussed in Chapter $5$.
$GHZ$ state is a tripartite (or three qubit) maximally entangled pure state, first defined by Greenberger \textit{et. al} \cite{ghzstate} in their paper titled $``$Bell's theorem without inequalities'', and is defined as follows\\
\begin{eqnarray}
\vert GHZ\rangle_{ABC} = \frac{1}{\sqrt{2}}\:(\vert 000\rangle_{ABC} + \vert 111\rangle_{ABC}),
\label{ghzstate}
\end{eqnarray}\\
where $A$, $B$ and $C$ are three parties Alice, Bob and Charlie. This state can be generated in the laboratory using entanglement swapping starting from three down converters \cite{zukowski1995} or can be demonstrated experimentally using two pairs of entangled photons \cite{pan1999}.\\\\
Given the role of $`$sender' to Alice, $`$receiver' to Bob and making Charlie as the $`$controller', the purpose of controlled dense coding, however, was that, Alice would send classical information to Bob, while Charlie would supervise  and not only decide what type of two qubit channel $`$sender' and $`$receiver' share between them, but also would control the transmission of bits between them. In order to transmit more information through the quantum channel, entanglement is needed. The trio must share a kind of tripartite entangled state among themselves before hand just like the state defined in eq. (\ref{ghzstate}).\\\\
Let us suppose that Charlie measures his qubit $C$ under the basis \\
\begin{eqnarray}
\lbrace\:\vert +\rangle_{C} = \cos\:\theta\:\vert 0\rangle_{C} + \sin\:\theta\:\vert 1\rangle_{C}\:,\:\nonumber\\
\vert -\rangle_{C} = \sin\:\theta\:\vert 0\rangle_{C} - \cos\:\theta\:\vert 1\rangle_{C}\:\rbrace.
\label{charliebasis}
\end{eqnarray}\\
Here it has been supposed that $|\:\sin\:\theta\:|\:\leq\:|\:\cos\:\theta\:|$. In the new basis $\lbrace\:\vert +\rangle_{C}\:,\:\vert -\rangle_{C}\:\rbrace$, the GHZ state (\ref{ghzstate}) can be re-written as\\
\begin{eqnarray}
\vert GHZ\rangle_{ABC} =\frac{1}{\sqrt{2}}\:(\:\vert \varphi\rangle_{AB}\:\vert +\rangle_{C}\: +\:\vert \phi\rangle_{AB}\:\vert -\rangle_{C}\:),
\label{ghzinnewbasis}
\end{eqnarray}\\
where \\
\begin{eqnarray}
\vert \varphi\rangle_{AB} = \cos\:\theta\:\vert 00\rangle_{AB} + \sin\:\theta\:\vert 11\rangle_{AB},\nonumber\\
\vert \phi\rangle_{AB} = \sin\:\theta\:\vert 00\rangle_{AB} - \cos\:\theta\:\vert 11\rangle_{AB}.
\label{varphiphi}
\end{eqnarray}\\
It is obvious that, the projective measurement of qubit $C$ by Charlie gives two readouts, either $\vert +\rangle_{C}$ or $\vert -\rangle_{C}$; each occurring with equal probability i.e. $\frac{1}{2}$. The state of qubits $A$ and $B$ collapses to $\vert \varphi\rangle_{AB}$ if the readout is $\vert +\rangle_{C}$ or it collapses to $\vert \phi\rangle_{AB}$, when the readout is $\vert -\rangle_{C}$. Now since $\vert \varphi\rangle_{AB}$ or $\vert \phi\rangle_{AB}$ is not maximally entangled, so the success probability of dense coding, with either of these two states as channels, shall be less than $1$.\\\\
At this juncture, Hao \textit{et. al} presented two schemes for dense coding in \cite{hao2001}. One of them will be discussed now.\\\\
Charlie sends his measurement result to Alice so that the latter has the information about which general entangled state she shares with Bob while Bob has no inkling of what he shares with Alice, since Charlie decides to inform only one of them. What Alice does now is that she introduces an auxiliary qubit, say $\vert 0\rangle_{aux}$, and performs a collective unitary operation under the basis $\lbrace \vert 00\rangle_{A\:aux}\:,\: \vert 10\rangle_{A\:aux}\:,\: \vert 01\rangle_{A\:aux}\:,\: \vert 11\rangle_{A\:aux}\:\rbrace$ on her qubit $A$ and the auxiliary qubit as well. If the general state shared by Alice and Bob be $\vert \varphi\rangle_{AB}$, with proper choice of unitary operator $U$, it was shown in \cite{hao2001} that collective operation $U\:\otimes\:I$ transformed the state $(\cos\:\theta\:\vert 00\rangle_{AB} + \sin\:\theta\:\vert 11\rangle_{AB})\:\otimes\:\vert 0\rangle_{aux}$ to\\
\begin{eqnarray}
\vert \psi\rangle_{ABaux} = \sqrt{2}\:\sin\:\theta\:\left\{\:\frac{1}{\sqrt{2}}\:(\vert 00\rangle_{AB} + \vert 11\rangle_{AB})\:\right\}\:\otimes\:\vert 0\rangle_{aux}\nonumber\\
+\cos\:\theta\:\sqrt{1-\frac{\sin^{2}\:\theta}{\cos^{2}\:\theta}}\:\vert 10\rangle_{AB}\:\otimes\:\vert 1\rangle_{aux},
\label{newstate}
\end{eqnarray}\\
where, $\vert 10\rangle_{AB}$ is the un-entangled state of the two qubits $A$ and $B$. Now when Alice performs projective measurement on auxiliary qubit with two possible outcomes $\lbrace\: \vert 0\rangle_{aux}, \vert 1\rangle_{aux}\:\rbrace$, and if in the due process she gets the outcome as $\vert 1\rangle_{aux}$, then it is clear from eq. (\ref{newstate}) that the procedure fails as Alice knows that her and Bob's qubit are un-entangled, whereas if the outcome of her measurement is $\vert 0\rangle_{aux}$, the two qubits $A$ and $B$ are maximally entangled. After performing one of the four operations $\lbrace\:I\:,\:\sigma_{x}\:,\: \sigma_{y}\:,\:\sigma_{z}\:\rbrace$, she sends her qubit to Bob. Bob, subsequently, carries out controlled NOT (CNOT) operation on his qubit to get $2$ bits of information from Alice. Hence, on an average \\
\begin{eqnarray}
I_{trans} = 1 + 2\:|\:\sin\:\theta\:|^{2}
\label{transmitted}
\end{eqnarray}\\
bits are transmitted from Alice to Bob at the cost of one GHZ state defined in (\ref{ghzstate}). Since the maximal value of $|\:\sin\:\theta\:|$ is $\frac{1}{\sqrt{2}}$, which corresponds to the maximally entangled Bell state $\vert \Psi^{+}\rangle_{AB}$, the success probability is unity and two bits are transmitted by one qubit.\\
\subsection{\textbf{Secret sharing:}}
Another striking feature of quantum information theory will now be discussed, which is called $`$secret sharing'. Just like its classical counterpart, this will play a remarkable role in $`$quantum cyber ethics'.\\\\
 \textit{$``$The process of splitting a message into two or several parts in a way such that no subset of parts is sufficient to read the message while the reading of entire set is needed, is known as \textit{Secret Sharing}''}\cite{hillery1999}.\\\\ Hillery \textit{et. al} \cite{hillery1999} set the motivation for secret sharing in the following way. Cliff, who is in one part of the world, is separated from two of her accomplices Alice and Bob who are situated at some other parts of the world. Cliff wants them to do some task on behalf of him. For this, he could have sent the whole message to both Alice and Bob. But there lies one problem. Cliff is not sure about the honesty of either Alice or Bob. So he did not send the whole message to both of them as the dishonest one (if there be any) could get hold of the message and could have sabotaged the plan. Rather he decided to break it into two halves and sends each one of it to Alice and Bob. The plan can only be carried out once Alice, Bob and Cliff jointly participate in the task.\\\\
 Although classical cryptography has already an answer to the problem \cite{brucebook}, the classical procedure becomes more and more untrustworthy, when the number of subsets of the original set of message increases. An alternative solution was then given by Hillery \textit{et. al} using quantum mechanics \cite{hillery1999}. For this they have used the three particle maximally entangled state (GHZ state) defined in eq. (\ref{ghzstate}). The beauty of $`$secret sharing' protocol lies in the fact that neither Alice nor Bob can decode the final message without the help from one another. The pictorial repesentation of the protocol is shown below.\\
 \begin{figure}[hbtp]
\centering
\resizebox{10.5cm}{4.5cm}{\includegraphics{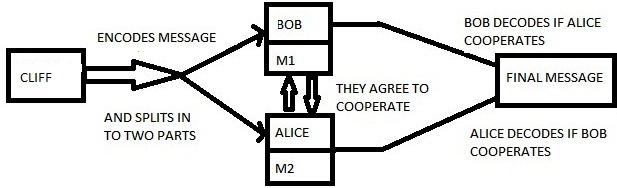}}

\caption{After feeding the message into the quantum channel Cliff splits it into two parts $M_{1}$ and $M_{2}$ and sends the respective parts to Bob and Alice each. Bob and Alice agree to cooperate with each other to decode the message. }
\end{figure}\\
Cliff, Alice and Bob, each one of them chooses randomly a particular direction along which they decide to measure. The choices of direction is announced publicly by them but they abstain from disclosing their measurement result. Cliff can develop a joint key with Alice and Bob as half the time Alice and Bob, by combining the results of their measurements, can determine what the result of Cliff's measurement is. That joint key can be used by Cliff to send his message. Corresponding to the choice of directions, one can define the eigenstates. So if the choice of direction is either $x$ or $y$, the $x$ and $y$ eigenstates will be defined as\\
\begin{eqnarray}
\vert x^{+}\rangle = \frac{1}{\sqrt{2}}\:(\vert 0\rangle + \vert 1\rangle), ~~~~~~~\vert y^{+}\rangle = \frac{1}{\sqrt{2}}\:(\vert 0\rangle + i\:\vert 1\rangle),\nonumber\\
\vert x^{-}\rangle = \frac{1}{\sqrt{2}}\:(\vert 0\rangle - \vert 1\rangle), ~~~~~~~\vert y^{-}\rangle = \frac{1}{\sqrt{2}}\:(\vert 0\rangle - i\:\vert 1\rangle).
\label{eigenstates}
\end{eqnarray}\\
With the help of these eigenstates the GHZ state of the form (\ref{ghzstate}) can then be re-written as\\
\begin{eqnarray}
\vert GHZ\rangle_{ABC} = \frac{1}{2\:\sqrt{2}}\:\lbrace(\:\vert x^{+}\rangle_{A}\:\vert x^{+}\rangle_{B} + \vert x^{-}\rangle_{A}\:\vert x^{-}\rangle_{B} \:)\:(\:\vert 0\rangle_{C} + \vert 1\rangle_{C}\:)\nonumber\\
+ (\:\vert x^{+}\rangle_{A}\:\vert x^{-}\rangle_{B} + \vert x^{-}\rangle_{A}\:\vert x^{+}\rangle_{B} \:)\:(\:\vert 0\rangle_{C} - \vert 1\rangle_{C}\:)\rbrace.
\label{ghzoneigenstate}
\end{eqnarray}\\
The justification of announcing the measurement directions by the parties involved in the scheme is because when Cliff and Alice both measure their particles in a specific direction, Bob will also have to measure his in the same direction as that of Cliff and Alice. This will help Bob to identify whether measurement results of Cliff and Alice are correlated or anti-correlated. Otherwise if Bob chooses his direction different from that of Cliff and Alice, he gains nothing (i.e no information).  The announcement part is done in the following way. Alice and Bob both send to Cliff the direction of their measurements, who then sends all three measurement directions to Alice and Bob.\\\\
In the above scheme another problem may arise. The problem of eavesdropping. That may happen in two ways. Either a fourth entity other than Cliff-Alice-Bob trio, gets stuck with the information processing scheme to get hold of the information en-route, without being noticed or any one of Alice-Bob (recipients of partial information) is actually dishonest and somehow gains access to both of Cliff's transmission. Eavesdropping problem can however be tackled with quantum cryptographic protocols\cite{bennett1984,ekert1991}. Eavesdropping has also been discussed with secret sharing protocol by Hillery \textit{et. al} in \cite{hillery1999}.\\\\
In Chapter $6$ a modified secret sharing protocol has been presented where quantum cloning (or simply cloning) played an important role.\\
\subsection{\textbf{Cloning:}}
$`$Does a classical photocopying machine have any quantum counterpart?' was a fundamental question in quantum information science, until Wootters and Zurek proved that the answer to this question was $`$NO!'. \textit{$``$No apparatus exists which will amplify an arbitrary polarization...''} was the claim by Wootters and Zurek in their seminal paper of $1982$, $`$Single quantum cannot be cloned'\cite{wootters1982}, famously known as \textit{No cloning theorem}. More than a decade later, Buzek and Hillery showed that quantum copying is indeed possible but copy will not be an exact replica of the original \cite{buzek1996}, rather an approximate one can be obtained. They designed \textit{universal quantum copying machine} (UQCM) to study the possibility of copying an arbitrary state of spin-$\frac{1}{2}$ particle\footnote{ All known elementary fermions have a spin of $\frac{1}{2}$.} and they succeeded. In $1998$ Buzek-Hillery wrote one another paper in which they designed an universal optimal cloning machine which was meant for copying states in arbitrary dimensions \cite{buzek1998}. 
\subsection*{\textbf{Universal optimal cloning machine for copying quantum states in arbitrary dimension:}}
The quantum machine constructed by Buzek and Hillery in \cite{buzek1998} is an $n-$ dimensional quantum system where the Hilbert space of the cloning machine has an orthonormal basis $\vert X_{i}\rangle_{x}$, $i=1,2,...,n$. The cloner is initially prepared in a particular state $\vert X\rangle_{x}$. Unitary transformation acting on the basis vectors of the tensor product space of the original quantum system $\vert \Psi_{i}\rangle_{a}$, the copier, an additional $n-$dimensional system which becomes the copy (initially prepared in a specific state $\vert 0\rangle_{b}$) together constitute the action of the cloning transformation. Thus the transformation of the basis vectors is given by
\begin{eqnarray}
\vert \Psi_{i}\rangle_{a}\:\vert 0\rangle_{b}\:\vert X\rangle_{x} \rightarrow \:c\:\vert \Psi_{i}\rangle_{a}\:\vert \Psi_{i}\rangle_{b}\:\vert X_{i}\rangle_{x}\:+\:d\:\sum_{j\:\neq\:i}^{n}\:(\:\vert \Psi_{i}\rangle_{a}\:\vert \Psi_{j}\rangle_{b}\nonumber\\
+\vert \Psi_{j}\rangle_{a}\:\vert \Psi_{i}\rangle_{b}\:)\:\vert X_{j}\rangle_{x}\:,
\label{buzekcloning}
\end{eqnarray}\\
where $c$ and $d$ are real coefficients satisfying the condition
\begin{eqnarray}
c^{2}\:+\:2\:(\:n-1\:)\:d^{2} = 1\:.
\label{cdrelation1}
\end{eqnarray}\\
Using the transformation (\ref{buzekcloning}) it is seen that the particles $a$ and $b$ at the output part of the cloner are in the same state (i.e. have the same reduced density matrices). The density operators are therefore obtained as
\begin{eqnarray}
\hat{\rho}_{a}^{(out)}=\hat{\rho}_{b}^{(out)}=\sum_{i=1}^{n}\:|\:\alpha_{i}\:|^{2}\:[\:c^{2}\:+\:(\:n-2\:)\:d^{2}\:]\:\vert \Psi_{i}\rangle\langle \Psi_{i}\vert\:\nonumber\\
~~~~~~~~+\:\sum_{i,j=1\:,\:i\neq j}^{n}\alpha_{i}\:\alpha_{j}^{*}\:[\:2\:c\:d\:+\:(n-2)\:d^{2}\:]\:\vert \Psi_{i}\rangle\langle \Psi_{j}\vert\:+\:d^{2}\:\hat{1}.
\label{aboutput}
\end{eqnarray}\\
Now for a universal cloning transformation, it is expected that the quality of the cloning should not depend on the state to be cloned. For this the output reduced density matrix should be in the following form\\
\begin{eqnarray}
\hat{\rho^{(out)}_{j}}=\:s\:\hat{\rho_{j}^{(id)}} \: + \:\frac{1-s}{n}\:\hat{1}\:,
\label{scalingfactor}
\end{eqnarray}\\
where $\hat{\rho_{j}^{(id)}}=\vert \Phi\rangle\langle \Phi\vert$ is the density operator describing the original state which is going to be cloned and $s$ is the scaling factor. Comparing the eqs. (\ref{aboutput}) and (\ref{scalingfactor}) it is found that the real coefficients $c$ and $d$ must satisfy the following relation\\
\begin{eqnarray}
c^{2}\:=\:2\:c\:d.
\label{cdrelation2}
\end{eqnarray}\\
Taking into account the normalization condition in eq. (\ref{cdrelation1}), it has been shown in \cite{buzek1998} that\\
\begin{eqnarray}
c^{2}\:=\:\frac{2}{n+1},~~~~~~~~~~~~~
d^{2}\:=\:\frac{1}{2\:(n+1)}\nonumber\\\nonumber\\\nonumber\\
s\:=\:c^{2} + (n-2)\:d^{2}\:=\:\frac{n+2}{2\:(n+1)}.
\label{cdrelation3}
\end{eqnarray}\\
For qutrit system (or $3\otimes 3$ dimensional system, i.e. when $n=3$), the values of the parameters $c$ and $d$ are respectively given by $c^{2}=\frac{1}{2}$ and $d^{2}=\frac{1}{8}$ so that the maximum possible value of the scaling factor ($s$) is $\frac{5}{8}$. The optimality of the cloner described by the unitary transformation (\ref{buzekcloning}) has been numerically tested in \cite{buzek1998} and also has been independently confirmed by Werner in \cite{werner1827}.
\chapter{MEMS and NMEMS in Teleportation}
\label{ch:tpp}
\textit{$``$Besides the fact that teleportation does not work the way science-fiction authors imagine, we have learned something much more important. We have learned something whose relevance goes far
beyond teleportation and science fiction."}
\begin{flushright}
- Anton Zeilinger, (extracts from $``$Dance of Photon'')
\end{flushright}
\vskip1cm
\section{\textbf{Introduction:}}
This chapter\footnote{The Chapter is mainly based on our work\\\
\textsc{S. Adhikari, A. S. Majumdar, S. Roy, B. Ghosh and N. Nayak}, \textbf{$'$Teleportation via maximally and non-maximally entangled mixed states'}, \textsc{Quantum Information and Computation, Vol. 10, No. 5 and 6, 0398-0419, (2010), Rinton Press}.} mainly concerns with $`$maximally entangled mixed states (MEMS)' and $`$non-maximally entangled mixed states (NMEMS)' as a resource for teleportation. Both of these types of states have already been defined in section $2.10$ of Chapter $2$. It is a known fact that not every mixed entangled state is useful for teleportation \cite{mhorodecki1996}. Therefore the following questions have been addressed here.
\begin{itemize}
\item Is every maximally entangled mixed state useful for teleportation?
\item Is there a relation between the amount of entanglement for a state and its efficiency as a teleportation channel?
\item Among all the MEMS discussed in section $2.10$ of Chapter $2$, which of them can be regarded as the best suitable for teleportation?
\item Is there any other forms of non-maximally entangled mixed states, other than Werner derivative (already discussed in section $2.10.5$ of Chapter $2$) which acts better as teleportation channel?
\item Is there any NMEMS state which does not violate the Bell-CHSH inequality but is still useful for teleportation?
\item Can one consider the magnitude of entanglement and violation of local inequalities to be good indicators  of their ability to perform quantum information processing tasks such as teleportation?
\end{itemize}
The answers to the above questions have been investigated in this chapter. For this,  a few more mathematical rudiments will first be described (which have earlier been skipped in Chapter $2$).
\section{\textbf{Teleportation Fidelity}}
The efficiency of a quantum channel used for teleportation is measured in terms of its \textit{average teleportation fidelity} given in \cite{horodecki541996} as,\\
\begin{eqnarray}
f^{T}_{opt}\:(\rho_{\phi})=\int_{S}\:d\:M(\rho_{\phi})\:\sum_{k}\:p_{k}\:Tr\:(\rho_{k}\:\rho_{\phi}),
\label{avtelefid}
\end{eqnarray}\\
where, $\rho_{\phi}$ is the input pure state and $\rho_{k}$ is the output state provided the outcome $k$ is obtained by Alice (Alice is the sender who possesses the unknown qubit which will supposed to be sent to Bob in the teleportation protocol). The quantity $Tr\:(\rho_{k}\:\rho_{\phi})$, which is a measure of how the resulting state is similar to the input one, is averaged over the probabilities of outcomes $p_{k}$ and then over all possible input states ($M$ denotes the uniform distribution\footnote{\textit{Geometry of Quantum States: An Introduction to Quantum Entanglement}, - I. Bengtsson and K. Zyczkowski, Cambridge University Press.} of all the input states on the Bloch sphere $S$\footnote{The geometrical representation of the pure state space of the two - level quantum mechanical system is called \textit{Bloch Sphere}. It is a unit three dimensional sphere, the state vector of the qubit is depicted as a point on the surface of the sphere and such a vector is known as \textit{Bloch vector}.}).\\

\noindent In some instances of teleportation, the teleportation fidelity depends upon the input states. It gives better results for some input states and worse for some other input states. For specific cases of input states however, it is possible to perform a calculation for the best (or the worst) teleportation fidelity (rather than the average optimum) as we illustrate here now.\\\\
If one considers the input state to be teleported is of the form\\
\begin{eqnarray}
\rho^{in} = \left(%
\begin{array}{cc}
x & y\\
y^{*}& 1-x\\
\end{array}%
\right)
\label{instate}, 
\end{eqnarray}\\
and if the teleportation channel is given by $\rho_{mjwk}$ of eq. (\ref{munromems}), then the teleported state (using the standard approach) after performing suitable unitary transformations corresponding to the four Bell - state measurement outcomes $\vert \Psi^{+}\rangle$, $\vert \Psi^{-}\rangle$, $\vert \Phi^{+}\rangle$ and $\vert \Phi^{-}\rangle$ is given (for the following two cases) as\\\\
\textbf{Case $1$}: $C \geq \frac{2}{3}$.\\
\begin{eqnarray}
\rho^{out}_{B_{1}} = \rho^{out}_{B_{2}} = \left(%
\begin{array}{cc}
\frac{x\:C}{2\:N} & \frac{y\:C}{2\:N}\\
\frac{y^{*}\:C}{2\:N}& \frac{x\:(2-3\:C)+C}{2\:N}\\
\end{array}%
\right)\nonumber\\
\rho^{out}_{B_{3}} = -\rho^{out}_{B_{4}} = \left(%
\begin{array}{cc}
\frac{(3\:x-2)\:C+2\:(1-x)}{2\:N_{1}} & \frac{y\:C}{2\:N_{1}}\\
\frac{y^{*}\:C}{2\:N_{1}}& \frac{(1-x)\:C}{2\:N_{1}}\\
\end{array}%
\right).
\label{instate1}
\end{eqnarray}\\
To determine the efficiency of the teleportation channel, the distances between the input and output state are calculated, using Hilbert Schmidt norm (already defined in section $2.14.3$) and they are given as\\
\begin{eqnarray}
D_{B_{1}}=D_{B_{2}}=x^{2}\:\lbrace1-\frac{C}{2\:N}\rbrace^{2} + 2\:|y|^{2}\:\lbrace1-\frac{C}{2\:N}\rbrace^{2}\nonumber\\\nonumber\\
+\lbrace\:(1-x)-\frac{x\:(2-3\:C)+C}{2\:N}\rbrace^{2}, \nonumber\\\nonumber\\
D_{B_{3}}=\lbrace x-\frac{(3\:x-2)\:C+2\:(1-x)}{2\:N_{1}}\rbrace^{2} + 2\:|y|^{2}\:\lbrace 1-\frac{C}{2\:N_{1}}\rbrace^{2} \nonumber\\+ (1-x)^{2}\:\lbrace 1-\frac{C}{2\:N_{1}}\rbrace^{2}, \nonumber\\\nonumber\\
D_{B_{4}}=\lbrace x+\frac{(3\:x-2)\:C+2\:(1-x)}{2\:N_{1}}\rbrace^{2} + 2\:|y|^{2}\:\lbrace 1+\frac{C}{2\:N_{1}}\rbrace^{2} \nonumber\\+ (1-x)^{2}\:\lbrace 1+\frac{C}{2\:N_{1}}\rbrace^{2},
\label{hcnormcase1}
\end{eqnarray}\\
where $N=x\:(1-\frac{C}{2})+\frac{(1-x)\:C}{2}$ and $N_{1}=\frac{x\:C}{2}+(1-x)\:(1-\frac{C}{2})$.\\\\
\textbf{Case $2$}: $C < \frac{2}{3}$.\\
\begin{eqnarray}
\rho^{out}_{B^{/}_{1}} = \rho^{out}_{B^{/}_{2}} = \left(%
\begin{array}{cc}
\frac{x}{3\:N} & \frac{y\:C}{2\:N}\\
\frac{y^{*}\:C}{2\:N}& \frac{1}{3\:N}\\
\end{array}%
\right)
\nonumber\\
\rho^{out}_{B^{/}_{3}} = -\rho^{out}_{B^{/}_{4}} = \left(%
\begin{array}{cc}
\frac{1}{3\:N_{1}} & \frac{y\:C}{2\:N_{1}}\\
\frac{y^{*}\:C}{2\:N_{1}}& \frac{(1-x)}{3\:N_{1}}\\
\end{array}%
\right).
\label{instate2}
\end{eqnarray}\\
In this case, the distances between input and output state by Hilbert Schmidt norm, are\\
\beq
D_{B^{/}_{1}}=D_{B^{/}_{2}}=x^{2}\:\lbrace 1-\frac{1}{3\:N^{/}}\rbrace^{2} + 2\:|y|^{2}\:\lbrace 1-\frac{C}{2\:N^{/}}\rbrace^{2}\nonumber\\\nonumber\\
+\lbrace\:(1-x)-\frac{1}{3\:N^{/}}\rbrace^{2}, \nonumber\\\nonumber\\
D_{B^{/}_{3}}=\lbrace x-\frac{1}{3\:N^{/}_{1}}\rbrace^{2} + 2\:|y|^{2}\:\lbrace 1-\frac{C}{2\:N^{/}_{1}}\rbrace^{2} \nonumber\\+ (1-x)^{2}\:\lbrace 1-\frac{1}{3\:N^{/}_{1}}\rbrace^{2}, \nonumber\\\nonumber\\
D_{B_{4}}=\lbrace x+\frac{1}{3\:N^{/}_{1}}\rbrace^{2} + 2\:|y|^{2}\:\lbrace 1+\frac{C}{2\:N^{/}_{1}}\rbrace^{2} \nonumber\\+ (1-x)^{2}\:\lbrace 1+\frac{1}{3\:N^{/}_{1}}\rbrace^{2},
\label{hcnormcase2}
\eeq\\
where, $N^{/}=\frac{2\:x}{3} + \frac{1-x}{3}$ and $N^{/}_{1}=\frac{x}{3}+\frac{2\:(1-x)}{3}$.\\\\
The teleportation fidelity $(F)$ can easily be calculated by using the formula $F=1-D$.\\\\ Clearly, the fidelity depends on the input state and hence one can easily calculate the best (or the worst) fidelity by choosing some particular input state. However, the purpose of the present chapter is to evaluate the average performance of various forms of MEMS and NMEMS class as teleportation channel. To this end, it is therefore better to stick to the $`$optimal teleportation fidelity' to carry on with the comparative study.\\\\ It has been shown in \cite{mhorodecki1996} that if a state $\rho$ is useful for standard teleportation, the optimal teleportation fidelity can be expressed as\\
\begin{eqnarray}
f^{T}_{opt}\:(\rho)=\frac{1}{2}\:\left\{ 1 + \frac{N\:(\rho)}{3}\right\}
\label{opttelfid},
\end{eqnarray}\\
where $N(\rho)=\sum^{3}_{i=1}\:\sqrt{u_{i}}$ and $u_{i}$'s are the eigenvalues of the matrix $T^{\dagger}T$. The elements of the matrix $T$ are given by\\
\begin{eqnarray}
t_{nm}=Tr\:(\rho\:\sigma_{n}\:\otimes\:\sigma_{m})
\label{elementofT},
\end{eqnarray}\\
where $\sigma_{i}$'s are the Pauli spin matrices.  Now, in terms of the quantity $N(\rho)$, a general result \cite{mhorodecki1996} holds that any mixed spin-$\frac{1}{2}$ state is useful for (standard) teleportation if and only if \\
\begin{eqnarray}
N\:(\rho)\:>\:1.
\label{nrho>1}
\end{eqnarray}\\
Again the relation between the optimal teleportation fidelity $f^{T}_{opt}\:(\rho)$ and the maximal singlet fraction defined in eq. (\ref{singletfraction}) is given in \cite{horodecki601999} as\\
\begin{eqnarray}
f^{T}_{opt}\:(\rho) = \frac{2\:F(\rho)+1}{3}
\label{relsingletopttel}.
\end{eqnarray}\\
From equations (\ref{opttelfid}) and (\ref{relsingletopttel}) it follows that,\\
\begin{eqnarray}
F\:(\rho)=\frac{1+N\:(\rho)}{4}
\label{relation1}.
\end{eqnarray}\\
Using the inequality \cite{verstraete2002}\\
\begin{eqnarray}
F\:\leq\:\frac{1+N}{2}\:\leq \frac{1+C}{2}
\label{inequality1},
\end{eqnarray}\\
where $N$  denotes the \textit{negativity} of the state (defined as eq. (\ref{negativity1}) in  section $2.12.5$) and $C$ is the concurrence (defined in eq. (\ref{concurrence}) of section $2.12.2$), the following inequality is obtained.\\
\begin{eqnarray}
N(\rho)\:\leq\:1+2\:N.
\label{inequality2}
\end{eqnarray}\\
Another important aspect is the violation of Bell - CHSH inequality by mixed states. Any state described by the density operator $\rho$ violates Bell - CHSH inequality \cite{chsh1969} if and only if the inequality\\
\begin{eqnarray}
M\:(\rho)=\:max_{i\:>\: j}\:(u_{i}+u_{j})\:>\:1
\label{inequality3},
\end{eqnarray}\\
holds, where $u_{i}$'s are the eigenvalues of the matrix $T^{\dagger}\:T$ \cite{mhorodecki1996}.\\\\
Now to attain the fulfilment of the objectives of this chapter, MEMS class will be examined first. 
\section{\textbf{Werner state as a teleportation channel}}
The Werner state is a convex combination of a pure maximally entangled state and a maximally mixed state. Ishizaka and Hiroshima \cite{ishizaka2000} showed that the entanglement of formation \cite{wootters1998} of the Werner state cannot be increased by any unitary transformation. Therefore, the Werner state can be regarded as a maximally entangled mixed state.\\\\
Werner state can be represented in various ways. The particular form of Werner state given in eq. (\ref{wernerstate}) was actually proposed by Ishizaka and Hiroshima \cite{ishizaka2000}. In computational basis, the density matrix of the state can explicitly be written as\\
\begin{eqnarray}
\rho_{werner} = \left(%
\begin{array}{cccc}
\frac{1-F_{w}}{3}&0&0&0\\
0&\frac{1+2\:F_{w}}{6}&\frac{1-4\:F_{w}}{6}&0\\
0&\frac{1-4\:F_{w}}{6} &\frac{1+2\:F_{w}}{6}&0\\
0&0&0&\frac{1-\:F_{w}}{3}
\end{array}%
\right)
\label{wernerstate2}, 
\end{eqnarray}\\
where $\vert \Phi^{-}\rangle$, $F_{werner}$  have their usual meanings as already discussed in section $2.9$ of chapter $2$. In the matrix representation defined in (\ref{wernerstate2}), for simplicity, $F_{werner}$ has been denoted by $F_{w}$. $F_{werner}$ is also related to the linear entropy $S_{L}$ of eq. (\ref{linearentropy2}) as\\
\begin{eqnarray}
F_{werner} = \frac{1+3\:\sqrt{1-S_{L}}}{4}.
\label{fwlinearrel}
\end{eqnarray}\\
Using the formula given in eq. (\ref{concurrence}), the concurrence of the Werner state $\rho_{werner}$ is found to be\\
\begin{eqnarray}
C\:(\rho_{werner}) &=& \mbox{max}\:\lbrace\:0\:,\:2\:F_{werner}-1\:\rbrace, \nonumber\\\nonumber\\
             &=& \left\{\begin{array}{cccc} 0
& & & 0\leq F_{werner} \leq \frac{1}{2}\\
2F_{werner}-1 & & & \frac{1}{2}< F_{werner} \leq 1
\end{array}
\right.
\label{wernerconcur}.
\end{eqnarray}\\
When the Werner state is used as a quantum channel for teleportation, the average optimal teleportation fidelity is given by \cite{horodecki601999,badziag2000,mista2002}.\\
\begin{eqnarray}
f^{T}_{opt}\:(\rho_{werner})=\frac{2\:F_{werner}+1}{3},\nonumber\\
\frac{1}{2}\:<F_{werner}\:\leq 1
\label{inequality4}.
\end{eqnarray}\\
The relation between the teleportation fidelity and the concurrence of the Werner state is thus obtained as\\
\begin{eqnarray}
f^{T}_{opt}\:(\rho_{werner})=\frac{2+C(\rho_{werner})}{3} .
\label{wertelcon}
\end{eqnarray}\\
In terms of the linear entropy $S_{L}$, eq. (\ref{inequality4}) can be re-written as\\
\begin{eqnarray}
f^{T}_{opt}\:(\rho_{werner})= \frac{1+\sqrt{1-S_{L}}}{2},\nonumber\\
0\:\leq \:S_{L}\: <\frac{8}{9}.
\label{inequality5}
\end{eqnarray}\\
Further, it is noticed that\\
\beq
F(\rho_{werner}) = \frac{1+N(\rho_{werner})}{4}.
\label{singletnegwer}
\eeq \\
Now using the inequality (\ref{inequality2}) and eq. (\ref{singletnegwer}), the following is observed.\\
\beq
F(\rho_{werner}) \leq \: \frac{1}{2}\:\lbrace 1 + N^{\rho_{werner}}\:\rbrace ,
\label{inequality6}
\eeq\\
which is the upper bound of the singlet fraction for the Werner state in terms of negativity ($N^{\rho_{werner}}$ is the negativity of the Werner state).\\\\
The status of the violation of the Bell-CHSH inequality by the Werner state is now reviewed below.\\\\ 
Using eq. (\ref{elementofT}), the eigenvalues of the matrix $(T^{\dagger}_{\rho_{werner}}\:T_{\rho_{werner}})$ are given by $u_{1}=u_{2}=u_{3}=\frac{(4F_{werner}-1)^{2}}{9}$, where $(t_{w})_{nm}=Tr\:(\rho_{werner}\: \sigma_{n}\otimes \sigma_{m})$ denotes the elements  of the matrix $T_{\rho_{werner}}$. The Werner state violates Bell-CHSH inequality if and only if $M(\rho_{werner})>1$, where $M(\rho_{werner})$ is given by\\
\beq
M(\rho_{werner})=2\:\frac{(4F_{werner}-1)^{2}}{9}.
\label{mwerner}
\eeq\\
Using eq. (\ref{wernerconcur}) it follows that the Werner state satisfies the Bell-CHSH inequality although it is entangled when the maximal singlet fraction $F_{werner}$ lies within the range\\
\beq
\frac{1}{2}\:\leq F_{werner}\:\leq \frac{3+\sqrt{2}}{4\:\sqrt{2}}.
\label{inequality7}
\eeq\\
The optimal teleportation fidelity  of the Werner state in terms of $M(\rho_{werner})$ is then given by\\
\beq
f^{T}_{opt}\:(\rho_{werner}) = \frac{\sqrt{\frac{M(\rho_{werner})}{2}}+1}{2}.
\label{mwernertel}
\eeq\\
Thus, from eqs. (\ref{inequality4}) and (\ref{mwernertel}) it follows that the Werner state can be used as a quantum teleportation channel (average optimal teleportation fidelity exceeding $\frac{2}{3}$) even without violating the Bell - CHSH inequality in the said domain.
\section{\textbf{Munro-James-White-Kwiat (MJWK) state in teleportation}}
The maximally entangled mixed state suggested by Munro \textit{et. al} has already been defined in (\ref{munromems}) of Chapter $2$. The form of linear entropy of (\ref{munromems}) is\\
\beq
S_{L}= \left\{\begin{array}{cccc} \frac{8}{3}\:(C-C^{2})
& & & C\geq\frac{2}{3}\\
\frac{2}{3}\:(\frac{4}{3}-C^{2}) & & & C < \frac{2}{3}
\end{array}
\right.
\label{munroentropy}.
\eeq\\
The performance of the MEMS state (\ref{munromems}) as a teleportation channel will now be analyzed, for which it is necessary to know the fidelity of teleportation channel. The maximal singlet fraction of the state described by the density operator $\rho_{mjwk}$ using the definition (\ref{singletfraction}) is found to be\\
\beq
F_{mjwk} &=& max\: \left\{\: h(C)+\frac{C}{2}, h(C)-\frac{C}{2}, \frac{1}{2}-h(C), \frac{1}{2}-h(C)\:\right\}\nonumber\\
&=& h(C)+\frac{C}{2}
\label{munrosinglet}.
\eeq\\
Taking into account the result given in eq. (\ref{opttelfid}) relating the optimal teleportation fidelity and the singlet fraction of a state $\rho$ and thereby using eqs. (\ref{opttelfid}) and (\ref{munrocon}), the optimal teleportation fidelity is given by,\\
\beq
f^{T}_{opt}\:(\rho_{mjwk}) = \left\{\begin{array}{cccc} \frac{2\:C+1}{3}
& & & C\geq\frac{2}{3}\\
\frac{5+3\:C}{9} & & & C < \frac{2}{3}
\end{array}
\right.
\label{opttelmunro1}.
\eeq\\
Now inverting the relation (\ref{munroentropy}), i.e. expressing $C$ in terms of $S_{L}$, one can re-write eq. (\ref{opttelmunro1}) in terms of the linear entropy ($S_{L}$) as\\
\beq
f^{T}_{opt}\:(\rho_{mjwk}) =\left\{\begin{array}{cccc} \frac{2}{3}+\frac{\sqrt{2-3\:S_{L}}}{3\:\sqrt{2}}
& & & 0\:\leq\: S_{L}\:\leq \frac{16}{27}\\
\frac{5}{9}+\frac{\sqrt{8-9\:S_{L}}}{3\:\sqrt{6}} & & & \frac{16}{27}< S_{L} \leq \frac{8}{9}
\end{array}
\right.
\label{opttelmunro2}.
\eeq\\
It follows that the Munro-class of MEMS can be used as a faithful teleportation channel when the mixedness of the state is less than the value $S_{L}=\frac{22}{27}$.\\\\
The non-local properties of the Munro-class of MEMS $\rho_{mjwk}$ will now be analyzed.\\\\
Wei \textit{et. al} \cite{wei2003} have studied the state $\rho_{mjwk}$ from the perspective of Bell's - inequality violation. Here, the parametrization of the state given in (\ref{munromems}) has been focussed. The range of concurrence where the Bell-CHSH inequality is violated, is demarcated over here. In order to use the result (\ref{inequality3}), the matrix $T_{\rho_{mjwk}}$ is constructed first as follows.\\
\begin{eqnarray}
T_{\rho_{mjwk}}=\left(%
\begin{array}{ccc}
h(C)+C&0&0\\
0&-C&0\\
0&0 &4h(C)-1\\
\end{array}%
\right)
\label{TMJWK1},
\end{eqnarray}\\
The eigenvalues of the matrix $(T^{\dagger}_{\rho_{mjwk}}\:T_{\rho_{mjwk}})$  are given by\\
\beq
u_{1} = (h(C)+C)^{2}, ~~~~~~u_{2}=C^{2}, ~~~~~~u_{3}=(4\:h(C)-1)^{2}.
\label{eigenvaluetmjwk}
\eeq\\
In accordance with eq. (\ref{munrocon}), the eigenvalues (\ref{eigenvaluetmjwk}) take two different forms which are discussed separately now.\\\\
\textbf{Case $I$:} When $h(C)=\frac{C}{2}$, ~~~~~~$\frac{2}{3}\:\leq C\: \leq 1$.\\\\
The eigenvalues (\ref{eigenvaluetmjwk}) reduce to\\
\beq
u_{1} = \frac{9\:C^{2}}{4}, ~~~~~~u_{2}=C^{2}, ~~~~~~u_{3}=(2\:C-1)^{2}.
\label{eigenvaluetmjwk1}
\eeq\\
When $C\:\geq \frac{2}{3}$, the eigenvalues can be arranged as $u_{1}\:>\:u_{2}\:>\:u_{3}$. So, we have\\
\beq
M(\rho_{mjwk}) = u_{1}+ u_{2} = \frac{13\:C^{2}}{4}.
\label{sumeigenvalue1}
\eeq\\
One can easily see that $M(\rho_{mjwk})\:>\:1$ when $C\:\geq\:\frac{2}{3}$ and hence, in this case, the state $\rho_{mjwk}$ violates the Bell-CHSH inequality.\\\\
\textbf{Case $II$:} When $h(C)=\frac{1}{3}$,~~~~~~~~~~~$0\:\leq\:C\:<\frac{2}{3}$.\\\\
The eigenvalues given by eq. (\ref{eigenvaluetmjwk}) reduce to\\
\beq
u_{1}=\frac{(3\:C+1)^{2}}{9},~~u_{2}=C^{2},~~\mbox{and}~~u_{3}=\frac{1}{9}.
\label{eigenvaluetmjwk2}
\eeq\\
The interval $0\:\leq C\:<\frac{2}{3}$ is now split into two sub-intervals, one is $0\:\leq C\:\leq \frac{1}{3}$ and the other one is $\frac{1}{3}\:<C\:<\frac{2}{3}$, where the ordering of the eigenvalues are different.\\\\
\textbf{Subcase $(i)$: $0\:\leq C\:\leq \frac{1}{3}$}.\\\\
The ordering of the eigenvalues are $u_{1}\:>\:u_{3}\:>\:u_{2}$. In this case one has\\
\beq
M(\rho_{mjwk})-1 = u_{1} + u_{3} -1 = \frac{9\:C^{2}+6\:C-7}{9}.
\label{eigenvaluetmjwk3}
\eeq\\
From eq. (\ref{eigenvaluetmjwk3}) it is clear that $M(\rho_{mjwk})<1$, when $0\:\leq C\: \leq \frac{1}{3}$. Hence, the Bell-CHSH inequality is satisfied by $\rho_{mjwk}$.\\\\
\textbf{Subcase $(ii)$: $\frac{1}{3}\:< C\:< \frac{2}{3}$}.\\\\
The ordering of the eigenvalues here are $u_{1}\:>\:u_{2}\:>\:u_{3}$. Therefore, the expression for $(M(\rho_{mjwk})-1)$ is given by\\
\beq
M(\rho_{mjwk})-1 = u_{1} + u_{2} - 1 = \frac{2\:(9\:C^{2}+3\:C-4)}{9}.
\label{eigenvaluetmjwk4}
\eeq\\
From eq. (\ref{eigenvaluetmjwk4}), it follows that $M(\rho_{mjwk})>1$, when, $\frac{\sqrt{153}-3}{18}\:<C\:<\frac{2}{3}$ and hence the state $\rho_{mjwk}$ violates the Bell-CHSH inequality. On the contrary, $M(\rho_{mjwk})\leq 1$ when $\frac{1}{3} < C \leq \frac{\sqrt{153}-3}{18}$, and subsequently the state $\rho_{mjwk}$ satisfies the Bell-CHSH inequality although it is entangled. It was noticed earlier \cite{munro482001} that the MJWK state needs a much higher degree of entanglement to violate the Bell-CHSH inequality compared to the Werner states. The above observations revalidate this fact.\\
\section{\textbf{Wei class of MEMS in teleportation}}
The maximally entangled mixed state proposed by Wei \textit{et. al} has already been defined in the section $2.10.4$ and is defined by equation (\ref{weimems}) there. The entanglement of $\rho_{wei}$ is quantified as\\
\beq
C(\rho_{wei}) = max\:(\gamma - 2\:\sqrt{a\:b},\:,0).
\label{weiconcur}
\eeq\\
Therefore, the state $\rho_{wei}$ is entangled only if $\gamma >2\:\sqrt{a\:b}$. The correlation matrix for $\rho_{wei}$ is given by\\
\begin{eqnarray}
T_{\rho_{wei}}=\left(%
\begin{array}{ccc}
\gamma&0&0\\
0&-\gamma&0\\
0&0 &x+y+\gamma -a-b\\
\end{array}%
\right)
\label{twei}.
\end{eqnarray}\\
The eigenvalues of the symmetric matrix $(T^{+}_{wei}\:T_{wei})$ are then found as $v_{1}=\gamma^{2}$, $v_{2}=\gamma^{2}$ and $v_{3}=(x+y+\gamma -a-b)^{2}=(1-2\:a-2\:b)^{2}$. Now, for the quantity $M(\rho_{wei})$ two cases may arise.\\\\
\textbf{Case $a$:} $M(\rho_{wei})=2\:\gamma^{2}$, ~~when either~~ $\gamma\:>2\:(a+b)-1$~~ and ~~$\gamma >\:2\:\sqrt{a\:b}$, ~~for ~~ $a+b\:>\:\frac{1}{2}$ or $\gamma\:>\:1-2(a+b)$ ~~and ~~$\gamma > \:2\:\sqrt{a\:b}$~~ for~~ $a+b\:<\frac{1}{2}$.\\\\
\textbf{Case $b$:} $M(\rho_{wei})=\gamma^{2}+(1-2\:a-2\:b)^{2}$ ~~when either~~$2(a+b)-1\:<\gamma\:<2\:\sqrt{a\:b}$, ~~for~~$a+b\:>\frac{1}{2}$ or $2\:\sqrt{a\:b}\:<\gamma\:<1-2\:(a+b)$ ~~for~~$a+b\:<\:\frac{1}{2}$.\\\\
In either case the Bell-CHSH inequality is violated if $M(\rho_{wei})\:>\:1$.\\\\
To find the condition when the state $\rho_{wei}$ can be used as a teleportation channel it is essential to find the condition under which $N(\rho_{wei})>1$. Now $N(\rho_{wei})$ is given by\\
\beq
N(\rho_{wei}) = \sqrt{v_{1}}+\sqrt{v_{2}}+\sqrt{v_{3}} = 1+2\:(\gamma -a - b).
\label{nrhowei}
\eeq\\
Therefore, we have \\
\beq
N(\rho_{wei})\:>\:1\:\nonumber\\\Rightarrow \gamma\:>a+b\:>2\:\sqrt{a\:b}.
\label{nweiineq}
\eeq\\
It follows from eq. (\ref{nrhowei}) that\\
\beq
f^{T}_{opt}(\rho_{wei})&=&\frac{1}{2}\:\left\{\:1+\frac{N(\rho_{wei})}{3}\:\right\}\:\nonumber\\&=&\frac{2}{3} + \frac{1}{3}\:(\gamma\:-a-b).
\label{opttelwei}
\eeq\\
Writing the optimal teleportation fidelity in the above form enables a useful comparison with the teleportation capability of the Werner state as channels. It was noted that for either $a=0$ ~~or~~$b=0$, one has $C(\rho_{werner})=\gamma$. Hence, it follows that the average optimal teleportation fidelity of the Werner state can be written as \\
\beq
f^{T}_{opt}\:(\rho_{werner})=\frac{2}{3} + \frac{\gamma}{3}. 
\label{opttelwerwei}
\eeq\\
From eqs. (\ref{opttelwei}) and (\ref{opttelwerwei}) it immediately follows that\\
\beq
f^{T}_{opt}(\rho_{wei})\:<\: f^{T}_{opt}(\rho_{werner})
\label{werweirel}.
\eeq\\
This shows that the Werner state performs better as a teleportation channel than the general MEMS.\\\\
After much have been talked about MEMS as teleportation channel, it is now time to look into the realms of NMEMS (as teleportation channels).
\section{\textbf{Werner derivative (an NMEMS) in teleportation}}
Hiroshima and Ishizaka \cite{hiroshima2000} proposed the NMEMS known as Werner derivative (defined in the section $2.10.5$). The aim here is to study the efficiency of the Werner derivative as a teleportation channel. To begin with, the matrix $T_{\rho_{wd}}$ for the state $\rho_{wd}$ is first formed.\\
\begin{eqnarray}
T_{\rho_{wd}}=\left(%
\begin{array}{ccc}
\frac{2\:\sqrt{a\:(1-a)}\:(4\:F_{w}-1)}{3}&0&0\\
0&-\frac{2\:\sqrt{a\:(1-a)}\:(4\:F_{w}-1)}{3}&0\\
0&0 &\frac{(4\:F_{w}-1)}{3}\\
\end{array}%
\right).
\label{twerder}
\end{eqnarray}\\
In the matrix, as before, $F_{w}=F_{werner}$.The eigenvalues of the matrix $(T^{\dagger}_{\rho_{wd}}\:T_{\rho_{wd}})$ are $u_{1}=u_{2}=\frac{4\:a\:(1-a)\:(4\:F_{werner}-1)^{2}}{9}$,~~ $u_{3}=\frac{(4\:F_{werner}-1)^{2}}{9}$. The Werner derivative can be used as a teleportation channel if and only if it satisfies eq. (\ref{nrho>1}), i.e. $N(\rho_{wd})\:>\:1$, where\\
\beq
N(\rho_{wd})&=&\sqrt{u_{1}}+\sqrt{u_{2}}+\sqrt{u_{3}}\nonumber\\
&=& \frac{(4\:F_{werner}-1)\:[1+4\:\sqrt{a\:(1-a)}]}{3}.
\label{nwd}
\eeq\\
It follows that the Werner derivative can be used as a teleportation channel if and only if 
\beq
16\:a^{2}-16\:a+\alpha^{2}\:<\:0,
\label{inequality8}
\eeq\\
where $\alpha = \frac{4(1-F_{werner})}{4F_{werner}-1}$. Solving (\ref{inequality8}) for the parameter $a$, we get\\
\beq
\frac{1}{2}\:\leq a \:<\frac{1}{2} + \frac{\sqrt{4-\alpha^{2}}}{4}\nonumber\\\nonumber\\
\Rightarrow \frac{1}{2}\:\leq a \:<\frac{1}{2}\:\left\{1+\frac{\sqrt{3\:(4\:F_{werner}^{2}-1)}}{4\:F_{werner}-1}\right\}.
\label{inequality9}
\eeq\\
Therefore teleportation can be done faithfully with the state $\rho_{wd}$ when the parameter $a$ satisfies the inequality (\ref{wenerderivative2}) of section $2.10.5$.\\\\
The fidelity of teleportation is thus given by\\
\beq
f^{T}_{opt}(\rho_{wd}) &=& \frac{1}{2}\:\left\{1+\frac{1}{3}\:N(\rho_{wd})\right\}\nonumber\\
&=&\frac{1}{18}\:[9+(4F_{werner}-1)\:(1+4\:\sqrt{a\:(1-a)})].
\label{opttelwd}
\eeq\\
When $a=\frac{1}{2}$, the Werner derivative reduces to the Werner state and the teleportation fidelity also reduces to that of the Werner state given in eq. (\ref{inequality4}). From eq. (\ref{opttelwd}), it is clear that $f^{T}_{opt}(\rho_{wd})$ is a decreasing function of $a$ and hence from eq. (\ref{inequality9}), one obtains\\
\beq
\frac{2}{3} < f^{T}_{opt}(\rho_{wd}) \leq \frac{2\:F_{werner}+1}{3}.
\label{inequality10}
\eeq\\
Further, we can express the teleportation fidelity $f^{T}_{opt}(\rho_{wd})$ given in eq. (\ref{opttelwd}) in terms of linear entropy $S_{L}$ as\\
\beq
f^{T}_{opt}(\rho_{wd})
= \frac{9+3\sqrt{1-S_{L}}\:(1+4\:\sqrt{a\:(1-a)})}{18},\nonumber\\ 
0\:\leq S_{L}\:<\frac{8}{9}
\label{inequality11}.
\eeq\\
It will now be investigated whether the state $\rho_{wd}$ violates the Bell-CHSH inequality or not. The real valued function $M(\rho)$ for the Werner derivative state is given by\\
\beq
M(\rho_{wd}) &=& u_{2} + u_{3}\nonumber\\ &=& \frac{(1+4a-4a^{2})\:(4F_{werner}-1)^{2}}{9}
\label{mrhowd1}.
\eeq\\
It follows that\\
\beq
M(\rho_{wd})-1 = \frac{-(4F_{werner}-1)^{2}}{9}\:\left\{(a-\beta)\:(a-\gamma)\right\}
\label{mrhowd2},
\eeq\\
where\\
\beq
\beta &=& \frac{1}{2}\:\left\{\:1-\frac{\sqrt{2\:(4F_{werner}-1)^{2}-9}}{4F_{werner}-1}\:\right\}, \nonumber\\\nonumber\\
\gamma &=& \frac{1}{2}\:\left\{\:1+\frac{\sqrt{2\:(4F_{werner}-1)^{2}-9}}{4F_{werner}-1}\:\right\}
\label{mrhowd3}.
\eeq\\
For $\beta$ and $\gamma$ to be real, one must have $\frac{3+\sqrt{2}}{4\:\sqrt{2}}\:\leq\:F_{werner}\leq 1$. From the expression of $\beta$ and eq. (\ref{wernerderivative1}), it is clear that $\beta\:\leq\:\frac{1}{2}\:\leq a <\frac{1}{2}\:\left\{1+\frac{\sqrt{3\:(4F^{2}_{werner}-1)}}{4F_{werner}-1}\right\}$, as $\frac{3+\sqrt{2}}{4\:\sqrt{2}}\:\leq F_{werner}\:\leq 1$. Hence $a -\beta\:\geq 0$. On the other hand, from the expression of $\gamma$, it follows that $\gamma\:\leq \frac{1}{2}\:\left\{1+\frac{\sqrt{3\:(4F^{2}_{werner}-1)}}{4F_{werner}-1}\right\}$.\\\\
The following three cases are now considered.\\\\
\textbf{Case $1$:} If $\gamma\:< a <\frac{1}{2}\:\left\{1+\frac{\sqrt{3\:(4F^{2}_{werner}-1)}}{4F_{werner}-1}\right\}$ and $\frac{3+\sqrt{2}}{4\:\sqrt{2}}\:< F_{werner}\:\leq 1$, then $M(\rho_{wd})-1\:<0$. In this case the Bell-CHSH inequality is obeyed by the state $\rho_{wd}$ although the state is entangled there.\\\\
\textbf{Case $2$:} If $\frac{1}{2}\:\leq a \:<\gamma$ and $\frac{3+\sqrt{2}}{4\:\sqrt{2}}\:< F_{werner}\:\leq 1$, then $M(\rho_{wd})-1\:>\:0$. Thus in this range of the parameter $a$ the Bell-CHSH inequality is violated by the state $\rho_{wd}$.\\\\
\textbf{Case $3$:} In this situation however, when $F_{werner}=\frac{3+\sqrt{2}}{4\:\sqrt{2}}$, then $\beta = \gamma = \frac{1}{2}$ and hence $M(\rho_{wd})\:\leq \: 1$ holds for $\frac{1}{2}\:\leq\:a\:<\frac{1}{2}\:\left\{1+\frac{1+\sqrt{1+2\:\sqrt{2}}}{2}\right\}$. The equality sign is achieved when $a\:=\:\beta\:=\:\gamma\:=\frac{1}{2}$. Therefore, in the case when $F_{werner}=\frac{3+\sqrt{2}}{4\:\sqrt{2}}$, the Werner derivative satisfies the Bell-CHSH inequality although it is entangled.\\
\section{\textbf{A new class of NMEMS as teleportation channel}}
A two-qubit density matrix $\rho_{nmems}$ is now constructed by taking a convex combination of a separable density matrix $\rho^{G}_{AB} = Tr_{C}\:(\vert GHZ\rangle_{ABC})$ and an inseparable density matrix $\rho^{W}_{AB}=Tr_{C}\:(\vert W\rangle_{ABC})$, where $\vert GHZ\rangle$  and $\vert W\rangle$ denote the three qubit $GHZ$ state \cite{ghzstate} and the $W-$state \cite{wstate} respectively. This construction is somewhat similar in spirit to the Werner state which is a convex combination of a maximally mixed state and a maximally entangled pure state. The properties, that the $GHZ$ state and the $W-$state are two qubit separable and inseparable states, respectively, when a qubit is lost from the corresponding three qubit states, have been exploited over here. By constructing this type of a non-maximally entangled mixed state, the aim is to show that it can act as a better teleportation channel compared to the Werner derivative state.\\\\
The two qubit state described by the density matrix $\rho_{nmems}$ can be explicitly written as \\
\begin{eqnarray}
\rho_{nmems}=p\rho^{G}_{AB}+(1-p)\rho^{W}_{AB},~~~~0\leq p \leq 1.
\label{new1}
\end{eqnarray}\\
The density matrix of the state $\rho_{nmems}$ in the computational basis is given by\\
\begin{eqnarray}
\rho_{nmems} = \left(%
\begin{array}{cccc}
\frac{p+2}{6} & 0 & 0 & 0 \\
0& \frac{1-p}{3} & \frac{1-p}{3} & 0\\
0 & \frac{1-p}{3}&\frac{1-p}{3} & 0 \\
 0 & 0 & 0 & \frac{p}{2}
\end{array}%
\right)
\label{new2}.
\end{eqnarray}\\
Since the state described by the density matrix (\ref{new2}) is of the form\\
\begin{eqnarray}
\tau = \left(%
\begin{array}{cccc}
a & 0 & 0 & 0 \\
0& b & c & 0\\
0 & c^{*}& d & 0 \\
 0 & 0 & 0 & e
\end{array}%
\right)
\label{new3},
\end{eqnarray}\\
its amount of entanglement \cite{bruss332003} is given by\\
\begin{eqnarray}
C(\rho_{nmems}) = C(\tau) &=& 2\:\mbox{max}\:(\:|c|-\sqrt{a\:e},\:0)\nonumber\\\nonumber\\
&=& 2\:\mbox{max}\:\left\{\:\frac{1-p}{3}-\sqrt{\frac{p(p+2)}{12}},0\:\right\}
\label{newcon}.
\end{eqnarray}\\
Therefore $\rho_{nmems}$ is entangled only if $\frac{1-p}{3}-\sqrt{\frac{p(p+2)}{12}}>0$, i.e. when
$0\leq p<0.292$.\\\\
In the limiting case of $p=0$, however, the state $\rho_{nmems}$ reduces to \\
\beq
\rho^{W}_{AB} = \frac{1}{3}\:\vert 00\rangle\langle 00\vert + \frac{2}{3}\:\vert \Phi^{+}\rangle\langle \Phi^{+}\vert,
\label{new4}
\eeq\\
where $\vert \Phi^{+}\rangle$ is one of the Bell states defined in (\ref{bellstates}). The state $\rho^{W}_{AB}$ is maximally entangled since it can be put into Ishizaka and Hiroshima's \cite{ishizaka2000} proposed class of MEMS. The concurrence of this state is $\frac{2}{3}$. When this state is used as a teleportation channel, the teleportation fidelity becomes  $f^{T}_{opt}\:(\rho^{W}_{AB})= \frac{7}{9}$. Moreover, it can be checked that the state $\rho^{W}_{AB}$ satisfies the Bell-CHSH inequality although it is an entangled state.\\\\
To obtain the teleportation fidelity for the state $\rho_{nmems}$, the matrix $T_{\rho_{nmems}}$ is now built using eq. (\ref{elementofT}), which is given by,\\
\beq
T_{\rho_{nmems}}=\left(%
\begin{array}{ccc}
\frac{2\:(1-p)}{3}&0&0\\
0&\frac{2\:(1-p)}{3}&0\\
0&0 &\frac{(4\:p-1)}{3}\\
\end{array}%
\right).
\label{Tnew}
\eeq\\
The eigenvalues of $(T^{\dagger}_{\rho_{new}}\:T_{\rho_{new}})$ are given by $u_{1}=u_{2}=\frac{4\:(1-p)^{2}}{9}$ and $u_{3}=\frac{(4\:p-1)^{2}}{9}$.\\\\ When $p>\frac{1}{4}$, one has $N\:(\rho_{nmems})=\sqrt{u_{1}} + \sqrt{u_{2}} + \sqrt{u_{3}}=1$. Therefore, the teleportation fidelity becomes $f^{T}_{opt}\:(\rho_{nmems})=\frac{1}{2}\:\left\{\:1+\frac{1}{3}\:N(\rho_{nmems})\right\}=\frac{2}{3}$. Hence for $p>\frac{1}{4}$, the state $\rho_{nmems}$ cannot be used as an efficient teleportation channel since it does not overtake the classical fidelity. But when $0\:\leq\:p\:<\frac{1}{4}$, $N\:(\rho_{nmems})=\frac{5-8\:p}{3}\:>1$ and hence $\rho_{nmems}$ can be used as an efficient teleportation channel. In this case, the average optimal teleportation fidelity is given by,\\
\beq
f^{T}_{opt}\:(\rho_{nmems}) = \frac{7-4\:p}{9},~~~~~~~~0\:\leq p\:<\frac{1}{4},
\label{opttelnew}
\eeq\\
and it follows that,\\
\beq
\frac{2}{3}\:<\:f^{T}_{opt}\:(\rho_{nmems})\:\leq \frac{7}{9}.
\label{inequality12}
\eeq\\
Consequently it is interesting to follow that the state $\rho_{nmems}$ cannot be used as an efficient teleportation channel when $0.25\:<p\:<0.292$, although the state is entangled there.\\\\
When $\rho_{nmems}$ is used as a quantum teleportation channel the mixedness of the state is given by,\\
\beq
S_{L}=\frac{2}{27}\:(8+14\:p-13\:p^{2}),\nonumber\\
 0\:\leq p\: <\frac{1}{4}.
\label{slnew1}
\eeq\\
Therefore, the teleportation fidelity $f^{T}_{opt}\:(\rho_{nmems})$, in terms of $S_{L}$ is given by,\\
\beq
f^{T}_{opt}\:(\rho_{nmems}) = \frac{7\:-\:\frac{4}{26}\:(14\:-\:\sqrt{612\:-\:702\:S_{L}}\:)\:}{9},\nonumber\\
\frac{208}{351}\:\leq\:S_{L}\:<\frac{2223}{2808}
\label{slnew2}.
\eeq\\\\
The question as to whether the state $\rho_{nmems}$ violates the Bell-CHSH inequality will be addressed now.\\The real valued function $M\:(\rho_{nmems})$ for the state $\rho_{nmems}$ for the two following cases is now calculated separately for two different domains of the state parameter $p$.\\\\
\textbf{Case $A$:} When $0\:\leq\:p\:<\:\frac{1}{2}$, $M\:(\rho_{nmems})=u_{1}+u_{2}=\frac{8+8\:p^{2}-16\:p}{9}$. Substituting the values of $p$ in the above range it is easy to see that $M\:(\rho_{nmems})\:\leq\:1$, i.e. the Bell-CHSH inequality is satisfied.\\\\
\textbf{Case $B$:} When $\frac{1}{2}\:\leq\:p\:<\:1$, $M\:(\rho_{nmems})=u_{1}+u_{3}=\frac{20\:p^{2}-16\:p\:+\:5}{9}$. It easily follows that for the given range of values of $p$, one has $M\:(\rho_{nmems})\:\leq\:1$.\\\\
Therefore, it is concluded that in any case (i.e. $0\:\leq p\:\leq\:1$), the constructed state $\rho_{nmems}$ does not violate the Bell-CHSH inequality although it is entangled for $0\:\leq p\:<\:0.292$.\\\\
The above analysis on the new class of NMEMS can be shown pictorially below in figure $3.1$.
\begin{figure}[hbtp]
\centering
\resizebox{7.5cm}{5.5cm}{\includegraphics{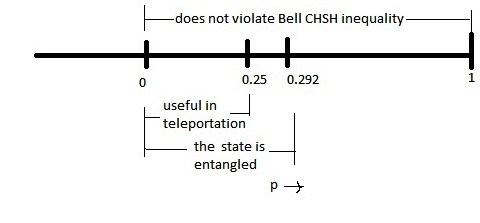}}
\caption{The figure clearly shows that for $0\:\leq\:p\:<0.292$, the state $\rho_{nmems}$ is only entangled, whereas when $p\:\in (0,\:0.25)$, then the state is useful in teleportation. However the state satisfies Bell - CHSH inequality for entire domain of the parameter $p$ i.e. for $p\:\in\:[0,1]$.}
\end{figure} 
\section{\textbf{Comparison of teleportation fidelities of different mixed states}}
So far the teleportation capacities of the various types of maximally as well as non-maximally entangled mixed channels have been discussed. It would be now actually interesting to compare their performances in terms of the average optimal fidelities corresponding to their respective magnitudes of entanglement, mixedness and also in relation to their non-locality properties manifested by the violations of the Bell-CHSH inequality. \\\\
In this respect, the two MEMS, Werner ($\rho_{werner}$) and Munro state ($\rho_{mjwk}$) are first taken into consideration. The average optimal teleportation fidelities of these two classes of MEMS in terms of their respective concurrences are given by (eq. (\ref{wertelcon}) and (\ref{opttelmunro1})) are given by,\\
\beq
f^{T}_{opt}\:(\rho_{werner})&=&\frac{2+C\:(\rho_{werner})}{3},\nonumber\\
f^{T}_{opt}\:(\rho_{mjwk}) &=& \left\{\begin{array}{cccc} \frac{2\:C+1}{3}
& & & C\geq\frac{2}{3}\\
\frac{5+3\:C}{9} & & & C < \frac{2}{3}
\end{array}
\right.
\label{comparison1}.
\eeq\\
These optimal fidelities $f^{T}_{opt}$ versus $C$ respectively for these two MEMS are now plotted in figure $3.2$.  One can see that the Werner state performs better as a teleportation channel compared to the Munro state for any given amount of entanglement. It is also to be noted that, the MJWK state is useful for teleportation ($f^{T}_{opt}(\rho_{mjwk})\:>\:\frac{2}{3}$) only when $C>\frac{1}{3}$. whereas the Werner state is able to serve as a quantum channel for teleportation for any amount of its entanglement.\\
\begin{figure}[hbtp]
\centering
\resizebox{7.5cm}{5.5cm}{\includegraphics{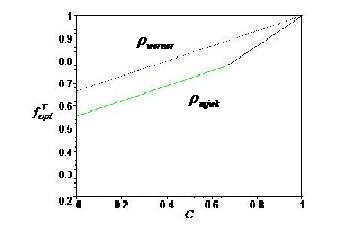}}

\caption{The average optimal teleportation fidelities for the channels $\rho_{werner}$ and $\rho_{mjwk}$ are plotted with respect to their respective magnitudes of entanglement $C$. Munro class of MEMS performs as a quantum channel only for $C>\frac{1}{3}$.}
\end{figure}\\
Next, the efficiency of teleportation of the two MEMS with respect to their non-locality properties have been compared. The average teleportation fidelities corresponding to the Werner state and the MJWK state are plotted versus the function $M(\rho)$ in figure $3.3$. We know that, $M(\rho)>1$ signifies the violation of Bell-CHSH inequality. Although, it can be seen from the figure that the Bell-CHSH inequality is being violated by MJWK state, yet depending upon certain region of parameter space, the MJWK state can be used as teleportation channel. This is in contrast to the behaviour of the Werner state which can act as a quantum teleportation channel though it satisfies the Bell-CHSH inequality.\\
\begin{figure}[hbtp]
\centering
\resizebox{7.5cm}{5.5cm}{\includegraphics{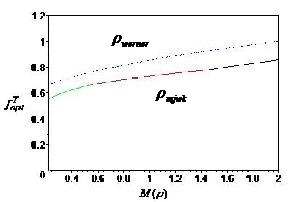}}

\caption{The average optimal teleportation fidelities for the channels $\rho_{werner}$ and $\rho_{mjwk}$ are plotted with respect to the quantity $M(\rho)$ indicating the non-local property of the channel. $M(\rho)>1$ signifies the violation of the Bell - CHSH inequality.}
\end{figure}\\
In figure $3.4$, however, a comparison of the Werner state and the generalized MEMS state (i.e. Wei class of state $\rho_{wei}$) has been brought forward, by respectively plotting their average teleportation fidelities versus the function $M(\rho)$. It is observed there that the Werner state and $\rho_{wei}$ are both useful for teleportation whether they violate the Bell - CHSH inequality or not. But the Werner state always performs better as a teleportation channel compared to the general MEMS $\rho_{wei}$, except at the value of $M(\rho)=1.7672$ where the teleportation fidelities for both the states are the same. \\
\begin{figure}[hbtp]
\centering
\resizebox{7.5cm}{5.5cm}{\includegraphics{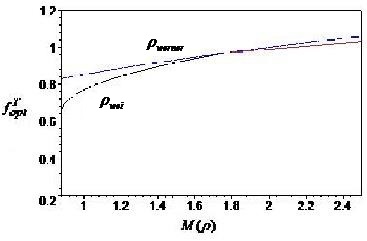}}

\caption{The average optimal teleportation fidelities for the channels $\rho_{werner}$  and $\rho_{wei}$  are plotted with respect to the quantity $M(\rho)$ indicating the non-local property of the channel. $M(\rho)>1$ signifies the violation of the Bell-CHSH inequality.}
\end{figure}\\
The relationship between the mixedness of a channel and its ability to perform quantum teleportation has been one of the focal points of exploration in this chapter. For this purpose, once again the expressions (\ref{inequality5}), (\ref{opttelmunro2}), (\ref{inequality11}) and (\ref{slnew2}) for the teleportation fidelities in terms of the linear entropy for all the four types of states (Werner state $\rho_{werner}$, Munro class of MEMS $\rho_{mjwk}$, Werner derivative $\rho_{wd}$ and proposed NMEMS $\rho_{nmems}$) are written down in eqs. (\ref{recapitulation1}) and (\ref{recapitulation2}) respectively.\\
\beq
f^{T}_{opt}(\rho_{mjwk})  = \left\{\begin{array}{cccc} \frac{2}{3}\:+\frac{\sqrt{2-3\:S_{L}}}{3\:\sqrt{2}}
& & & 0\:\leq S_{L}\:\leq\:\frac{16}{27}\\
\frac{5}{9}+\frac{\sqrt{8-9\:S_{L}}}{3\:\sqrt{6}} & & & \frac{16}{27}\:\leq\:S_{L}\:\leq\:\frac{8}{9}
\end{array}
\right.\nonumber\\\nonumber\\
f^{T}_{opt}(\rho_{werner}) = \frac{1+\sqrt{1-S_{L}}}{2}\:.~~~~~~~~~~
0\:\leq\:S_{L}\:<\frac{8}{9}.
\label{recapitulation1}
\eeq\\
\beq
f^{T}_{opt}(\rho_{nmems}) = \frac{7-\frac{4}{26}(14-\sqrt{612-702\:S_{L}})}{9}\:,\nonumber\\
\frac{208}{351}\:\leq S_{L}\:<\frac{2223}{2808}.\nonumber\\\nonumber\\
f^{T}_{opt}(\rho_{wd})=\frac{9+3\sqrt{1-S_{L}}\:(1+4\:\sqrt{a\:(1-a)})}{18}\:,\nonumber\\
0\:\leq\:S_{L}\:<\frac{8}{9}\nonumber\\\nonumber\\
\label{recapitulation2}
\eeq
First the comparison between the two maximally entangled mixed states, viz. the Werner state and MJWK state has been made. From the above expressions of $f^{T}_{opt}$ for these two states it follows that $f^{T}_{opt}(\rho_{werner})=f^{T}_{opt}(\rho_{mjwk})$ only for $S_{L}=0$. For all the other finite degrees of mixedness, $f^{T}_{opt}(\rho_{werner})\:>\:f^{T}_{opt}(\rho_{mjwk})$. The two respective fidelities are plotted versus the linear entropy in figure $3.5$. The MJWK state can be used as a quantum teleportation channel only when the mixedness of the state is less than $\frac{22}{27}$. Although both these states could be used as quantum teleporatation channels for a range of values of mixedness, one observes that the Werner state outperforms the MJWK state for all finite values of mixedness even though the latter is more entangled for specific values of linear entropy \cite{wei2003}. This is an interesting result implying the fact that all the entanglement of the MJWK class of states is less useful as a resource for teleportation.\\
 \begin{figure}[hbtp]
\centering
\resizebox{7.5cm}{5.5cm}{\includegraphics{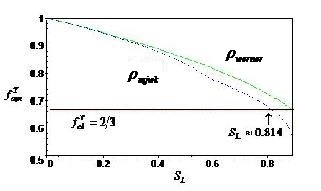}}

\caption{The average optimal teleportation fidelities for the channels $\rho_{werner}$ and $\rho_{mjwk}$ are plotted with respect to the linear entropy $S_{L}$. The horizontal line represents the maximum classical fidelity.}
\end{figure}\\
From the above eqs. (\ref{recapitulation1}) and (\ref{recapitulation2}), the two non-maximally entangled mixed states $\rho_{wd}$ and $\rho_{nmems}$, which have been investigated in this chapter, can also be easily compared. Using the relationship between the teleportation fidelity and the quantity $N(\rho)$ \cite{mhorodecki1996} given by eq. (\ref{opttelfid}), the ranges for the parameters for which the condition $N(\rho_{nmems})>N(\rho_{wd})$ holds such that the teleportation fidelity of the state $\rho_{nmems}$ will be greater than that of $\rho_{wd}$, are derived. It is already shown that\\
\beq
N(\rho_{wd}) = \frac{(4F_{werner}-1)\:(1+4\:\sqrt{a\:(1-a)})}{3},\nonumber\\
\frac{1}{2}<\:F_{werner}\:\leq 1,
\label{recapitulation2a} 
\eeq\\
\mbox{and}
\beq
N(\rho_{nmems}) = \frac{5-8\:p}{3}, \nonumber\\
0\:\leq p\:<\frac{1}{4},
\label{recapitulation3}
\eeq\\
where the parameter $a$ of eq. (\ref{recapitulation2a}) lies within the range specified in eq. (\ref{wenerderivative2}). The state $\rho_{nmems}$ performs better as a quantum teleportation channel compared to the state $\rho_{wd}$ only when $N(\rho_{nmems})>N(\rho_{wd})$ from which using eqs. (\ref{recapitulation2a}) and (\ref{recapitulation3}), it follows that,\\
\beq
p\:<\:1-\:\left\{\frac{1+2F_{werner}}{4}+\frac{(4F_{werner}-1)\:\sqrt{a(1-a)}}{2}\right\}.
\label{recapitulation3a}
\eeq\\
It can be easily verified that the condition (\ref{recapitulation3a}) on the value of $p$ is compatible with the upper bound on $p$ in the expression of eq. (\ref{recapitulation3}). However, consistency with the lower bound ($p>0$) imposes the following conditions on the parameters $F_{werner}$ and $a$.\\
\beq
\frac{1}{2}+\frac{\sqrt{(F_{werner}+1)\:(3F_{werner}-2)}}{4F_{werner}-1}\:<\:a\:<\:\frac{1}{2}\:\left\{1+\frac{\sqrt{3(\:4F_{werner}^{2}-1)}}{4F_{werner}-1}\right\}\:,\nonumber\\
F_{werner}\:>\:\frac{2}{3}.\nonumber\\
\label{recapitulation4}
\eeq\\
Therefore, when the parameters $F_{werner}$, $a$ and $p$ satisfy the relations given in equations (\ref{recapitulation3a}) and (\ref{recapitulation4}), then one can say that $f^{T}_{opt}(\rho_{nmems})>f^{T}_{opt}(\rho_{wd})$.\\\\
It can therefore be concluded that, for $\rho_{nmems}$ to perform better than $\rho_{wd}$,  one must have $f^{T}_{opt}(\rho_{nmems})\:>\:f^{T}_{opt}(\rho_{wd})$, which in turn implies that $N(\rho_{nmems})>N(\rho_{wd})$, since $N(\rho)$ is related to the teleportation fidelity of a channel by $f^{T}_{opt}(\rho)=\frac{1}{2}\:[1+\frac{N(\rho)}{3}]$. It has also been proved that $f^{T}_{opt}(\rho_{nmems})\:>\:f^{T}_{opt}(\rho_{wd})$ holds true only when the inequality (\ref{recapitulation3a}) is satisfied with appropriate choices of $a$ and $F_{werner}$ which are the parameters of the state $\rho_{wd}$ and also with the choice of $p$, the parameter of the newly proposed NMEMS state $\rho_{nmems}$.
Another perspective of this comparative study is that, $\rho_{wd}$ violates the Bell-CHSH inequality but $\rho_{nmems}$ satisfies it and that in this case too, the teleportation fidelity $f^{T}_{opt}(\rho_{wd})$ could still be less than the teleportation fidelity $f^{T}_{opt}(\rho_{nmems})$.\\\\
In the beginning of this chapter a few issues regarding the maximally and non-maximally entangled mixed states, their utility in teleportation and the nature of their Bell-violation were raised. So before bringing the curtain down to this chapter the answers to those questions have been summarized below.\\
\begin{itemize}
\item The maximally entangled mixed states (MEMS) are not always useful for teleportation. One of the examples is Munro class of MEMS, $\rho_{mjwk}$ \cite{munro2001}, which is not useful as quantum teleportation channel when its mixedness exceeds certain bound.
\item The results show that a state which is less entangled for a given degree of mixedness, e.g. the Werner state \cite{werner401989}, could act as a more efficient teleportation channel compared to a state that is more entangled like Munro class of MEMS. The result that the Werner state acts better as quantum channel for teleportation compared to the other MEMS class of states can also be understood in terms of their respective negativities $N^{\rho}$, i.e. $N^{\rho_{mjwk}}<N^{\rho_{werner}}$ \cite{wei2003}.
\item The average teleportation fidelities in terms of the respective concurrences of the states like Werner and MJWK have been shown here. In spite of the fact that both these states fall into the category of maximally entangled mixed states, one of them viz. the Werner state, outperforms the other viz. the Munro state for either any fixed degree of mixedness or any specified magnitude of entanglement.
\item A new class of non-maximally entangled mixed state $\rho_{nmems}$ have been proposed here. This NMEMS is a convex combination of a separable state and an entangled state. The Werner derivative state which is also an NMEMS, can act as an efficient quantum teleportation channel (with its average fidelity of teleportation exceeding the classical bound of $\frac{2}{3}$). The new class of NMEMS, however, outperforms the Werner derivative in terms of their efficiency as a teleportation channel.
\item For MEMS state, the analysis shows that the Werner state satisfies the Bell-CHSH inequality and yet performs as a quantum teleportation channel efficiently for a certain range of parameter space. In this context the conditions on the parameters for which the Werner derivative state satisfies the criterion of non-locality by violating the Bell-CHSH inequality is derived. It has been found then that the constructed NMEMS ($\rho_{nmems}$) could perform as a quantum teleportation channel in spite of satisfying the Bell-CHSH inequality. Moreover, the state $\rho_{nmems}$ yields a higher teleportation fidelity compared to the Werner derivative even for a range of parameter values where the latter violates the Bell-CHSH inequality.
\item With all the above observations in mind, it can easily be concluded that, for both maximally and non-maximally entangled mixed states neither the magnitude of entanglement nor the violation of local inequalities may be good indicators of their ability to perform quantum information processing tasks such as teleportation.
\end{itemize}
\chapter{Cloning a qutrit and Information Processing Tasks}
\label{ch:cqp}
\textit{$``$Computers are famous for being able to do complicated things starting from simple programs."}
\begin{flushright}
- Seth Lloyd
\end{flushright}
\vskip1cm
\section{\textbf{Introduction:}} 
Quantum teleportation was first proposed in 1993 where quantum inseparability could be used as a resource. The basic idea of the task was to use a pair of particles in a singlet state shared by distant partners Alice and Bob to perform successful communication of an unknown qubit from the sender Alice to the receiver Bob. In the previous chapter, this interesting direction has been studied by taking into account two qubit maximally and non-maximally entangled mixed states as teleportation channels. Mixed states are obtained in various ways. One such way to generate the mixed states is via quantum cloning machine. In this case, when one tries to get approximate copy by passing a pure input state through the cloning machine, the state interacts with the surroundings, which are nothing but the machine parameters here. The output thus obtained by tracing out the machine parameters is a mixed state. With the advent of quantum cloning machines \cite{buzek1996,gisin1997,buzek1998,bru1998,duan1998,zanardi1998,fan2001,delgado2007} and with the knowledge of the existing entanglement between the input and the output states in hand, it is quite expected to presume the idea of utilising these cloned output states as quantum teleportation channels. In this direction the first step was taken by Adhikari \textit{et.al} \cite{adhikari2008}, in which they showed that the two qubit entangled state which comes as an output from the Buzek-Hillery (BH) cloning machine \cite{buzek1996} can be efficiently used as quantum teleportation channel. In this chapter\footnote{The Chapter is mainly based on our work\\\
\textsc{S. Roy, N. Ganguly, A. Kumar, S. Adhikari and A. S. Majumdar}, \textbf{$'$A cloned qutrit and its utility in information processing tasks'}, \textsc{Quantum Information Processing, Vol. 13, No. 3, 629-638, (2014), Springer}.} it has been shown that indeed in higher dimensional system (in this case it is $3\otimes3$), the entangled output from a universal quantum cloning machine \cite{buzek1998} can also be used as a resource for quantum teleporation  \cite{bennett1993}. Since teleportation and dense coding are inverse procedures to one another, therefore these qutrit output states have also been considered in studying dense coding \cite{bennett1992}. Buzek's universal quantum cloning machine (BH-UQCM) designed for arbitrary dimensions \cite{buzek1998} has been used for the purpose. The following questions have therefore been addressed accordingly.\\
\begin{itemize}
\item What are the different types of outputs one obtains from the BH-UQCM?
\item Can one use these types  of outputs in teleportation as well as in dense coding?
\item Can these outputs be distilled?
\item After distillation, which of the distilled output types become suitable for information processing tasks such as teleportation and dense coding?
\end{itemize}
Central to this investigation, is the usefulness of mixed states of two qutrits, obtained as an output from BH-UQCM\cite{buzek1998}, as resources for quantum teleportation as well as for dense coding. Analysis of teleportation and dense coding using higher dimensional systems provides some interesting aspects about the outputs obtained from the cloning machine when used as information processing resources.\\
\section{\textbf{Two qutrit output state from BH-UQCM and their entanglement properties:}}
The transformation structure provided by the BH-UQCM has already been discussed in section $2.16.6$ of chapter $2$. If  a single qutrit input
$\vert \varphi\rangle = \frac{1}{\sqrt{3}}(\vert 0\rangle + \vert 1\rangle + \vert 2\rangle)$ is fed into the cloning machine, the corresponding two qutrit output obtained after tracing out ancilla states is given by\\
\begin{eqnarray}
\rho_{ab}^{out} = \frac{(1-4\:d^{2})}{3}\:\sum_{i=0}^{2}\:\vert i,i\rangle\langle i,i\vert + \frac{2}{3}\:d^{2}\:\sum_{i\:\neq\:j}^{2} \vert i,j\rangle\:\left\{\:\langle i,j\vert + \langle j,i\vert\:\right\} \nonumber\\
+\frac{(\sqrt{1-4\:d^{2}})}{3}\: d\:\left\{\sum_{i\: \neq \: j}^{2}\:\vert i,i\rangle\:(\langle i,j\vert\: + \langle j, i\vert\:)+\sum_{i\:\neq\:j}^{2}\:\vert i,j\rangle\:(\langle i,i\vert + \langle j,i\vert)\right\}\nonumber\\ + \frac{d^{2}}{3}\:\left\{\sum_{i\:\neq\:j}^{2}\:\vert i,j\rangle\:(\:\langle i,j+1\vert + \langle j+1,i\vert + \langle i+1,j+1\vert +\langle j+1, i+1\vert\:)\:(mod\:2)\:\right\} \nonumber\\
\label{qutritoutput}.
\end{eqnarray}\\
Positive partial transposition criteria \cite{peres1996} shows that, with respect to the system $a$, at least one of the two eigenvalues\\
\begin{eqnarray}
e_{1} = \frac{1+4d^{2}}{6}-\frac{1}{6}\sqrt{1+24\:d^{2}-104\:d^{4}+32\:\sqrt{-(2d-1)(2d+1)}\:d^{3}},
\label{eigenvalue1}
\end{eqnarray}\\
and \\
\begin{eqnarray}
e_{2} = \frac{1-5d^{2}}{6}-\frac{1}{6}\sqrt{1-6\:d^{2}+25\:d^{4}-16\sqrt{-(2d-1)(2d+1)}\:d^{3}}
\label{eigenvalue2},
\end{eqnarray}\\
are always negative when the machine parameter $d\:\in\:(0,\frac{1}{2}\:]$. The eigenvalue $e_{1}$ is negative when $d\:\in\:(0,\frac{6+\sqrt{2}}{17}\:)$ and the eigenvalue $e_{2}$ is negative when either $d\:\in\:(0,\frac{1}{2\:\sqrt{2}}\:]$ or $d\:\in\:(\frac{1}{2\:\sqrt{2}},\frac{1}{2}\:]$. Thus the state (\ref{qutritoutput}) is an (negative partial transposition) NPT state for $d\:\in\:(0,\frac{1}{2}]$, which consequently motivates one to analyze the state's utility for quantum information tasks. It has already been pointed out in section $2.16.6$ that for $d^{2}=\frac{1}{8}$, the output is of optimal type while for other values of parameter $d$ in the range $(0, \:\frac{1}{2}]$, the output will be treated as of non-optimal type.
\section{\textbf{Output states of BH-UQCM in teleportation and dense coding:}}
\subsection{\textbf{Optimal output:}}
The two qutrit output state, (which is denoted here as $\rho_{ab}^{opt}$) from the cloning machine, shown in the previous section, will be an optimal one, for $d^{2}=\frac{1}{8}$ and using equation (\ref{qutritoutput}) and reference \cite{peres1996} one can easily interpret that such an output state would be an NPT state.\\\\
A qutrit system, however, can be used as a teleportation channel, if its fully entangled fraction is greater than $\frac{1}{3}$ \cite{horodeckis1999}. Now for an arbitrary state $\rho$, the fully entangled fraction or sometimes known as maximal singlet fraction, has been defined in equation (\ref{singletfraction}) of section $2.9$ of Chapter $2$.\\\\
The maximally entangled orthonormal basis for two qutrit system can be represented as \cite{karimipour2006} \\
\begin{eqnarray}
\vert \phi_{x,\:y}\rangle = \frac{1}{\sqrt{3}}\:\sum_{j=0}^{2}\:\xi^{jy}\:\vert j,j+x\rangle ,\:x,y = 0, 1, 2,
\label{basiskarimi}
\end{eqnarray}\\
where, $\xi:=e^{\frac{2\:\pi\:i}{3}}$ and $\lbrace \vert 0\rangle, \vert 1\rangle, \vert 2\rangle\rbrace$ is an orthonormal basis for the space of one qutrit. The states (\ref{basiskarimi}) are maximally entangled and are mutually orthogonal.\\\\
Now considering the optimal output state $\rho_{ab}^{opt}$ and using (\ref{singletfraction}) and (\ref{basiskarimi}) and the following property,\\
\begin{eqnarray}
\langle \phi_{ij}\vert \rho \vert \phi_{ij}\rangle = Tr\:(\:\rho\:\vert \phi_{ij}\rangle\langle \phi_{ij}\vert\:),
\label{property}
\end{eqnarray}\\
it is found that the fully entangled fraction of $\rho_{ab}^{opt}$, i.e. $F(\rho_{ab}^{opt})=\frac{1}{6}\:<\:\frac{1}{3}$. Hence the state $\rho_{ab}^{opt}$ cannot be used as a resource for teleportation.\\\\
Also for any $3\otimes3$ dimensional system, the capacity of dense coding for any arbitrary given shared state $\rho_{ab}$ is defined as\\
\begin{eqnarray}
\chi = \log_{2}3 + S_{V}(\rho_{b})-S_{V}(\rho_{ab})
\label{capacityofdensecoding},
\end{eqnarray}\\
where, $S_{V}(\rho_{b})$ is the von-Neumann entropy  (already defined in section $2.13.1$ of Chapter $2$) of the reduced system $\rho_{b}$ and $S_{V}(\rho_{ab})$ is the von-Neumann entropy for the joint state $\rho_{ab}$. Now the states which can effectively be used in dense coding are known as \textit{dense-codeable} states \cite{lewenstein2004}. It is also known that  a shared quantum state $\rho_{ab}$ is said to be \textit{dense-codeable}, if the corresponding capacity $\chi$ is more than $\log_{2}(3)$. It is easily implied from (\ref{capacityofdensecoding}) that such states are precisely those for which $S_{V}(\rho_{b})-S_{V}(\rho_{ab})>0$.\\\\
It is also observed that, for the state $\rho_{ab}^{opt}$, the quantity $S_{V}(\rho_{b}^{opt})-S(\rho_{ab}^{opt})=-0.43872\:<\:0$ and therefore, the state $\rho_{ab}^{opt}$ is not dense-codeable.
\subsection{\textbf{Non-optimal Output:}}
The output generated from BH-UQCM depends on the machine parameters $c$ and $d$. For all the values of the parameter $d\:\in\:(0,\frac{1}{2}]$, (except for $d^{2}=\frac{1}{8}$), the output state $\rho_{ab}^{non-opt}$ is of non-optimal type. Although in this range of the parameter $d$ ($d^{2} \ne \frac{1}{8}$), the state $\rho_{ab}^{non-opt}$ is always an NPT state, yet when $d\:\in\:(0,\frac{1}{2}]$, the fully entangled fraction (\ref{singletfraction}) of the state $F(\rho_{ab}^{non-opt})=\frac{4\:d^{2}}{3}$ which consequently never exceeds $\frac{1}{3}$. Therefore the non-optimal state $\rho_{ab}^{non-opt}$ is also not useful as quantum teleportation channel.\\\\
An alternative approach to show that the state $\rho_{ab}^{non-opt}$ being not useful as a quantum channel for teleportation is to use a teleportation witness operator. If a hermitian operator $W$ is a teleportation witness operator, then for all the states $\sigma$ which are not useful for teleportation, one must have $Tr(W\:\sigma)\geq 0$ \cite{nirman1,nirman2}. Hence for the qutrit system, one can use the following teleportation witness operator,
\beq
W = \frac{I}{3} - \vert\phi^{+}\rangle\langle \phi^{+}\vert, \label{telepw}
\eeq 
where, $\vert\phi^{+}\rangle = \frac{1}{3}\sum_{i=0}^{2}\vert ii\rangle$. Thus,for the non-optimal output of two-qutrit system, we find that $Tr(W\:\rho_{ab}^{non-opt})=\frac{4\:d^{2}}{3}$, which is always positive for $0 <d <\frac{1}{2}$. This proves that the state $\rho_{ab}^{non-opt}$ cannot be used in teleportation.\\\\
Also, the state $\rho_{ab}^{non-opt}$ is not useful in dense coding as the quantity $S_{V}(\rho_{b}^{non-opt})-S_{V}(\rho_{ab}^{non-opt})\:<\:0$, where $\rho_{b}^{non-opt}$ is the reduced density operator of the two qutrit output state $\rho_{ab}^{non-opt}$ and is given by\\
\beq
\rho^{non-opt}_{b} = \left(%
\begin{array}{ccc}
  \frac{1}{3} & \frac{1}{3}\:k & \frac{1}{3}\:k \\
  \frac{1}{3}\:k & \frac{1}{3} & \frac{1}{3}\:k \\
  \frac{1}{3}\:k & \frac{1}{3}\:k & \frac{1}{3} \\
\end{array}%
\right),
\label{reducedrho}
\eeq\\
where, $k=d\:(2\:\sqrt{1-4\:d^{2}}+d)$.\\\\
$S_{V}(\rho_{b}^{non-opt})-S_{V}(\rho_{ab}^{non-opt})$ against the machine parameter $d$ is plotted in the figure $4.1$.
\begin{figure}[hbtp]
\centering
\resizebox{8.5cm}{5.5cm}{\includegraphics{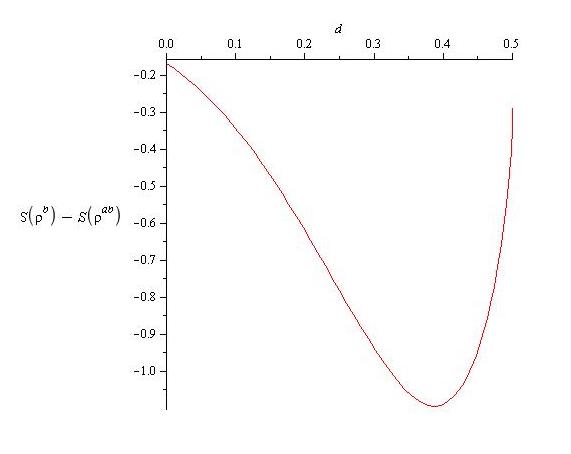}}

\caption{The figure shows that the difference $S(\rho_{b}^{non-opt})-S(\rho_{ab}^{non-opt})$ always lies in the fourth quadrant of the cartesian plane with respect to the values $d$ in the range $(0,\frac{1}{2}]$.}
\end{figure}
It is clear from figure $4.1$, that the two qutrit non-optimal output state $\rho_{ab}^{non-opt}$ is not useful for dense coding when $d\:\in\:(0,\frac{1}{2}]$.\\\\
From the above two observations, it is perceived that, neither the optimal nor the non-optimal form of the output directly obtained from the cloning machine is useful either as teleportation channel or in dense coding.\\\\
It is known that a given inseparable mixed state may conatin entanglement. To benefit from the entanglement present in the mixed state, one must convert it into the active siglet form by means of LOCC between the parties sharing the pairs of particles in the mixed state. Such a process is known as \textit{purification} or \textit{distillation} of the noisy entanglement \cite{bennett761996,deutsch771996,bennett541996,horodeckis781997}. Can the output states of BH-UQCM be distilled? \textit{Reduction Criteria} has the answer to this question, which states that $`$any state of $n \otimes n$ system violating the criteria can effectively be distilled '\cite{horodeckis1999}.
\subsection{\textbf{Teleportation and dense coding with distilled optimal state:}}
Horodecki \textit{et. al} \cite{horodeckis1999} showed that any state violating the reduction criteria is distillable. In  reduction criteria, given a state $\rho_{ab}$,  the state $\rho_{a} = Tr_{b}(\rho_{ab})$, is the reduction of the state of interest. To check whether a given state satisfies this criteria, one should check the non negativity of the eigenvalues of the operator i.e. one should have the conjunction of the following two conditions to be satisfied.\\
\begin{eqnarray}
\rho_{a} \otimes I - \rho_{ab} \geq 0 \label{redcrit1},
\end{eqnarray}\\
and
\begin{eqnarray}
I \otimes \rho_{b} - \rho_{ab} \geq 0 \label{redcrit2}.
\end{eqnarray}\\
It is easy to find that the optimal state $\rho_{ab}^{opt}$ violates the reduction criteria and hence is distillable.\\\\
The distilled $\rho_{ab}^{opt}$ can be obtained by calculating the eigenvector corresponding to the suitable negative eigenvalue of the state $(\rho_{a}^{opt} \otimes I - \rho_{ab}^{opt})$ and thereby subjecting the state to the appropriate filter. For this, one needs to calculate the eigenvector corresponding to the negative eigenvalue of the operator $(\rho_{a}^{opt} \otimes I - \rho_{ab}^{opt})$. If the form of such an eigenvector is $\vert \Phi\rangle = \sum_{i,j=1}^{N}\:a_{ij}\:\vert i\rangle\:\vert j\rangle$, then the filter $A$ is nothing but an operator which can simply be represented using a matrix where the element of the matrix can be given as $A_{ij}=\sqrt{N}\:a_{ij}$ \cite{horodeckis1999}. \\\\
Also, it is known that if $\rho$ is any state and $A$ is a filter then the distilled form of the state $\rho$ (say, $\rho_{distilled}$) is given by\\
\begin{eqnarray}
\rho_{distilled}= \frac{A^{\dagger}\: \otimes \:I \:\rho A \:\otimes \: I}{Tr (\:\rho A\:A^{\dagger}\: \otimes \: I)} \label{filteredform}.
\end{eqnarray}\\
By following the procedures described in \cite{horodeckis1999}, the filter for the optimal state $\rho_{ab}^{opt}$ is now constructed. The filter is denoted by $A_{opt}$ and is given by\\
\begin{eqnarray}
A_{opt} = \left(%
\begin{array}{ccc}
 \sqrt{3}(\frac{3}{2}-\frac{\sqrt{29}}{2}) &  \sqrt{3}(-\frac{7}{2}+\frac{\sqrt{29}}{2}) &  -\sqrt{3}\\
 \sqrt{3}(\frac{7}{2}-\frac{\sqrt{29}}{2}) &  \sqrt{3}(-\frac{3}{2}+\frac{\sqrt{29}}{2}) &  \sqrt{3}\\
 \sqrt{3}(\frac{5}{2}-\frac{\sqrt{29}}{2}) &  \sqrt{3}(-\frac{5}{2}+\frac{\sqrt{29}}{2}) &  0\\
\end{array}%
\right)\label{optfilter}.
\end{eqnarray}\\
Using (\ref{optfilter}), the distilled form of the optimal state $\rho^{opt}_{ab}$ is attained. Let it be denoted by $\rho^{opt}_{distilled^{ab}}$ and consequently it is of the form\\
\begin{eqnarray}
\rho^{opt}_{distilled^{ab}} = \frac{A_{opt}^{\dagger}\: \otimes I \:\rho^{opt}_{ab} \:A_{opt}\: \otimes I}{Tr (\rho^{opt}_{ab} \:
A_{opt} A_{opt}^{\dagger}\: \otimes I)} \label{filteredoptrhoout}.
\end{eqnarray}\\
Interestingly, it is seen that the fully entangled fraction  of  $\rho^{opt}_{distilled^{ab}}$ i.e. $ F(\rho^{opt}_{distilled^{ab}}) = 0.38789 > \frac{1}{3}$ implying that the distilled optimal output state (\ref{filteredoptrhoout}) is now useful in teleportation. Moreover, the optimal teleportation fidelity for any arbitrary state $\rho$ in $n - $ dimensional system is defined as \cite{horodeckis1999}\\
\begin{eqnarray}
f(\rho) = \frac{n\: F(\rho) + 1}{n + 1} \label{telefidy}.
\end{eqnarray}\\
Therefore in this case the optimal teleportation fidelity of the state $\rho^{opt}_{distilled^{ab}}$ is obtained as\\
\begin{eqnarray}
f(\rho^{opt}_{distilled^{ab}}) =  \frac{3 \:F\left\{\rho^{opt}_{distilled^{ab}}\right\} + 1}{4}=0.5409 \label{valueopttelefidy}.
\end{eqnarray}\\
On the other hand, using (\ref{capacityofdensecoding}), one can easily verify that  $S_{V}(\rho^{opt}_{distilled^{a}}) - S_{V}(\rho^{opt}_{distilled^{ab}}) = -0.3327 < 0$. Therefore, the distilled optimal state (\ref{filteredoptrhoout}) is still not useful for dense coding.\\
\subsection{\textbf{Teleportation and dense coding with distilled non-optimal state:}}
Similar to the optimal output state, the non optimal form of the output state i.e. $\rho_{ab}^{non-opt}$ violates the reduction criteria also \cite{horodeckis1999}. So as before one can  distill $\rho_{ab}^{non-opt}$ by calculating the eigenvector corresponding to the suitable negative eigenvalue of the state, $(\rho_{a}^{non-opt} \otimes I - \rho_{ab}^{non-opt})$, subjecting the state to the appropriate filter. The eigenvalue which is negative for any values of parameter $d \in (\frac{6+\sqrt{2}}{17}, \frac{1}{2}]$ is found and is given by\\
\beq
e =\frac{1-3d^{2}}{6}+\frac{1}{3}\:\sqrt{-(2d-1)(2d+1)}\:d\nonumber\\-\frac{1}{6}\:\sqrt{1-18d^{2}++4\sqrt{-(2d-1)(2d+1)}\:d+113\:d^{4}-44\:d^{3}\:\sqrt{-(2d-1)(2d+1)}}.\nonumber\\
\label{eigenvaluenonopt}
\eeq\\
By taking $d \in (\frac{6+\sqrt{2}}{17}, \frac{1}{2}]$ and following the procedures described in \cite{horodeckis1999} a filter is constructed then, which is shown in the following,\\
\begin{eqnarray}
A_{non-opt} =  \sqrt{3}\:\left(%
\begin{array}{ccc}
 1 &  -r &  -r\\
-r &   1 &  -r\\
-r &  -r &   1 \\
\end{array}%
\right) \label{nonoptfilter1},
\end{eqnarray}\\\\
where $r= \frac{11\:d^{2}-2\:\sqrt{1-4d^{2}}\:d + \sqrt{1- 18d^{2}+ 4\:\sqrt{1 - 4d^{2}}\:d+113\:d^{4}-44\:d^{3}\sqrt{1-4\:d^{2}}}}{4\:d^{2}}$.\\\\ The filter (\ref{nonoptfilter1}) will transform the non optimal state $\rho^{non-opt}_{ab}$ to its distilled form $\tau^{non-opt}_{distilled^{ab}}$, which is given by\\
\begin{eqnarray}
\tau^{non-opt}_{distilled^{ab}} = \frac{A_{non-opt}^{\dagger}\: \otimes I\: \rho^{non-opt}_{ab} \: A_{non-opt}\otimes\: I}{Tr (\rho^{non-opt}_{ab}\: A_{non-opt}\:A_{non-opt}^{\dagger}\: \otimes \:I)} \label{filterednonoptrhoout1}.
\end{eqnarray}\\
Now the fully entangled fraction of $\tau^{non-opt}_{distilled^{ab}}$ is given as\\
\begin{eqnarray}
F(\tau^{non-opt}_{distilled^{ab}}) = \nonumber\\\nonumber\\
\frac{4[d^{2}(2(1-t_{1})+d^{2}(22t_{1}-31)+t_{2}(10-110d^{2}-6t_{1})+198d^{4})]}{3[(1-k)+t_{2}(6-68d^{2}+94d^{4}-4t_{1}+12t_{1}d^{2})+d^{2}(6t_{1}-9-5d^{2}+23t_{1}d^{2})]},\nonumber\\ \label{feffilnonopt1}
\end{eqnarray}\\
where $t_{2}=d\:\sqrt{1-4\:d^{2}}$ and $t_{1}=\sqrt{1-18\:d^{2}+4\:t_{2}+113\:d^{4}-44\:d^{2}\:t_{2}}$.\\\\ It is evident from the above eq. (\ref{feffilnonopt1}) that, for $\frac{6+\sqrt{2}}{17} < d \leq \frac{1}{2}$, the state $\tau^{non-opt}_{distilled^{ab}}$ is not suitable for teleportation even after distillation of original non-optimal form of the output, since in this case too it is found that $F (\tau^{non-opt}_{distilled^{ab}}) < \frac{1}{3}$. This is also clear from the following figure $4.2$.\\
\begin{figure}[hbtp]
\centering
\resizebox{7.5cm}{5.5cm}{\includegraphics{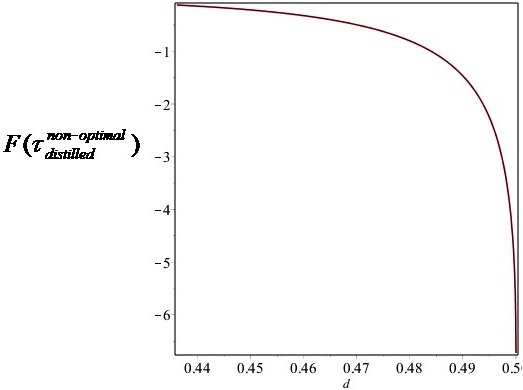}}

\caption{The ordinate of the figure represents the maximal singlet fraction of the distilled non-optimal output state which has been plotted against the parameter $d\in (\frac{6+\sqrt{2}}{17},\frac{1}{2}]$.}
\end{figure}\\
However, it is interesting to see that though the state $\tau^{non-opt}_{distilled^{ab}}$ is not useful for teleportation, the state $\tau^{non-opt}_{distilled^{ab}}$ can be used in dense coding, which is clear from the figure $4.3$.\\\\
A plot of capacity of dense coding of the state $\tau^{non-opt}_{distilled^{ab}}$ against the parameter $d$ is shown in figure $4.3$. The figure confirms that the non-optimal output from BH-UQCM is not dense-codeable while after the distillation process the distilled form of the non-optimal output state can, however, be used in dense coding when $d=\frac{1}{2}$, as the capacity of dense coding $\chi(\tau^{non-opt}_{distilled^{ab}})\:>\:2$.\\
\begin{figure}[hbtp]
\centering
\resizebox{7.5cm}{5.5cm}{\includegraphics{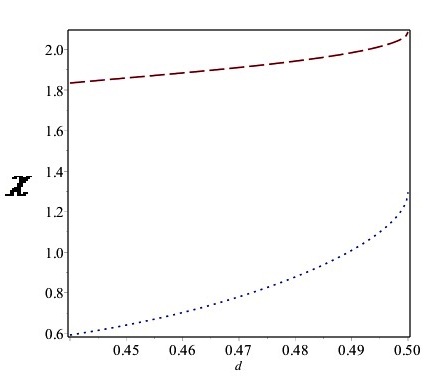}}

\caption{In the figure the dotted line represents the capacity of dense coding i.e. $\chi(\tau^{non-opt}_{{ab}})$ of the non-optimal state before filtering and dashed line represent the capacity of dense coding  i.e. $\chi(\tau^{non-opt}_{distilled^{ab}})$ of the distilled non-optimal state with respect to the state parameter $d\in (\frac{6+\sqrt{2}}{17},\frac{1}{2}]$.}
\end{figure}\\
The answers to the questions that were raised in the beginning, are now summarized below.\\
\begin{itemize}
\item The output states generated from Buzek - Hillery universal quantum cloning machine are indeed negative partial transposed states (the characteristic of the state depends on the machine parameters), which implied that these states could be tested in the quantum information processing tasks like teleportation and dense coding.
\item Two types of outputs were focussed here viz. the optimal form and the  non-optimal form of the output. The utility of both these types were examined in information processing.
\item Both the optimal and non-optimal forms of the output directly obtained from BH-UQCM were initially not useful in teleportation and dense coding.
\item It was then shown that both optimal and non-optimal cloned outputs from the machine violated reduction criteria \cite{horodeckis1999} and that they could be distilled. To distill these states, appropriate forms of the filters were found which were shown in eqs. (\ref{optfilter}) and (\ref{nonoptfilter1}).
\item The distilled forms of the outputs (optimal and non-optimal) can then be shown to be entangled resource for the information tasks for a certain range of machine parameters i.e. for $d\in (\frac{6+\sqrt{2}}{17},\frac{1}{2}]$. The distilled form of the optimal output is shown to be useful in teleportation but cannot be used in dense coding. It is however, interesting to note down that, though  the distilled non-optimal output is still not useful in teleportation for $d\in (\frac{6+\sqrt{2}}{17},\frac{1}{2}]$, yet for $d = \frac{1}{2}$, the state (\ref{filterednonoptrhoout1}) is dense-codeable.
\end{itemize}
\chapter{Controlled Dense Coding}
\label{ch:cdc}
\textit{$``$... And a new philosophy emerged called quantum physics, which suggests that the individual's function is to inform and to be informed. You really exist only when you are in a field sharing and exchanging information. You create the realities you inhabit..."}
\begin{flushright}
- Timothy Francis Leary, (American Psychologist)
\end{flushright}
\vskip1cm
\section{\textbf{Introduction:}}
In earlier chapters, the discussions about information processing schemes such as teleportation and dense coding have been made, where efficacies of teleportation channels portrayed  by states like MEMS, NMEMS (in Chapter $3$) were proved. Also it was shown how the mixed entangled outputs from a cloning machine could be utilized in the domains of teleportation and dense coding (detailed discussion was made in Chapter $4$). In all these cases one thing was common i.e. the entanglement present in the states shared by the parties involved in the quantum information processing tasks. The entanglement was used as a resource for doing those tasks.\\\\ One of the many surprising applications of shared entanglement in the domain of quantum information is dense coding originally proposed by Bennett \textit{et. al} \cite{bennett1992}. In the protocol, two parties Alice and Bob shared initially a pair of entangled qubits in a Bell-state. Once Alice performed any of the four unitary operations (identity $I$ or the Pauli spin operators $\sigma^{x}$, $\sigma^{y}$ and $\sigma^{z}$) on her qubit, the initial state of two qubits was mapped to a different member of Bell-basis. These Bell-basis states are mutually orthogonal and are fully distinguishable which were used to encode two bits of classical information. Then after encoding her qubit Alice sent it to Bob, who on his end would extract those two bits of classical information by performing a joint measurement on his qubit and the qubit which he got from Alice.\\\\
The question is, instead of two parties, if there are three parties who share a pure (tripartite) maximally entangled state, then can any one of these three sharing the state control the environment in such a way so that he will modulate the entanglement retained by the other two parties, thereby controlling the information transfer between them as well as the success probability of the scheme? The scheme of sharing bits of classical information between sender and the receiver under such modulated environment by a third party is known as \textit{Controlled dense coding} scheme. Likewise any teleportation scheme performed in controlled settings is known as controlled teleportation scheme. Many works have been carried out on controlled teleportation so far \cite{karlsson1998,yan2003,ting2005,lxhan2007,man2007,song2008,gao2008,zha2013}. Although the concept of controlled dense coding (CDC) is not new to the domain of quantum information processing, some more study regarding this protocol need to be discussed to get the essence of how entanglement plays central role in the theory of quantum information.\\\\
First pioneering step on CDC was initiated by Hao \textit{et. al} \cite{hao2001}, where the three parties Alice, Bob and Cliff shared a tripartite pure maximally entangled state (called GHZ state). The details of their scheme have already been discussed in section $2.16.4$. Some other notable works have also been done in this arena of CDC \cite{peng2002,yynie2008,wangwu2009,sixcdc2011,cavitycdc2013}.\\\\
In this chapter\footnote{The Chapter is mainly based on our work.
\textsc{S. Roy and B. Ghosh}, \textbf{$'$Study of Controlled Dense Coding with some Discrete Tripartite and Quadripartite States'}, \textsc{International Journal of Quantum Information, Vol. 13, No. 5, 1550033-(1-20), (2015), World Scientific}.} a few further questions have been raised.
\begin{itemize}
\item Are there any other types of multi-partite entangled states that can be used in CDC?
\item Can W - class of states be used in controlled dense coding?
\item How does CDC work in $3\otimes3$ system?
\end{itemize}
\section{\textbf{$3$- and $4$- qubit GHZ states:}}
\subsection{\textbf{CDC with GHZ-class of states:}}
Controlled dense coding was successfully achieved with Greenberger-Horne-Zeilinger state by Hao \textit{et. al}\cite{hao2001}. Various other orthogonal states can be created on application of unitary operations on initial GHZ state $\frac{\vert 000\rangle + \vert 111\rangle}{\sqrt{2}}$ \cite{pathakbook}. One can call such states obtained from initial GHZ state as members of GHZ-class of states. Some of these states are illustrated below.\\
\begin{center}
\begin{tabular}{|c|c|}
\hline State& Obtained after orthogonal transformation to the GHZ state  \\
\hline $G_{1}$& $\frac{1}{\sqrt{2}}(\vert 010 \rangle _{ABC} + \vert 101 \rangle_{ABC})$\\
\hline $G_{2}$&  $\frac{1}{\sqrt{2}} (\vert 010 \rangle _{ABC} - \vert 101 \rangle_{ABC})$\\
\hline $G_{3}$&  $\frac{1}{\sqrt{2}}(\vert 001 \rangle _{ABC} - \vert 110 \rangle_{ABC})$\\
\hline $G_{4}$&  $\frac{1}{\sqrt{2}}(\vert 001 \rangle _{ABC} + \vert 110 \rangle_{ABC})$\\
\hline $G_{5}$&  $\frac{1}{\sqrt{2}}(\vert 100 \rangle _{ABC} - \vert 011 \rangle_{ABC})$\\
\hline $G_{6}$&  $\frac{1}{\sqrt{2}}(\vert 100 \rangle _{ABC} + \vert 011 \rangle_{ABC})$\\
\hline $G_{7}$&  $\frac{1}{\sqrt{2}}(\vert 000 \rangle _{ABC} - \vert 111 \rangle_{ABC})$\\
\hline
\end{tabular},
\end{center}\
\newline
where subscripts $A$, $B$ and $C$ are the parties Alice, Bob and Cliff respectively. Following the scheme that was followed throughout the work of \cite{hao2001} and which was reviewed in section $2.16.4$ of Chapter $2$, proper choice of measurement basis $\lbrace \vert +\rangle_{C}, \vert -\rangle_{C}\rbrace$ by controller Cliff decided whether sender Alice, indeed, was capable of transmitting bits to the receiver Bob. After confirming from Cliff about his measurement outcome, Alice  performed a joint unitary operation on her qubit from the shared state and on the auxiliary qubit (this qubit was introduced into the system by Alice herself) together. Alice's von-Neumann measurement on the auxiliary qubit, the sending of the qubits then to Bob and Bob's subsequent CNOT\footnote{CNOT operation is also known as CONTROLLED NOT operation and has been discussed in Appendix.} operation on his qubit paved the way for the success of CDC.
In the same way, however, it can easily be shown that the average number of bits transmitted for states like $G_{1}$, $G_{4}$ and $G_{6}$ is $1+2\:\vert \sin\:\theta\:\vert^{2}$ and for states like $G_{2}$, $G_{3}$, $G_{5}$ and $G_{7}$, it is $1+2\:\vert \cos\:\theta\:\vert^{2}$. So the Hao \textit{et. al} scheme is also successful with members of GHZ class of states. This is clear from the following pictorial representation (figure $5.1$) also.\\
\begin{figure}[hbtp]
\centering
\resizebox{7.5cm}{5.5cm}{\includegraphics{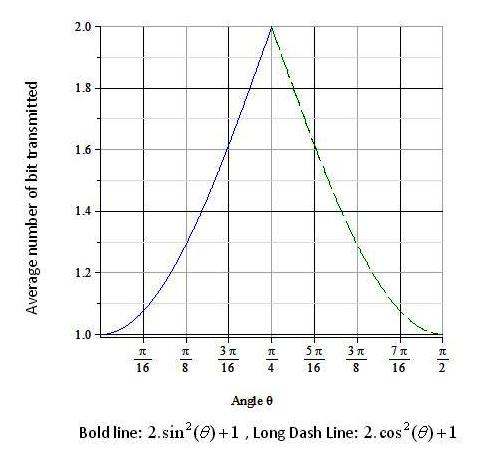}}

\caption{The figure represents average number of bits transmitted against the angle $\theta$. The solid line represents $(1 + 2\sin^{2}\theta)$ bits are transmitted for some GHZ forms for $0 \leqslant \theta \leqslant \frac{\pi}{4}$. The dashed line represents $(1 + 2\cos^{2}\theta)$ bits are transmitted for some other GHZ forms for $\frac{\pi}{4} \leqslant \theta \leqslant \frac{\pi}{2}$ whereas optimum value of bit transmission is $2$.}
\end{figure}\\
When Alice shares Bell-states with Bob, then she can send $2-bits$ of classical information to Bob by sending a single qubit, while if they share GHZ state then Alice is able to send only $\log_{2}\:8=3$ bits of classical message to Bob by sending two qubits, instead of $4-bits$ of message which was quite expected \cite{cereceda2001}. This proves that dense coding with GHZ state is not as efficient as dense coding with Bell-state. But, controlled dense coding with GHZ state and with its various classifications are constructive from the view point of modulation of controlled environment by any of the parties.
\subsection{\textbf{GHZ type states in CDC:}}
It is a well known fact that the states like $\frac{1}{\sqrt{3}}\: (\:\sqrt{2}\:\vert 000 \rangle _{ABC} + \vert 111 \rangle_{ABC})\:$ cannot be used for perfect super dense coding \cite{liqiu2007}. Pati et. al showed in reference \cite{patiparaagra2005} that with such states dense coding is performed probabilistically.  A general form of such states is now considered here.\\
\beq
\vert P\rangle_{ABC} = L\:(\vert 000\rangle_{ABC} + l\:\vert 111\rangle_{ABC}),
\label{patistate1}
\eeq\\
where, $L=\frac{l}{\sqrt{1+l^{2}}}$ and $l>0$ (considering $l$ to be real) are the parameters of the state. The normalization condition however shows that the desirable value of the parameter $l$ is $1$ and consequently $L=\frac{1}{\sqrt{2}}$ so that the state (\ref{patistate1}) takes the form of well-known GHZ state.\\\\ 
Cliff here chooses his measurement basis from eq. (\ref{charliebasis}). With respect to these basis, the state (\ref{patistate1}) can be expressed as\\
\beq
\vert P \rangle_{ABC} = \frac{1}{\sqrt{1+l^{2}}}\:(\:\vert p_{1}\rangle_{AB} \vert +\rangle_{C}\:
+\vert p_{2}\rangle_{AB}\:\vert -\rangle_{C}\:)
\label{patistate2},
\eeq \\
where\\
\beq
\vert p_{1}\rangle_{AB} = \cos\:\theta\:\vert 00\rangle_{AB} + l\:\sin\:\theta\:\vert 11\rangle_{AB},\nonumber\\
\vert p_{2}\rangle_{AB} = \sin\:\theta\:\vert 00\rangle_{AB} - l\:\cos\:\theta\:\vert 11\rangle_{AB}
\label{patistate3}.
\eeq
\\
If von-Neumann measurement of Cliff gives the readout as $\vert +\rangle_{C}$, the non-maximally entangled state shared between Alice and Bob will be $\vert p_{1}\rangle_{AB}$ (while for the readout $\vert -\rangle_{C}$, it is $\vert p_{2}\rangle_{AB}$). After Alice introduces auxiliary qubit $\vert 0\rangle_{aux}$ and considers the unitary operator \\\\ 
\beq
U_{1} = \left(%
\begin{array}{cccc}
\frac{\sin\:\theta}{\cos\:\theta}& 0& \sqrt{1-\frac{\sin^{2}\:\theta}{\cos^{2}\:\theta}}& 0\\
0& 1& 0& 0\\
0& 0& 0& -1\\
\sqrt{1-\frac{\sin^{2}\:\theta}{\cos^{2}\:\theta}}& 0& -\frac{\sin\:\theta}{\cos\:\theta}& 0
\end{array}%
\right),
\label{unitarymatrix2}
\eeq\\
the collective unitary operation $U_{1}\otimes I_{B}$ transforms the state $\vert p_{1}\rangle_{AB} \otimes \vert 0\rangle_{aux}$ to\\
\beq
\vert p_{1}\rangle_{A\:B\:aux} = \frac{\sin\:\theta}{L}\:\left\{\:L\:(\vert 00\rangle_{AB} + l\:\vert 11\rangle_{AB}\:)\:\right\}\:\otimes\:\vert 0\rangle_{aux}
+ \nonumber\\ \sqrt{1-\frac{\sin^{2}\:\theta}{\cos^{2}\:\theta}}\:\cos\:\theta\:\vert 10\rangle_{AB} \otimes \vert 1\rangle_{aux}.
\label{patistate3}
\eeq\\
Alice's von-Neumann measurement of $\vert 1\rangle_{aux}$ shows that the qubits of Alice and Bob are un-entangled and only one bit is transferred from Alice to Bob. The measurement outcome of $\vert 0\rangle_{aux}$ however justifies that Alice and Bob shares $L\: (\:\vert 00 \rangle _{AB} + l\:\vert 11 \rangle_{AB})$. The average number of bits transmitted is then $1 + \frac{\vert \sin\:\theta \vert^{2}}{L^{2}}$, and the scheme is successful only if $\theta\:=\sin^{-1}\:(\pm\frac{1}{\sqrt{1+l^{2}}})$. Alice applies any one of the four unitary operators $\lbrace I, \sigma^{x}, i\sigma^{y},\sigma^{z} \rbrace$. The shared state between Alice and Bob then may take anyone of the following forms, given as\\
\begin{eqnarray}
L(\vert 00\rangle_{AB}+l\:\vert 11\rangle_{AB}),\nonumber{}\\ 
L(\vert 10\rangle_{AB}+l\:\vert 01\rangle_{AB}),\nonumber{}\\ 
L(-\vert 10\rangle_{AB}+l\:\vert 01\rangle_{AB}),\nonumber{}\\ 
L(\vert 00\rangle_{AB}-l\:\vert 11\rangle_{AB}). 
 \label{basis}
\end{eqnarray}\\
Alice sends her qubit to Bob. Bob performs a projection on to the basis spanned by the basis states $\lbrace \vert 00 \rangle , \vert 11\rangle \rbrace$ and $\lbrace \vert 01 \rangle , \vert 10\rangle \rbrace$. Bob can extract two bits of classical information with a success probability of $\frac{2\:l^{2}}{1+l^{2}}$. Thus for maximally entangled state, i.e. for $l=1$, the success probability is unity. It is also clear that for maximally entangled state, Cliff's measurement angle is therefore $\pm \frac{\pi}{4}$. The above analysis is shown in the table below\\
\begin{center}
\begin{tabular}{|c|c|c|c|}
\hline $l$&  $\theta$&  shared state&  Success probability = $\frac{2\:l^{2}}{1+l^{2}}$\\
\hline $0$ &  $\pm \frac{\pi}{2}$ &  $\vert 00 \rangle_{AB}$& $0$ \\
\hline $1$ &  $\pm \frac{\pi}{4}$ &   $\frac{1}{\sqrt{2}}[\vert 00\rangle_{AB} + \vert 11\rangle_{AB}]$& $1$\\
\hline
\end{tabular}\\
\end{center}\vskip0.5cm
\noindent It is seen that with states of the form (\ref{patistate1}), controlled dense coding is thus achieved with unit probability of success only when Cliff chooses a proper measurement angle $\theta$ and with proper choice of state parameter (which is in this case $l=1$). A relation between the parameter $l$ and angle $\theta$ can then be formulated as,\\
\beq
\theta\:=\tan^{-1}\frac{1}{l}.\: \label{thetal}
\eeq\\
The relation (\ref{thetal}) has been shown graphically in figure $5.2$.\\
\begin{figure}[hbtp]
\centering
\resizebox{7.5cm}{5.5cm}{\includegraphics{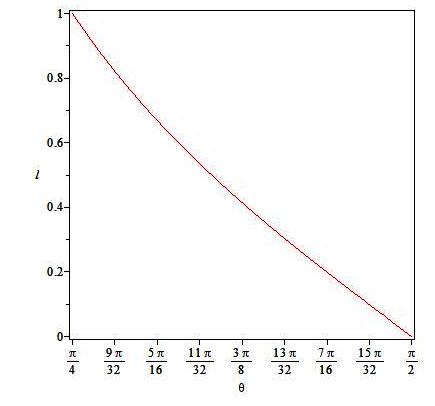}}

\caption{The figure shows that for $\frac{\pi}{4}\:\leq\:\theta\:\leq\:\frac{\pi}{2}$, we have $1\:\geq l\:\geq\: 0$. The scheme of CDC is successful when $\theta=\frac{\pi}{4}$ and $l=1$.}
\end{figure}\\
The entanglement of a bipartite state is given by concurrence, a measure which was first introduced in \cite{wootters1998} and has been discussed in the section $2.12.2$. Using eq. (\ref{concurrence}), the concurrence $C$ of the shared state $L\:(\:\vert 00\rangle_{AB}+l\:\vert 11\rangle_{AB})$ is found to be\\
\beq
C=\vert 2\:L^{2}\:l\:\vert
\label{concurrencel}.
\eeq\\
Also using (\ref{thetal}) and (\ref{concurrencel}), a relation between $C$ and $\theta$ is obtained that is given by\\
\beq
C = \vert \sin\:2\:\theta\:\vert
\label{Candtheta}.
\eeq\\
Figure $5.3$ is hence plotted to show the relation (\ref{Candtheta}) pictorially.\\
\begin{figure}[hbtp]
\centering
\resizebox{7.5cm}{5.5cm}{\includegraphics{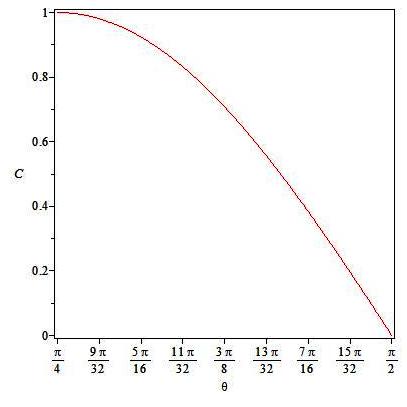}}

\caption{The figure shows that for $\frac{\pi}{4}\:\leq\:\theta\:\leq\:\frac{\pi}{2}$, we have $1\:\geq C\:\geq\: 0$.}
\end{figure}\\
The above analysis shows that, before dense coding is executed between Alice and Bob, what state will they share, is also controlled by Cliff. The protocol of dense coding is done successfully provided Cliff fixes his measurement angle to $\theta\:=\frac{\pi}{4}$, as in this case only Alice and Bob shares maximally entangled state and the success probability is therefore $1$.\\
\subsection{\textbf{Four particle GHZ state and CDC:}}
Fu \textit{et.al} have utilised the protocol of controlled dense coding with a non-maximally entangled state of the form shown in \cite{fuxializh2005}. In this section Fu protocol of CDC is discussed for four particle GHZ state. The state is defined as follows:\\
\beq
\vert GHZ\rangle_{PABC}\:= \frac{1}{\sqrt{2}}\lbrace\:\vert 0000\rangle_{PABC}+ \vert 1111 \rangle_{PABC}\:\rbrace.
\label{ghz4}
\eeq\\
where $P$, $A$, $B$ and $C$ respectively represents Paul, Alice, Bob and Cliff. Cliff chooses his measurement basis as given in (\ref{charliebasis}). The state (\ref{ghz4}) is therefore expressed as\\
\beq
\vert GHZ\rangle_{PABC}\:= \frac{1}{\sqrt{2}}[\:\vert \varsigma\rangle_{PAB}\vert +\rangle_{C}+ \vert \tau \rangle_{PAB}\vert -\rangle_{C}\:]
\label{ghz4a},
\eeq\\
such that\\
\begin{eqnarray}
\vert \varsigma\rangle_{PAB}\:= \:\cos\:\theta \:\vert 000\rangle_{PAB} + \:\sin\:\theta \:\vert 111 \rangle_{PAB},\nonumber{}\\
\vert \tau\rangle_{PAB}\:= \:\sin\:\theta \:\vert 000\rangle_{PAB} - \:\cos\:\theta \:\vert 111 \rangle_{PAB}.
\label{ghz4a}.
\end{eqnarray}\\
When Cliff chooses a basis\\
\begin{eqnarray}
\lbrace\:\vert +\rangle _{C} &=& \cos \varepsilon\:\vert 0 \rangle_{C}\: + \sin \varepsilon \: \vert1 \rangle_{C},\nonumber\\
\vert - \rangle_{C} &=& \sin \varepsilon\:\vert 0 \rangle_{C}\: - \cos \varepsilon \: \vert 1 \rangle_{C}\:\rbrace,\ \label{paulbasis2}
\end{eqnarray}\\
and carries out a unitary operation on his qubit and if Cliff's measurement result gives $\vert +\rangle_{C}$, then Paul, Alice and Bob shares the state $\vert \varsigma\rangle_{PAB}$. If a similar basis like (\ref{paulbasis2}) is chosen by Paul, then the state, $\vert \varsigma\rangle_{PAB}$ can be expressed in the form\\
\beq
\vert \varsigma\rangle_{PAB}\:= \vert \mu \rangle_{AB}\otimes \vert +\rangle_{P} + \vert \nu \rangle_{AB}\otimes \vert -\rangle_{P}
\label{ghz4b},
\eeq\\
where\\
\begin{eqnarray}
\vert \mu \rangle_{AB}=\cos\:\theta\:\cos\:\varepsilon \vert 00\rangle_{AB}+\sin\:\theta\:\sin\:\varepsilon \vert 11\rangle_{AB},\nonumber{}\\
\vert \nu \rangle_{AB}=\cos\:\theta\:\sin\:\varepsilon \vert 00\rangle_{AB}-\sin\:\theta\:\cos\:\varepsilon \vert 11\rangle_{AB}.
\label{ghz4b1}
\end{eqnarray}\\
For Paul's local measurement result $\vert +\rangle_{P}$, the shared state between Alice and Bob is $\vert \mu \rangle_{AB}$. Alice then introduces an auxiliary qubit $\vert 0\rangle_{aux}$ and considering the unitary operator\\
\begin{eqnarray}
U_{2} = \left(%
\begin{array}{cccc}
\frac{\sin\:\theta\:\sin\:\varepsilon}{\cos\:\theta\:\cos\:\varepsilon}& 0& \sqrt{1-\frac{\sin^{2}\:\theta\:\sin^{2}\:\varepsilon}{\cos^{2}\:\theta\:\cos^{2}\:\varepsilon}}& 0\\
0& 1& 0& 0\\
-\sqrt{1-\frac{\sin^{2}\:\theta\:\sin^{2}\:\varepsilon}{\cos^{2}\:\theta\:\cos^{2}\:\varepsilon}}& 1& \frac{\sin\:\theta\:\sin\:\varepsilon}{\cos\:\theta\:\cos\:\varepsilon}& 0\\
0& 0& 0& -1\\
\end{array}%
\right)\label{unitarymatrix3},
\end{eqnarray}\\
and henceforth using the collective unitary operation $U_{2}\otimes I_{B}$,\: she transforms $\vert \mu \rangle_{AB} \otimes\: \vert 0\rangle_{aux}$ to the following\\
\begin{eqnarray}
\vert \mu \rangle_{A\:B\:aux}\:=\:\sin\:\theta\:\sin\:\varepsilon\:[\vert 00\rangle_{AB}+\vert 11\rangle_{AB}]\otimes \vert 0\rangle_{aux}\nonumber{}\\\nonumber{}\\
-\:\:\sqrt{1-\frac{\sin^{2}\:\theta\:\sin^{2}\:\varepsilon}{\cos^{2}\:\theta\:\cos^{2}\:\varepsilon}}\:\cos\:\theta\:\cos\:\varepsilon\:\vert 00\rangle_{AB}\otimes \vert 1\rangle_{aux}
\label{ghz4c}.
\end{eqnarray}\\
The von-Neumann measurement outcome $\vert 0\rangle_{aux}$ of Alice shows that the non-maximally entangled state shared between Alice and Bob is therefore\:\:\: $\sin\:\theta\:\sin\:\varepsilon\:(\vert 00\rangle_{AB}+\vert 11\rangle_{AB})$ (while the readout $\vert 1\rangle_{aux}$ shows qubits of Alice and Bob are un-entangled). \\\\
Using \cite{wootters1998}, the concurrence $C_{1}$ of the shared state $\sin\:\theta\:\sin\:\varepsilon\:(\vert 00\rangle_{AB}+\vert 11\rangle_{AB})$ is shown to be\\
\beq
C_{1}=\:2\:\sin^{2}\theta\:\sin^{2}\varepsilon.
\label{4ghzshared}
\eeq\\
The above relation (\ref{4ghzshared}) is shown graphically below in figure $5.4$.\\
\begin{figure}[hbtp]
\centering
\resizebox{6.5cm}{4.5cm}{\includegraphics{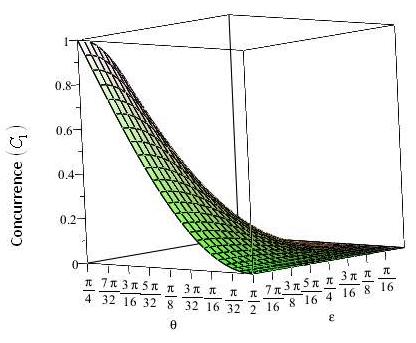}}

\caption{Here $0 \leq C_{1} \leq 1$ whereas $0 \leq \theta \leq \frac{\pi}{4}$ and $0 \leq \varepsilon \leq \frac{\pi}{2}$. It is clear that when $\theta = \frac{\pi}{4}$ and $\varepsilon = \frac{\pi}{2}$, concurrence is maximum i.e. $1$.}
\end{figure}\\
Hence when Paul fixes his measurement angle to $\varepsilon = \frac{\pi}{2}$, the state shared between Alice and Bob is ~~$\sin\:\theta\:[\vert 00\rangle_{AB} + \vert 11\rangle_{AB}]$ . When $\theta=\frac{\pi}{4}$, Alice and Bob shares maximally entangled state and two bits are transferred. What state will be shared by Alice and Bob is controlled both by Paul and Cliff.  The bit transmission between Alice and Bob is controlled by them too.\\
\section{\textbf{W-class of States:}}
The pure maximally entangled tripartite states can be classified into two categories with respect to their genuine tripartite entanglement viz. $GHZ$ state and $W$ state \cite{durr622000}. The entanglement properties of the $GHZ$ state are very fragile under loss of particles while entanglement of $W$ state is maximally robust under the loss of any one of the three qubits. The utility of $GHZ$ state has already been proved in controlled dense coding. Will Alice, Bob and Cliff be able to utilize the entanglement characteristic of $W$ state in controlled dense coding?\\\\ 
The following two states belong to $W-$ class of states viz.\\
\beq
\vert W^{(1)} \rangle_{ABC}&=&\frac{1}{\sqrt{3}}\left\{\vert 100\rangle_{ABC} + \vert 010\rangle_{ABC} + \vert 001\rangle_{ABC}\right\},\nonumber\\
\vert W^{(2)} \rangle_{ABC}&=&\frac{1}{2}\left\{\vert 100\rangle_{ABC} + \vert 010\rangle_{ABC} + \sqrt{2}\:\vert 001\rangle_{ABC}\right\}
\label{w1}.
\eeq
It is very interesting to note down that although $\vert W^{(1)} \rangle_{ABC}$ (prototypical $W-$ state) cannot be used in teleportation and dense coding, $\vert W^{(2)} \rangle_{ABC}$ (non-prototypical $W-$ state) can, however, be \cite{agrapati2006} used in both teleportation as well as in dense coding. Below it has been shown that the prototype $W-$ state i.e. $\vert W^{(1)} \rangle_{ABC}$ is also not useful in CDC.\\
\subsection{\textbf{CDC with W state:}}
When Cliff chooses his basis as given in (\ref{charliebasis}), the prototype $W-$ state can be expressed as\\
\begin{eqnarray}
\vert W^{(1)} \rangle_{ABC}=\frac{1}{\sqrt{3}}[\:\vert +\rangle_{C}\vert \varpi^{1}\rangle_{AB}
+\:\vert -\rangle_{C}\vert \varpi^{2}\rangle_{AB}]
\label{w2},
\end{eqnarray}\\
where\\
\begin{eqnarray}
\vert \varpi^{1}\rangle_{AB}=\cos\:\theta\: \vert 10\rangle_{AB}+\cos\:\theta\: \vert 01\rangle_{AB}+\sin\:\theta\: \vert 00\rangle_{AB}\nonumber{}\\
\vert \varpi^{2}\rangle_{AB}=sin\:\theta \:\vert 10\rangle_{AB}+\sin\:\theta\: \vert 01\rangle_{AB}-\cos\:\theta \:\vert 00\rangle_{AB}
\label{w3}.
\end{eqnarray}\\
If Cliff's von-Neumann measurement readout is $\vert +\rangle_{C}$, Alice and Bob will share non-maximally entangled state $\vert \varpi^{1}\rangle_{AB}$. Alice introduces auxiliary qubit $\vert 0\rangle_{aux}$  as before and she considers unitary operator defined in equation (\ref{unitarymatrix2}), such that the collective unitary operation $U_{1} \otimes I_{B}$ transforms the state $\vert \varpi^{1}\rangle_{AB} \otimes \vert 0\rangle_{aux}$ to\\
\begin{eqnarray}
\vert \varpi^{1}\rangle_{A\:B\:aux}=\:\left\{\sin\:\theta\:\vert 01\rangle_{AB}+\frac{\sin^{2}\theta}{\cos\:\theta}\:\vert 00\rangle_{AB}+\cos\:\theta\:\vert 10\rangle_{AB}\right\}\otimes \vert 0\rangle_{aux}\nonumber{}\\\nonumber{}\\
+ \left\{\sqrt{1-\frac{\sin^{2}\theta}{\cos^{2}\:\theta}}\cos\:\theta \vert 11\rangle_{AB} + \sqrt{1-\frac{\sin^{2}\theta}{\cos^{2}\:\theta}}\sin\:\theta \vert 10\rangle_{AB}\right\}\otimes \vert 1\rangle_{aux}
\label{w4}.
\end{eqnarray}
If now Alice gets her projective measurement outcome as $\vert 0\rangle_{aux}$, this will make Alice and Bob to share a state of the form $[\:\sin\:\theta\:\vert 01\rangle_{AB}+\frac{\sin^{2}\theta}{\cos\:\theta}\:\vert 00\rangle_{AB}+\cos\:\theta\:\vert 10\rangle_{AB}\:]$.
Using (\ref{concurrence}), the concurrence ($C_{2}$) of this shared state is measured and consequently it is found that\\
\beq
C_{2}= \sqrt{2}\:\vert\cos\:\theta\:\sin\:\theta\vert
\label{wconcur}
\eeq\\
It is clear from the figure below (figure $5.5$) that for $\theta$ assuming values from $\frac{\pi}{4}$ to $\frac{\pi}{2}$, concurrence of the shared state $[\:\sin\:\theta\:\vert 01\rangle_{AB}+\frac{\sin^{2}\theta}{\cos\:\theta}\:\vert 00\rangle_{AB}+\cos\:\theta\:\vert 10\rangle_{AB}\:]$ never reaches its maximum value i.e. $1$. When Cliff's measurement angle is $\frac{\pi}{4}$, the concurrence $C_{2}$ takes its highest value, which is $0.7$.  In other words Alice and Bob never share maximally entangled state for any value of $\theta$.  Hence the prototypical $W-$ state is not suitable for controlled dense coding.\\\\
\begin{figure}[hbtp]
\centering
\resizebox{7.5cm}{5.5cm}{\includegraphics{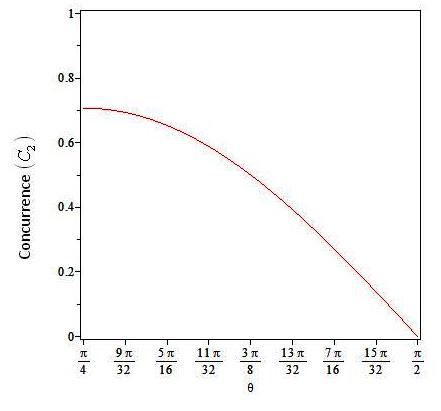}}

\caption{For $\theta$ belonging to the range of $[\frac{\pi}{4},\frac{\pi}{2}]$, we get $0.7 \geq C_{2} \geq 0$.}
\end{figure}\\
\subsection{\textbf{Four particle W-state:}}
It is known that all even qubit entangled states are not suitable for maximal dense coding e.g. maximal dense coding is not possible in $4-$ qubit prototype $W-$ state \cite{pathakbook}. This is also the scenario when such states are taken into consideration for controlled dense coding.\\\\
A four qubit $W-$ state of the form defined in \cite{liqiu2007} is shown below.\\
\beq
\vert W^{(1)} \rangle_{PABC}=\frac{1}{\sqrt{4}}[\vert 1000\rangle_{PABC} + \vert 0100\rangle_{PABC} + \vert 0010\rangle_{PABC} + \vert 0001\rangle_{PABC}],\nonumber\\
\label{fourpartyw1}
\eeq\\
where $P$, $A$, $B$ and $C$ respectively are Paul, Alice, Bob and Cliff as before. The state (\ref{fourpartyw1}) is then expressed in terms of Cliff's measurement basis $\lbrace\:\vert +\rangle_{C},\vert -\rangle_{C}\:\rbrace$ and is shown in the following\\
\beq
\vert W \rangle_{PABC}=\frac{1}{\sqrt{2}}[\vert t^{1}\rangle_{PAB}\:\vert +\rangle_{C} +\vert t^{2}\rangle_{PAB}\: \vert -\rangle_{C})].
\label{fourpartyw2}
\eeq\\
where\\
\beq
\vert t^{1}\rangle_{PAB}=\frac{\cos\theta(\vert 100\rangle_{PAB}+\vert 010\rangle_{PAB}+\vert 001\rangle_{PAB})+\sin\theta\:\vert 000\rangle_{PAB}}{\sqrt{2}},\nonumber{}\\
\vert t^{2}\rangle_{PAB}=\frac{\sin\theta(\vert 100\rangle_{PAB}+\vert 010\rangle_{PAB}+\vert 001\rangle_{PAB})-\cos\theta\:\vert 000\rangle_{PAB}}{\sqrt{2}}.\label{fourpartyw3}
\eeq\\
Again when Paul chooses  his basis $\lbrace\:\vert +\rangle_{P},\vert -\rangle_{P}\:\rbrace$ then the state (\ref{fourpartyw2}) can be re-expressed as,\\
\beq
\vert W^{1} \rangle_{PAB}=\frac{1}{\sqrt{2}}[\vert +\rangle_{P}\:\vert t^{3}\rangle_{AB}
+\vert -\rangle_{P}\vert t^{4}\rangle_{AB}]
\label{fourpartyw4},
\eeq\\
where\\
\begin{eqnarray}
\vert t^{3}\rangle_{AB} = \sin\:(\theta+\varepsilon) \vert 00\rangle_{AB} + \cos\:\theta\:\cos\:\varepsilon\:(\vert 10\rangle_{AB} + \vert 01\rangle_{AB}),\nonumber{}\\
\vert t^{4}\rangle_{AB}=-\cos\:(\theta+\varepsilon) \vert 00\rangle_{AB} + \cos\:\theta\:\sin\:\varepsilon\:(\vert 10\rangle_{AB} + \vert 01\
\rangle_{AB}).\nonumber{}\\
\label{fourpartyw5}
\eeq\\
The forms of the measurement basis chosen by both Cliff and Paul are of the type (\ref{charliebasis}), the only difference is that the measurement angle of Cliff is denoted by $\theta$ and that by Paul is $\varepsilon$. Further if Paul's von-Neumann measurement outcome results in $\vert +\rangle_{P}$, then the shared state between Alice and Bob is $\vert t^{3}\rangle_{AB}$. Alice introduces auxiliary qubit $\vert 0\rangle_{aux}$ and takes into consideration the unitary operator (\ref{unitarymatrix3}) such that the collective unitary operation $U_{2} \otimes I_{B}$ transforms the state $\vert t^{4}\rangle_{AB}\otimes \vert 0\rangle_{aux}$ as before and subsequently when Alice's von-Neumann measurement outcome is $\vert 0\rangle_{aux}$, then shared state between Alice and Bob is given by\\
\beq
\vert t^{3^(1)}\rangle_{AB} =\sin\:\theta\:\cos\:\varepsilon\:\vert 01\rangle_{AB} +\cos\:\theta\:\cos\:\varepsilon\:\vert 10\rangle_{AB}\nonumber{}\\
+ (\sin\:\theta\:\sin\:\varepsilon\: + \frac{\sin^{2}\theta\:\cos\:\varepsilon}{\cos\:\theta})\vert 00\rangle_{AB}.
\label{fourpartyw6}
\eeq\\
Using (\ref{concurrence}), the concurrence ($C_{3}$) of the state (\ref{fourpartyw6}) is calculated and is shown graphically below in figure $5.6$.\\
\begin{figure}[hbtp]
\centering
\resizebox{9.5cm}{4.5cm}{\includegraphics{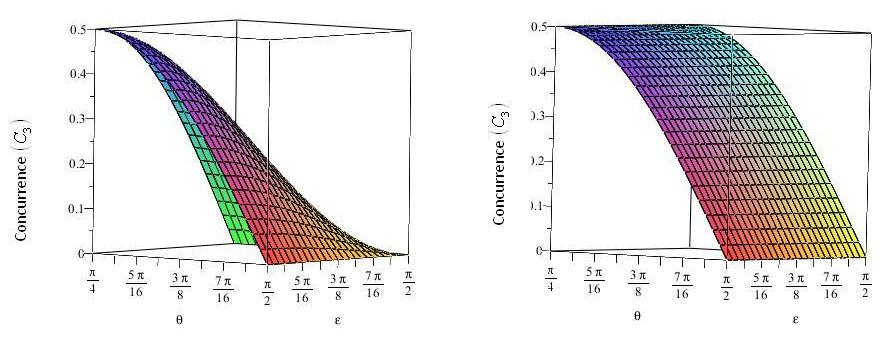}}
\caption{\footnotesize The range of the parameters $\theta$ and $\varepsilon$ are taken from $\frac{\pi}{4}$ to $\frac{\pi}{2}$ and consequently we have $0 \leq C_{3} \leq 0.5$.}
\end{figure}\\
In figure $5.6$, the first plot represents the concurrence $C_{3}$ against $\:\lbrace \varepsilon, \theta\:\rbrace$. The second plot shows the variation of $C_{3}$ against $\theta$ whereas $\varepsilon = \frac{\pi}{4}$.  It is clear from the figure that, the state (\ref{fourpartyw1}) is also not suitable  for controlled dense coding as because when Paul and Charlie varies their parameter $\varepsilon$ and $\theta$, the concurrence of the state shared between Alice and Bob reaches its maximum value $0.5$ and therefore the shared state is never maximally entangled. Hence both tri-partite and quadri-partite prototypical $W-$ states are not suitable in CDC.\\
\subsection{\textbf{Non- prototypical $W-$ state}}
Pati \textit{et. al} showed that \cite{agrapati2006} a particular class of $W-$ state is suitable for perfect teleportation and dense coding, which is of the following form\\
\beq
\vert W_{n}\rangle_{ABC} = \frac{1}{\sqrt{2+2\:n}}\:(\:\vert 100\rangle_{ABC}+\sqrt{n}\:e^{i\:\gamma}\:\vert 010\rangle_{ABC}+\sqrt{n+1}\:e^{i\:\delta}\:\vert 001\rangle_{ABC}), \nonumber\\
\label{wnpati1}
\eeq\\
The above state (\ref{wnpati1}) is a general form of non-prototypical $W-$ state. Li and Qiu in their paper \cite{liqiu2007} proved that, any state of $W-$ class is suitable for perfect teleportation and superdense coding if and only if it can be converted from state $\vert GHZ\rangle_{ABC}$ by a unitary operation which is the tensor product of a two qubit unitary operation and a one qubit unitary operation.\\
A special non-prototypical $W-$ state is now considered here. The state which is being considered in this section was defined by Li and Qiu\\
\beq
\vert W_{n}\rangle_{ABC}= \frac{1}{\sqrt{2}}[\:\vert \phi\rangle_{AB}\:\vert 0\rangle_{C}+\vert 00\rangle_{AB}\:\vert 1\rangle_{C}\:]\label{W1n},
\eeq\\
where\\
\beq
\vert \phi\rangle_{AB}= \frac{1}{\sqrt{n+1}}[\:\vert 10\rangle_{AB}+\sqrt{n}\vert 01\rangle_{AB}\:]\label{W2n}.
\eeq\\
This state can be converted from  $\vert GHZ\rangle_{ABC}=\frac{\vert 000\rangle_{ABC} + \vert 111\rangle_{ABC}}{\sqrt{2}}$ \cite{liqiu2007} as shown below\\
\beq
\vert W_{n}\rangle_{ABC}= (U_{AB} \otimes I_{C})\,\vert GHZ\rangle_{ABC}
\label{convert1},
\eeq\\
while $U_{AB}$ is a unitary operator acting on particles A and B given as \\
\beq
U_{AB}= \vert \phi\rangle\langle 00\vert + \vert 11\rangle\langle 01\vert + \vert \phi^{\bot}\rangle\langle 10\vert + \vert 00\rangle\langle 11\vert 
\label{convert2}.
\eeq\\
Phase factors have not been considered here.  Simple observation reveals that Pati states (\ref{wnpati1}) are indeed equivalent to the states defined in (\ref{W1n}).\\\\
The controller Cliff decides to inform about his measurement outcome to both sender (Alice) and the recipient (Bob). If Cliff gets his outcome as $\vert 1\rangle_{C}$ and informs both Alice and Bob, then they know that they share a separable state and only one bit can be transferred. If, however, Cliff gets his measurement outcome as $\vert 0\rangle_{C}$ and informs Alice and Bob, then they know that they share the state $\vert \phi\rangle_{AB}$ of eq. (\ref{W2n}). The state (\ref{W2n}) is equivalent to the state given in eq. (\ref{basis}). Such states have already been used in dense coding and have been discussed in section $5.2.2$. It is obvious that, when $n=1$, the shared state between Alice and Bob is the Bell-state.\\
\section{\textbf{Three party qutrit state and CDC:}}
A state of $N$ qudits of the form\\
\beq
\vert \Xi^{(N)}\rangle_{A_{1}\:A_{2}\:\cdots\:A_{N}} = \frac{1}{\sqrt{d}}\:[\:\vert 0\:0\:\cdots\:0\rangle + \vert 1\:1\:\cdots\:1\rangle + \cdots\: + \vert d\:d\:\cdots\:d\rangle\:],
\eeq\\
is a maximally entangled state for the $N$ particles, a special case of which is a
tripartite qutrit state given by \cite{liulongtongli2002}\\
\begin{equation}
\vert \Xi^{(3)}\rangle_{ABC} = \frac{1}{\sqrt{3}}\:[\:\vert 0\:0\:0\rangle_{ABC} +\vert 1\:1\:1\rangle_{ABC} + \vert 2\:2\:2\rangle_{ABC}\:]\label{qutritcdc1}.
\end{equation}\\
Here, three parties are $A$, $B$ and $C$, each of them holding three respective qubits $\vert 0\rangle$, $\vert 1\rangle$ and $\vert 2\rangle$ respectively. This state is basically a natural generalization of GHZ state in three level systems with full rank of all reduced density matrices \cite{mintertsalway2012}.\\\\
Party $3$ now considers the following basis\\
\beq
\vert \uparrow\rangle _{C} &=& \sin\:\theta\:\vert 0\rangle_{C} + \cos\:\theta\:\vert 2\rangle_{C},\nonumber{}\\
\vert \nearrow\rangle_{C}&=&\vert 1\rangle_{C},\nonumber{}\\
\vert \downarrow\rangle_{C} &=& \cos\:\theta\:\vert 0\rangle_{C} - \sin\:\theta\:\vert 2\rangle_{C}. \label{charliebasisforqutrit}
\eeq\\
Under this basis (\ref{charliebasisforqutrit}), the state (\ref{qutritcdc1}) takes the form\\
\beq
\vert \Xi^{(3)}\rangle_{ABC} = (\:\sin\:\theta\:\vert 00\rangle_{AB}+\cos\:\theta\:\vert 22\rangle_{AB}\:)\:\vert \uparrow\rangle_{C}\nonumber\\
+(\:\cos\:\theta\:\vert 00\rangle_{AB}-\sin\:\theta\:\vert 22\rangle_{AB}\:)\:\vert \downarrow\rangle_{C}+ \vert 11\rangle_{AB}\:\vert \nearrow\rangle_{C}.\label{qutritcdc2}
\eeq\\
When Cliff's projective measurement yields $\vert \nearrow\rangle_{C}$, Alice and Bob will share $\vert 11\rangle_{AB}$ and only one bit is transferred from Alice to Bob in that case. But when Cliff's measurement results in one from the set $\lbrace\:\vert \uparrow\rangle_{C}, \vert \downarrow\rangle_{C}\:\rbrace$, then  a non-maximally entangled state is shared between them, which is $(\sin\:\theta\:\vert 00\rangle_{AB}+\cos\:\theta\:\vert 22\rangle_{AB})$ (w.r.t $\vert \uparrow\rangle_{C}$) or $(\cos\:\theta\:\vert 00\rangle_{AB}-\sin\:\theta\:\vert 22\rangle_{AB})$ (for $\vert \downarrow\rangle_{C}$).\\\\ 
However, it is assumed that Cliff's measurement result is $\vert \uparrow\rangle_{C}$ and he informs about his outcome to Alice only while Bob is left to guess. Alice introduces an auxiliary qubit $\vert 0\rangle_{aux}$ and performs an unitary operation on her qubit A as well as on the auxiliary qubit with respect to the collective operation under the basis $\lbrace\:\vert i\:j\rangle\:\rbrace_{i,\:j=\:0}^{2}$.\\\\
The unitary operator which Alice chooses is the $9 \times 9$ unitary Braid matrix defined as  \cite{abdesselam2010}.\\
\begin{eqnarray}
V_{1} = \left(%
\begin{array}{ccccccccc}
\frac{\cos\:\theta}{\sin\:\theta}& 0& 0& 0& 0& 0& 0& 0& \sqrt{1-\frac{\cos^{2}\:\theta}{\sin^{2}\:\theta}}\\
0& 1& 0& 0& 0& 0& 0& 0& 0\\
0& 0& \frac{\sin\:\theta}{\cos\:\theta}& 0& 0& 0& \sqrt{1-\frac{\sin^{2}\:\theta}{\cos^{2}\:\theta}}& 0& 0\\
0& 0& 0& 1& 0& 0& 0& 0& 0\\
0& 0& 0& 0& 1& 0& 0& 0& 0\\
0& 0& 0& 0& 0& 1& 0& 0& 0\\
0& 0& \sqrt{1-\frac{\sin^{2}\:\theta}{\cos^{2}\:\theta}}& 0& 0& 0& -\frac{\sin\:\theta}{\cos\:\theta}& 0& 0\\
0& 0& 0& 0& 0& 0& 0& 1& 0\\
\sqrt{1-\frac{\cos^{2}\:\theta}{\sin^{2}\:\theta}}& 0& 0& 0& 0& 0& 0& 0& -\frac{\cos\:\theta}{\sin\:\theta}
\end{array}%
\right).\label{braidmatrix}
\end{eqnarray}\\
The collective unitary operation $V_{1} \otimes I_{B}$ transforms the state $(\sin\:\theta\:\vert 00\rangle_{AB}+\cos\:\theta\:\vert 22\rangle_{AB}) \otimes \vert 0\rangle_{aux}$ to the state\\
\beq
\vert \Phi\rangle_{AB\:aux}=(\cos\:\theta\:\:\vert 00\rangle_{AB}-\sin\:\theta\:\:\vert 22\rangle_{AB})\otimes \vert 0\rangle_{aux} + \nonumber{}\\\nonumber{}\\
\left\{\: \cos\:\theta\:\sqrt{1-\frac{\sin^{2}\theta}{\cos^{2}\theta}}\:\:\vert 02\rangle_{AB} + \sin\:\theta\:\sqrt{1-\frac{\sin^{2}\theta}{\cos^{2}\theta}}\:\:\vert 20\rangle_{AB}\:\right\} \otimes \vert 2\rangle_{aux}.\nonumber\\ \label{qutritcdc3}
\eeq\\
For Alice's von - Neumann readout $\vert 0 \rangle_{aux}$, the non-maximally entangled state shared between Alice and Bob is $\cos\:\theta\vert 00\rangle_{A_{1}\:A_{2}} - \sin\:\theta\vert 22\rangle_{A_{1}\:A_{2}}$. It immediately follows that when $\theta=\frac{\pi}{4}$, then Alice and Bob share the maximally entangled state $(\frac{\vert 00\rangle_{AB}-\vert 22\rangle_{AB}}{\sqrt{2}})$. After performing the operations identity $I$ or the Pauli spin operators $\sigma^{x}$, $\sigma^{y}$ and $\sigma^{z}$) on her qubit, she sends her qubit to Bob.\\\\
In other words, Alice applies the projection operators $\vert 0\rangle\langle 0\vert + \vert 2\rangle\langle 2\vert$, $\vert 0\rangle\langle 2\vert + \vert 2\rangle\langle 0\vert$, $\vert 0\rangle\langle 2\vert - \vert 2\rangle\langle 0\vert$ and $\vert 0\rangle\langle 0\vert - \vert 2\rangle\langle 2\vert$ respectively to the state $(\frac{\vert 00\rangle_{AB}-\vert 22\rangle_{AB}}{\sqrt{2}})$, and she obtains the following states.\\
\beq
\frac{\vert 00\rangle_{AB}-\vert 22\rangle_{AB}}{\sqrt{2}},\nonumber{}\\
\frac{\vert 20\rangle_{AB}-\vert 02\rangle_{AB}}{\sqrt{2}},\nonumber{}\\
\frac{\vert 02\rangle_{AB}+\vert 20\rangle_{AB}}{\sqrt{2}},\nonumber{}\\
\frac{\vert 00\rangle_{AB}+\vert 22\rangle_{AB}}{\sqrt{2}}.\label{qutritcdc5}
\eeq\\
Once she sends her qubit to Bob, he then uses a projection operator $\vert 00\rangle\langle 00\vert + \vert 22\rangle\langle 20\vert + \vert 02\rangle\langle 02\vert + \vert 20\rangle\langle 22\vert$, to his qubit.
In this way, usual dense coding will be performed by Alice and Bob as well.\\\\
If, however the shared state between Alice and Bob is $(\cos\:\theta\vert 00\rangle_{AB}-\sin\:\theta\vert 22\rangle_{AB})$, then Alice may consider the following form of the Braid matrix \cite{abdesselam2010}\\
\beq
V_{2} = \left(%
\begin{array}{ccccccccc}
\frac{\sin\:\theta}{\cos\:\theta}& 0& 0& 0& 0& 0& 0& 0& \sqrt{1-\frac{\sin^{2}\:\theta}{\cos^{2}\:\theta}}\\
0& 1& 0& 0& 0& 0& 0& 0& 0\\
0& 0& \frac{\cos\:\theta}{\sin\:\theta}& 0& 0& 0& \sqrt{1-\frac{\cos^{2}\:\theta}{\sin^{2}\:\theta}}& 0& 0\\
0& 0& 0& 1& 0& 0& 0& 0& 0\\
0& 0& 0& 0& 1& 0& 0& 0& 0\\
0& 0& 0& 0& 0& 1& 0& 0& 0\\
0& 0& \sqrt{1-\frac{\cos^{2}\:\theta}{\sin^{2}\:\theta}}& 0& 0& 0& -\frac{\cos\:\theta}{\sin\:\theta}& 0& 0\\
0& 0& 0& 0& 0& 0& 0& 1& 0\\
\sqrt{1-\frac{\sin^{2}\:\theta}{\cos^{2}\:\theta}}& 0& 0& 0& 0& 0& 0& 0& -\frac{\sin\:\theta}{\cos\:\theta}
\end{array}%
\right). \nonumber\\\label{braidmatrix2}
\eeq\\\\
Applying the protocol as described above, $2-$ bits of classical information again can be transferred from Alice to Bob.\\\\
The findings of this chapter may be summarized below.\\
\begin{itemize}
\item It has been observed here that just like the GHZ state, it's  similar other classifications can suitably be used in the controlled dense coding scheme. In all these cases the average number of bits transmitted is dependent on the Controller's choice of basis as well as on his measurement outcomes i.e. there is an inter-dependence between the average number of bits transmitted from the sender to the receiver and the measurements induced by the controller of the scheme. The general form of  GHZ type states like $L\:(\:\vert 000\rangle + l\:\vert 111\rangle\:)$ is then considered where,  $L=\frac{1}{\sqrt{1+l^{2}}}$ for real positive $l$. It is found that the parameter of the state is inversely proportional to the measurement angle of the controller. This means that for the successful completion of the scheme the parameter of the state should be so chosen that it has to be at par with polarization angle of the controller. More precisely with high values of the parameter $l$ the angle of polarization $\theta$ has to be decreased accordingly and vice - versa. However the success probability of the scheme of CDC is unity when $l=1$ and when Cliff modulates his polarization angle to $\frac{\pi}{4}$. The preparation of the state $L\:(\:\vert 000\rangle + l\:\vert 111\rangle\:)$ is made in a way by carefully choosing the parameter $l$ so that Cliff can manipulate his polarization angle $\theta$ so that the success probability, which is given by $\frac{2\:l^{2}}{1+l^{2}}$, is maximized. It has also been shown that four party GHZ state is also authentic to carry with the scheme of CDC.
\item Prototypical three and four party $W$ state are however not useful for controlled dense coding. These states are intrinsically different from $GHZ$ state with respect to their entanglement properties. It has been observed that for any sort of modulation from controller's end, the sender and the receiver never share bipartite maximally entangled state if the pre-assigned tripartite state shared among controller, sender and receiver is prototypical $W$ state. But if they share a special type of $W-$ class of states of the form (\ref{W1n}), then the scheme of controlled dense coding works. 
\item In $3\otimes3$ dimensional system, the 3 party qutrit state of the form (\ref{qutritcdc1}) can be taken into consideration in CDC. Using Hao protocol of CDC there and with a proper choice of unitary Braid matrix by the sender (in this case it was the Braid matrix defined in \cite{abdesselam2010}) as well as by the choices of the projection operators (both by sender and the receiver) it has been confirmed that controlled dense coding can be performed well with the state.\\\\
\end{itemize}
\chapter{Controlled Secret Sharing using Cloning}
\label{ch:css}
\textit{$``$Scientific discovery and scientific knowledge have been achieved only by those who have gone in pursuit of it without any practical purpose whatsoever in view."}
\begin{flushright}
- Max Planck
\end{flushright}
\vskip1cm
\section{\textbf{Introduction:}}
In chapter $4$ it was shown that the output of the cloning machines could be used in information processing protocols like teleportation and dense coding. This indicated an additional appeal to the cloning machine. Apart from this, cloning machines also play an important role in secret sharing. This feature of cloning machines will now be addressed.\\\\ 
The fundamental idea behind secret sharing has already been summarized in Chapter $2$ (section $2.16.5$).  Let us begin by recapitulating what has already been discussed. In a nutshell, in quantum secret sharing, quantum information encoded in a qubit is split among several parties such that only one of them is able to recover the information exactly, provided all the other parties agree to cooperate \cite{hillery1999}. Quantum secret sharing protocol was carried out using a bipartite pure entangled state in \cite{karlssonsecret1999} and using tripartite pure entangled states in \cite{bandyopadhyay2000,bagherinezad2003,lance2004,gordon2006,zhengsecret2006}. Quantum secret sharing has also been experimentally realized in \cite{tittel2001,schmidt2005,schmidt2006,bogdanski2008}. Although, a semi quantum secret sharing protocol was proposed using maximally entangled GHZ state that was secured against eavesdropping \cite{liqsecret2010}, however in real experimental set ups, the entangled resource shared by the users would be a mixed entangled state due to the noise induced by eavesdropping. 
The present context\footnote{The Chapter is mainly based on our work\\\
\textsc{S. Adhikari, S. Roy, S. Chakraborty, V. Jagadish, M. K. Haris and A. Kumar}, \textbf{$'$Controlled Secret Sharing Protocol using a Quantum Cloning Circuit'}, \textsc{Quantum Information Processing, Vol. 13, No. 9, 2071-2080, (2014)}.} proposes a situation in which there are two secret agents, one loyal and another treacherous. The treacherous one's motive is to pass classified data to an agency for which he works, while the loyal one, by any means, wants to prevent him in fulfilling his objective.\\
\begin{itemize}
\item Can the loyal agent stop the disloyal in sharing secret information to the unwanted recipients?
\item How can the disloyal agent send the information to his associates?  How does entanglement play as an aid to the loyal agent in achieving the goal?
\item What sort of quantum processing will the trustworthy agent apply in this case? What is the role of cloning machine in this respect?\\
\end{itemize}
These are the questions which we will try to answer by defining a new protocol of secret sharing, where quantum cloning machine will take pertinent role.
\section{\textbf{The Protocol}}
In the proposed secret sharing protocol there are four parties viz. Alice, Bob, Charlie and Cliff. Here Alice and Bob are the two accomplices of Charile (who is the dishonest guy). Charlie wants to prepare and encode the classical information in a maximally entangled pure state and consequently will try to send one qubit to one of the accomplices Alice and at the same time the other qubit to Bob. Cliff on the other hand will try to foil Charile's attempt and in the due process will take the help of quantum cloning circuit so that he can clone the individual qubits, those which are en-route through the channel to the un-intended recipients (Alice and Bob). Such an attempt by Cliff will create a noise into the system thus generating a mixed state.
\section{\textbf{Mixed state creation via quantum cloning circuit:}}
A bipartite pure state $\vert \psi\rangle^{in}$ in an $n\otimes n -$ dimensional system can be written in the Schmidt polar form as\\
\beq
\vert \psi\rangle^{in} = \sum^{n}_{i=1}\:\sqrt{\lambda_{i}}\:\vert i\rangle_{1}\:\otimes\:\vert i\rangle_{2},
\label{chap6eq1}
\eeq\\
where, $\lambda_{i}\:\geq\:0$, $i=1,2,...,n$ are the Schmidt coefficients and satisfy the condition $\sum_{i=1}^{n}\:\lambda_{i}=1$. Once Charlie prepares the bipartite entangled state, he sends particle $1$ to  Alice and particle $2$ to Bob through insecure channels where Cliff attempts to clone the particles $1$ and $2$ respectively.\\\\
Now the operation of the cloning circuit that Cliff applies to the individual particles can be accomplished by Buzek -Hillery universal quantum cloning machine\cite{buzek1998}. The characteristics of  the said machine have been reviewed in Chapter $2$ (section $2.16.6$). The expediency of the machine has already been proved in the domain of quantum information protocols like teleportation and dense coding in Chapter $4$.\\\\
Using the cloning transformation (\ref{buzekcloning}) it can be easily seen that, the state (\ref{chap6eq1}) transforms as follows\\
\beq
\vert \psi\rangle^{in}\:\rightarrow\: \vert \psi\rangle^{out}\:=\:c^{2}\:\sum_{i=1}^{n}\:\sqrt{\lambda_{i}}\:\lbrace\:\vert i,i\rangle_{13}\otimes\vert i,i\rangle_{24}\:\vert X_{i}\rangle\:\vert X_{i}\rangle\:\rbrace\nonumber{}\\ +c\:d\:\sum_{i\:\neq\:j}^{n}\:\sqrt{\lambda_{i}}\:\vert i,i\rangle_{13}\otimes\:\lbrace\:\vert i,j\rangle_{24} + \vert j,i\rangle_{24}\:\rbrace\:\vert X_{i}\rangle\:\vert X_{j}\rangle\nonumber{}\\ + c\:d\:\:\sum_{i\:\neq\:j}^{n}\:\sqrt{\lambda_{i}}\:\lbrace\:\vert i,j\rangle_{13} + \vert j,i\rangle_{13}\:\rbrace\otimes\:\vert i,i\rangle_{24}\:\vert X_{j}\rangle\:\vert X_{i}\rangle \nonumber{}\\ + d^{2}\:\sum_{i=1}^{n}\:\sqrt{\lambda_{i}}\:\left\{\:\sum_{i\:\neq\:j}^{n}\:(\:\vert i,j\rangle_{13}+\vert j,i\rangle_{13}\:)\otimes\:\sum_{i\:\neq\:l}\:(\:\vert i,l\rangle_{24}+\vert l,i\rangle_{24}\:)\:\vert X_{j}\rangle\:\vert X_{l}\rangle\:\right\}.\nonumber\\
\label{chap6eq2}
\eeq\\
where $\vert \rangle_{3}$ and $\vert \rangle_{4}$ are the qubits of the environment.\\\\
After tracing out the ancilla qubits, the four qubit states would be described by the density operator $\rho_{1324}$. Moreover, as the sent qubit $1\:(2)$ interacts with its corresponding cloned qubit $3\:(4)$, the state described by the density operator $\rho_{13}\:(\rho_{24})$ can be designated as local output(s) and would be given by,\\
\beq
\rho_{13}^{local}=\rho_{24}^{local}=c^{2}\:\sum_{i=1}^{n}\:\lambda_{i}\:\vert i,i\rangle\langle i,i\vert + d^{2}\:\sum_{i\:\neq\:j}^{n}\:\lbrace\: \vert i,j\rangle + \vert j,i\rangle\:\rbrace\:\lbrace \langle i,j\vert + \langle j,i\vert \:\rbrace. \nonumber\\
\label{chap6eq3}
\eeq\\
Since the state described by the density operator $\rho_{14}\:(\rho_{23})$ is formed between the original qubit $1\:(2)$ and the cloned qubit $4\:(3)$ which are located at two distant locations, the state can be termed as a non-local state such that\\
\beq
\rho_{14}^{non-local}&=&\rho_{23}^{non-local}\nonumber{}\\ &=&  P\:\sum_{i=1}^{n}\:\lambda_{i}\:\vert i,i\rangle\langle i,i\vert + Q\:\sum_{i\:\neq\:j}^{n}\:\sqrt{\lambda_{i}\:\lambda_{j}}\:\vert i,i\rangle\langle j,j\vert \nonumber{}\\ + && R\:\sum_{i\:\neq\:j}^{n}\:(\:\vert i,j\rangle\langle i,j\vert + \vert j,i\rangle\langle j,i\vert\:) + S\:\sum_{l,j\:\neq\:i}\:\lambda_{i}\:\vert j,l\rangle\langle j,l\vert,\nonumber\\
\label{chap6eq4}
\eeq\\
where,\\
\beq
P&=&(c^{2}+(n-1)\:d^{2})^{2},\nonumber{}\\Q&=&d^{2}\:(4\:c^{2}+4\:c\:d\:(n-2)+(n-2)\:d^{2}\:),\nonumber{}\\ R&=&d^{2}\:(c^{2}+(n-1)\:d^{2}\:),\nonumber{}\\S&=&d^{4}.
\label{chap6eq5}
\eeq\\
The following figure $6.1$ will make the situation clear.\\
\begin{figure}[hbtp]
\centering
\resizebox{9.5cm}{4.5cm}{\includegraphics{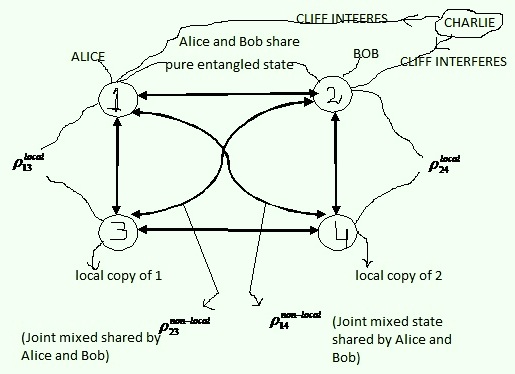}}
\caption{\footnotesize The figure depicts how Alice and Bob share their respective qubits sent to them by Charlie through insecure channels and Cliff attempts to foil it being a eavesdropper. The two local ($\rho_{13}\:/\rho_{24}$) and the two non-local  ($\rho_{14}\:/\rho_{23}$) outputs are generated.}
\end{figure}\\
After intercepting into the channel and cloning the qubits, Cliff resends the particles $1\:(2)$ and $4\:(3)$ to Alice and Bob respectively. Alice and Bob then will share a joint mixed state described by the density operator $\rho_{14}\:(\rho_{23})$. It is important to mention here that Cliff could also send the particles $1\:(3)$ and $2\:(4)$, respectively, to Alice and Bob. However, without any loss of generality, it can be assumed that for Alice and Bob to share an entangled state, Cliff sends particles $1$ and $4$ or $3$ and $2$ to Alice and Bob, respectively. This is obvious by the way local and non-local density operators are defined through equations (\ref{chap6eq3}) and (\ref{chap6eq4}).\\\\
In computational basis and for $2\otimes 2$ dimensional system, the local and non-local outputs of (\ref{chap6eq3}) and (\ref{chap6eq4}) can be re-expressed as,\\
\beq
\rho^{local}_{13} = \rho^{local}_{24}= \left(%
\begin{array}{cccc}
 c^{2}\:\lambda_{1} &  0 &  0 & 0\\
0 &   d^{2} &  d^{2} &0\\
0 &  d^{2} &   d^{2} & 0\\
0 & 0 & 0 &c^{2}\:\lambda_{2}\\
\end{array}%
\right)\label{chap6eq6},
\eeq\\
and\\
\beq
\rho^{non-local}_{14} = \rho^{non-local}_{23}= \left(%
\begin{array}{cccc}
 P\:\lambda_{1} + S\:\lambda_{2} &  0 &  0 & Q\:\sqrt{\lambda_{1}\:\lambda_{2}}\\
0 &   R &  0 &0\\
0 &  0 &   R & 0\\
Q\:\sqrt{\lambda_{1}\:\lambda_{2}} & 0 & 0 &P\:\lambda_{2} + S\:\lambda_{1}\\
\end{array}%
\right) \label{chap6eq7},
\eeq\\
where\\
\beq
P&=&(c^{2}+\:d^{2})^{2},\nonumber{}\\Q&=&4\:c^{2}\:d^{2},\nonumber{}\\ R&=&c^{2}\:d^{2}+d^{4},\nonumber{}\\S&=&d^{4}.
\label{chap6eq8}
\eeq\\
It is important however to note down that, when Cliff clones the qubits  en-route to Alice and Bob using the cloning circuit, the shared mixed state (\ref{chap6eq4}) may or may not be entangled. Using the concurrence \cite{wootters1998} and optimal witness operators \cite{bertlmanwitnes2005}, for two qubit systems, it is obtained that the shared state (\ref{chap6eq4}) would be entangled if there exists a critical value of concurrence which measures the initial entanglement present in the two qubit pure system. If the concurrence of initially prepared state is less than this critical value, then the shared state is separable.\\\\
Such an optimal witness operator for a two qubit system is given, in matrix form, as \cite{bertlmanwitnes2005}\\
\beq
W_{a}^{(1)}=\left(%
\begin{array}{cccc}
 0 &  0 &  0 & \frac{-1}{\sqrt{3}}\\
0 &   \frac{1}{\sqrt{3}} &  0 &0\\
0 &  0 &   \frac{1}{\sqrt{3}} & 0\\
\frac{-1}{\sqrt{3}} & 0 & 0 & 0\\
\end{array}%
\right) .
\label{chap6eq9}
\eeq\\
The non-local output $\rho_{14}^{non-local}=\rho_{23}^{non-local}$ would be entangled if and only if $Tr\:(W_{a}^{(1)}\:\:\rho_{14})\:<\:0$. It is easy to show that\\
\beq
Tr\:(W_{a}^{(1)}\:\:\rho_{14}^{non-local}) &=& Tr\:(W_{a}^{(1)}\:\:\rho_{23}^{non-local})\nonumber{}\\
&=& \left\{\frac{-2}{\sqrt{3}}\right\}\:(\:Q\:\sqrt{\lambda_{1}\lambda_{2}}-R)
\label{chap6eq9a}.
\eeq\\
Hence the condition of entanglement of the non-local output gives\\
\beq
Q\:\sqrt{\lambda_{1}\:\lambda_{2}} - R\:>\:0 \:\: &\Longrightarrow & 2\:\sqrt{\lambda_{1}\:\lambda_{2}} \nonumber\\&=& C\:(\vert \psi\rangle^{in})\nonumber\\ &>&\:C^{critical}\:(\vert \psi\rangle^{in}) = \frac{1+c^{2}}{4\:c^{2}},~~\frac{1}{\sqrt{3}}\:<\:c\:\leq\: 1. \nonumber\\
\label{chap6eq10}
\eeq\\
It is clear from (\ref{chap6eq10}) that the critical value of the concurrence is a decreasing function of the cloning parameter $c$ which means that as $c$ increases, the critical value of concurrence (i.e. $C^{critical}$) decreases . The lower value of concurrence (of initially prepared entangled state), however, would ensure that the non-local shared state is entangled if the quantum cloning circuit parameter $c$ tends towards unity.
On the other hand, the local shared state described by the density matrix $\rho_{13}^{local}=\rho^{local}_{24}$ is separable as $Tr\:(W_{a}^{(1)}\:\rho_{13}^{local})=Tr\:(W_{a}^{(1)}\:\rho_{24}^{local})=\frac{1}{3\:\sqrt{3}}\:>\:0$.\\\\
Using optimal witness operator \cite{sanperawitness} proposed by Sanpera \textit{et. al}, similar types of conclusion as above can be drawn. The witness is defined as \\
\beq
W_{a}^{(2)} = \frac{1}{2}\:(\:I-\:\iota),
\label{chap6eq11}
\eeq\\
while $\iota\:=\sigma^{x}\otimes\sigma^{x} - \sigma^{y}\otimes\sigma^{y} + \sigma^{z}\otimes\sigma^{z}$,\: ($\sigma^{x},\:\sigma^{y},\:\sigma^{z}$ are Pauli matrices.)\\\\
If Charlie initially prepares a maximally entangled state by taking $\lambda_{1}=\lambda_{2}=\frac{1}{2}$, then for a specific value of parameter $c\:=\:\sqrt{\frac{2}{3}}$, the shared state between Alice and Bob takes the form of a maximally entangled mixed state (MEMS) represented by,\\
\beq
\rho_{23}^{non-local} = \rho_{14}^{non-local} &=& \left(%
\begin{array}{cccc}
 \frac{13}{36} &  0 &  0 & \frac{4}{18}\\
0 &   \frac{5}{36} &  0 &0\\
0 &  0 &   \frac{5}{36} & 0\\
\frac{4}{18} & 0 & 0 & \frac{13}{36}\\
\end{array}%
\right)\nonumber\\\nonumber\\
&=& \frac{4}{9}\:\vert \Psi^{+}\rangle\langle \Psi^{+}\vert + \frac{5}{36}\:I_{4} .
\label{chap6eq12}
\eeq\\
$\vert \Psi^{+}\rangle$ is Bell-state from eq. (\ref{bellstates}). Thus if maximally entangled pure state sent through insecure channels is cloned by Cliff using transformations defined in (\ref{buzekcloning}), then there exists a value of the quantum cloning circuit parameter $c$, for which the maximally entangled pure state transforms to a maximally mixed entangled state that belongs to the family of Werner state \cite{werner401989}.\\
\section{\textbf{Two qubit bipartite mixed state in the quantum secret sharing protocol:}}
The secret sharing protocol, the main central point of investigation in this chapter, where Cliff wants to stop Charlie from leaking information (encoded in the two-qubit mixed entangled state (\ref{chap6eq4})) to his counterparts Alice and Bob, is now described here in this section.\\
\subsection*{\textbf{Maximally entangled pure state prepared by Charlie:}}
To split the information between Alice and Bob, Charlie prepares a two qubit maximally entangled pure state either in $\vert \Psi^{+}\rangle$ or in $\vert \Psi^{-}\rangle$ (the Bell-states from eq.(\ref{bellstates})) form. Which of these states to prepare is decided by Charlie classically, either by simply tossing a coin (or by seeing a cat $`$dead' or $`$alive' or by any classical means with two possible outcomes). If however, $`$head' (designated as $`$0') appears (or Cat comes out to be $`$dead') then Charlie decides to prepare $\vert \Psi^{+}\rangle$ and if $`$tail' (designated as $`$1') appears (or Cat is $`$alive') then he prepares $\vert \Psi^{-}\rangle$. Charlie, in this way, encodes one bit of information into the prepared state. Once the information is encoded, Charlie sends the qubits to Alice and Bob. Cliff then intercepts, and clones these qubits using a quantum cloning circuit. For both the qubits, Cliff applies the symmetric quantum cloning circuit described by (\ref{buzekcloning}). The protocol then proceeds with Cliff resending any one of the two qubits to Alice and Bob provided the state is entangled. The mixed state shared by Alice and Bob can thus be described by either of the following density operators ($\rho_{AB}^{+}$ or $\rho_{AB}^{-}$)\\
\beq
\rho^{\pm}_{AB} = \frac{P+S}{2}\:(\:\vert 00\rangle\langle 00 \vert + \vert 11\rangle\langle 11 \vert\:)\:\pm \frac{Q}{2}\:(\:\vert 00\rangle\langle 11 \vert + \vert 11\rangle\langle 00 \vert\:) 
\nonumber\\R\:(\:\vert 01\rangle\langle 01\vert + \vert 10\rangle\langle 10\vert\:),
\label{chap6eq13}
\eeq\\
where,\\
$P=(\:c^{2} + d^{2}\:)^{2}$, $Q=4\:c^{2}\:d^{2}$, $R=d^{2}\:(\:c^{2} + d^{2}\:)$, $S=d^{4}$ and $c^{2}+2\:d^{2}=1$.\\\\
\subsection*{\textbf{A single qubit measurement performed by Alice:}}
Alice  performs measurement on her qubit in the Hadamard basis $H_{B}=\:\left\{\:\frac{\vert 0\rangle + \vert 1\rangle}{\sqrt{2}}\:,\frac{\vert 0\rangle - \vert 1\rangle}{\sqrt{2}\:}\:\right\}$. The single qubit state received by Bob would depend on the measurement outcome of Alice's qubit.\\\\ 
If the shared state between Alice and Bob is $\rho^{+}_{AB}$ and Alice's measurement outcome is $\frac{\vert 0\rangle \pm  \vert 1\rangle}{\sqrt{2}}$, then Bob receives either $\rho_{B}^{+0}$ or $\rho_{B}^{+1}$ accordingly, where\\
\beq
\rho^{+\:0\:(1)}_{B} &=& Tr_{1}\left\{\left\{(\frac{\vert 0\rangle \pm \vert 1\rangle}{\sqrt{2}})(\frac{\langle 0\vert +\pm \langle 1\vert}{\sqrt{2}})\otimes I_{2}\right\}\rho^{+}_{AB}
\left\{(\frac{\vert 0\rangle \pm \vert 1\rangle}{\sqrt{2}})(\frac{\langle 0\vert \pm \langle 1\vert}{\sqrt{2}})\otimes I_{2}\right\}\right\}\nonumber\\\nonumber\\
&=& \frac{1}{2}[I_{2}\pm Q(\vert 0\rangle\langle 1 \vert + \vert 1\rangle\langle 0\vert)]
\label{chap6eq15}.
\eeq\\
If the shared state between Alice and Bob is $\rho_{AB}^{-}$ and Alice's measurement outcome is $\frac{\vert 0\rangle \pm \vert 1\rangle}{\sqrt{2}}$, then Bob receives either $\rho_{B}^{-0}$ or $\rho_{B}^{-1}$, which are as follows\\
\beq
\rho_{B}^{-0\:(1)} &=& Tr_{1}\left\{\left\{(\frac{\vert 0\rangle \pm \vert 1\rangle}{\sqrt{2}})(\frac{\langle 0\vert \pm \langle 1\vert}{\sqrt{2}})\otimes I_{2}\right\} \rho^{-}_{AB}
\left\{(\frac{\vert 0\rangle \pm \vert 1\rangle}{\sqrt{2}})(\frac{\langle 0\vert \pm \langle 1\vert}{\sqrt{2}})\otimes I_{2}\right\}\right\}\nonumber\\\nonumber\\
&=& \frac{1}{2}[I_{2}\mp Q(\vert 0\rangle\langle 1 \vert + \vert 1\rangle\langle 0\vert)]
\label{chap6eq17},
\eeq\\
where $I_{2}$ denotes the identity operator in $2 \otimes 2$ dimensional Hilbert space. Similarly, one can find the state obtained by Alice, if Bob chooses to perform measurement on his qubit. The equations (\ref{chap6eq15}) and (\ref{chap6eq17}) clearly explain that it is neither possible for Alice nor for Bob alone to decode Charlie's encoded information. They would only be able to decode Charlie's information if they both agree to collaborate with each other (which is quite expected from the very definition of $`$secret sharing'). Now if they convince themselves to collaborate, then the protocol may be brought forward to the next level.\\
\subsection*{\textbf{Declaration of measurement outcome by Alice:}}
Once Alice and Bob agree to cooperate with each other, Alice sends her measurement outcome to Bob.\\
\begin{enumerate}
\item If the measurement outcome is $\frac{\vert 0\rangle + \vert 1\rangle}{\sqrt{2}}$, then she sends Bob a classical bit $`$0' and 
\item If the measurement outcome is $\frac{\vert 0\rangle - \vert 1\rangle}{\sqrt{2}}$, then she sends classical bit $`$1' to Bob.\\
\end{enumerate}
\subsection*{\textbf{Positive operator valued measurement (POVM) performed by Bob:}}
At this stage the positive operators are introduced to unambiguously discriminate between Bob's mixed state $\rho_{B}^{+0}$ and $\rho_{B}^{-0}$ (or $\rho_{B}^{+1}$ and $\rho_{B}^{-1}$) corresponding to Alice's measurement outcome $\vert +\rangle$ (or $\vert -\rangle$). For this, three element positive operator -valued measurements (POVM), denoted by $E_{1}$, $E_{2}$ and $E_{3}$ are defined in the following\\
\beq
E_{1} &=& \left(%
\begin{array}{cc}
 \frac{Q}{2} &  1\\
0 &   \frac{Q}{2}\\
\end{array}%
\right),\nonumber\\
E_{2} &=& \left(%
\begin{array}{cc}
 \frac{Q}{2} &  -1\\
0 &   \frac{Q}{2}\\
\end{array}%
\right),\nonumber\\
E_{3} &=& 1-E_{1}-E_{2},
\label{chap6eq19}
\eeq\\
where, $Q\:\in [\:0,\frac{1}{2}\:]$. If Bob receives the classical bit $`$0', then Alice's measurement outcome would be $\vert +\rangle$ and correspondingly Bob will receive either $\rho_{B}^{+0}$ or $\rho_{B}^{-0}$. Using the operators defined in (\ref{chap6eq19}), Bob can successfully discriminate between the two states  $\rho_{B}^{+0}$ and $\rho_{B}^{-0}$ as\\
\beq
Tr\:(\:E_{1}\:\rho_{B}^{-0}\:)=Tr\:(\:E_{2}\:\rho_{B}^{+0}\:) &=& 0,\nonumber\\
Tr\:(\:E_{1}\:\rho_{B}^{+0}\:)=Tr\:(\:E_{2}\:\rho_{B}^{-0}\:) &=& Q
\label{chap6eq20}.
\eeq\\
In a similar way, if Bob receives the classical bit $`$1', then also he can discriminate the single qubit states $\rho_{B}^{+\:1}$ and $\rho_{B}^{-\:1}$ using POVM operators defined in (\ref{chap6eq19}).\\\\
The success probability of the protocol can thus be expressed as\\
\beq
P_{success} = \frac{1}{2}\:Tr\:[\:\rho^{+0}\:E_{1}\:]\:+\:\frac{1}{2}\:Tr\:[\:\rho^{-0}\:E_{2}\:].
\label{chap6eq21}
\eeq\\
The above equation can be re-expressed as\\
\beq
P_{success} = Q = 4\:c^{2}\:d^{2},
\label{chap6eq22}
\eeq\\
where, $Q\:\in\:[\:0,\:\frac{1}{2}\:]$. Clearly, the success probability depends on the cloning circuit parameters $c$ and $d$ such that $P_{success}\:\leq\:\frac{1}{2}$. So, the success probability $Q$ of the protocol can be controlled by Cliff and $Q\:\in\:[\:0,\:\frac{1}{2}\:]$.\\\\
For universal quantum cloning machine where the cloning parameters $c$ and $d$ assume the values $\frac{2}{\sqrt{3}}$ and $\frac{1}{\sqrt{6}}$, respectively, $P_{success} = 0.45$, which is very close to the maximum success probability $\frac{1}{2}$, that can be achieved using this protocol. However, if Cliff wants to stop Charlie from communicating the secret information to Alice and Bob, he will use Wootters-Zurek cloning machine \cite{wootters1982} where the cloning parameters $c$ and $d$ take values $1$ and $0$, respectively and hence, the success probability of the protocol would be zero. In this way Cliff can stop Charlie from leaking the secret information.\\\\
The findings can now be summarized below.\\
\begin{itemize}
\item A secret sharing protocol have been discussed here, where, the noise is introduced into the system through eavesdropping using a quantum cloning circuit. Certain interesting facts emerged out regarding the relation between the cloning parameters and the success probability of the protocol. Four parties were involved in this scheme. Out of them the loyal one (named $`$Cliff', the controller) would be able to stop the disloyal one (called, $`$Charlie', the encoder) from sharing a secret message with his accomplices ($`$Alice' and $`$Bob') and Cliff can do this with certainty using appropriately chosen cloning parameters.
\item The disloyal agent Charlie can achieve his goal by preparing a maximally entangled state which will be expected to be shared by his associates Alice and Bob. It has been shown that the bipartite state (sent by Charlie with the encoded classical information) received by Alice and Bob will be an entangled resource if and only if the concurrence of the initially prepared pure state surpasses a certain threshold value.
\item The trustworthy agent (Cliff) takes the help of quantum cloning machine (BH-UQCM) to clone the secret information encoded and sent to the un-intended recipients by un-trustworthy agent (Charlie). Interestingly, for a specific quantum cloning machine (like if he uses Wootters Zurek cloning machine), Cliff would succeed in preventing Charlie from doing any transgression with the secret information.
\end{itemize}
\chapter{Conclusions}
\label{ch:css}
In this chapter we summarize the findings of this thesis and discuss some possible future directions. The theory of quantum information promises to bring radical changes into the domain of technology. The central feature of this possibility relies on quantum entanglement. Quantum entanglement is the fundamental building block on which many information theoretic protocols such as teleportation, dense coding, secret sharing, cryptography and many more depend. A few aspects of these protocols which rely on the structure of entanglement present in various states have been discussed in this thesis. The properties of these states, used in quantum information theory, are interesting and need in depth analysis. \\\\
In \textit{\textbf{Chapter $3$}}, different types of mixed entangled states have been analysed with respect to their utility in teleportation protocols. The states which achieve maximum possible entanglement for specified degree of mixedness or vice versa are known as Maximally Entangled Mixed States or in short MEMS and the states which are not MEMS are Non Maximally Entangled Mixed States or NMEMS. The teleportation fidelities of a few kinds of MEMS (such as Werner state and Munro class of states) have been studied and a new class of NMEMS has been defined. A comparative study between the teleportation fidelity of the new NMEMS and that of an existing NMEMS viz. Werner derivative state has been made.\\\\
First of all, an immediate work which can be pursued is to study how noisy channels such as amplitude damping channel, phase damping channel etc may affect the capacity of teleportation of the constructed NMEMS.  Further, a class of two qutrit state, [specifically, mixtures of the maximally mixed state, like ($S_{9}\:=\frac{1}{9}\:I\otimes\: I$), with a maximally entangled state in qutrit system (as for example, $\vert \phi\rangle = \frac{1}{\sqrt{3}}\:(\:\vert 00\rangle + \vert 11\rangle + \vert 22\rangle\:)$] can be studied as well from information processing protocol's point of view. Such mixtures can be constructed by taking convex combination of $S_{9}$ and $\vert \phi\rangle$ e.g. $\rho_{c}=(1-c)\:S_{9} + c\:\vert \phi\rangle\langle \phi\vert$ and the separability criteria of this state has been discussed in \cite{caves9910001}. The state has been shown to be separable if and only if $c\:\leq\: \frac{1}{4}$. In contrast to this ($\rho_{c}$), the corresponding qubit state i.e. the Werner state is separable if and only if $c\:\leq\: \frac{1}{3}$. This clearly indicates that maximally entangled states of two qutrits are more entangled than maximally entangled states of two qubits\cite{caves9910001}. In qutrit system, the efficacy of such MEMS, as quantum teleportation channel, may be examined. The study can further be generalized to $d-$ dimensional system. Little has been accomplished in extending the characterization of MEMS because measures of genuine multi-partite entanglement have been identified recently \cite{bastian2011,zhma2011}. A first important step towards identifying MEMS for $n$ qubits ($n>2$) was pioneered by Rafsanjani et. al \cite{raf2012}. These multi-partite MEMS can also be taken into consideration to check their effectiveness in quantum information processing tasks like teleportation.\\\\
The mixed entangled states can be generated through cloning. In \textit{\textbf{Chapter $4$}}, therefore, Buzek-Hillery quantum cloning machine designed for higher dimensional ($n\otimes n$) system has been taken into consideration to generate mixed entangled states as outputs. By passing a single qutrit through this machine and by tracing out the ancilla states the two qutrit output state has been obtained. Both the optimal as well as non-optimal form of this output has been considered and their utilities in teleportation and in dense coding have been discussed. In $2 \otimes 2$ system Adhikari \textit{et. al } inspected the efficacy of the outputs of Buzek-Hillery quantum cloning machine in teleportation \cite{adhikari2008}.\\\\
Bru$\beta$ et al analyzed both universal $1\rightarrow 2$ quantum cloners and state dependent cloners \cite{bru1998}. There, the maximal fidelity of cloning was shown to be $\frac{5}{6}$ for universal cloners. Such a universal cloning machine can also be considered and the outputs obtained from the cloning machine when a qubit is used as an input state to the cloning machine may be checked in protocols like teleportation and dense coding. Several classes of state dependent quantum cloners for three-level systems was investigated by Cerf \cite{cerfjmo2011}. BH-UQCM which remained as the central figure throughout chapter $4$ is state-independent universal cloning machine. Cerf \cite{cerfjmo2011} spotlighted on state dependent cloning machine of non-universal type. In some cases the original qutrit may be prepared in a state that is selected from a known ensemble of states. Such a situation motivates to design state dependent cloners. Cerf cloners can also be taken into consideration to study the effectiveness of the outputs in information processing tasks. For multi-party system a study similar to that of chapter $4$ may be conducted. Multi-partite quantum states can be generated in a sequential manner . Universal sequential quantum cloning machine of this type which would produce $m\rightarrow n$ ($m\:\leq n$) copies was discussed in \cite{dang2008}. Another advantage of this type of machine is that, it can consider $d-$ level quantum states also. The outputs of this type of machine can also be checked in teleportation and dense coding.\\\\
\textit{\textbf{Chapter $5$}} focuses on the study of controlled dense coding using pure entangled states. Various types of tripartite and four partite entangled states like GHZ state, W - state etc. have been mainly studied with respect to this interesting protocol of quantum information science. CDC is a type of information processing where number of parties involved share an entangled state among them and one or many parties (who are known as controllers) will decide the fate of information which is supposed to be sent to the receiver by the sender. For tripartite scenario one party acts as the controller whereas for quadripartite case two parties are made controllers of the scheme. In line with this a qutrit case has also been discussed with respect to CDC.\\\\
It would be interesting to generalize the concept of controlled dense coding when the number of parties involved is $N\:\geq 3$ or when the bases of the parties involved are considered in arbitrary $d-$ dimensions. When in a quantum dense coding protocol there are many senders and receivers the protocol is generally known as \textit{distributed quantum dense coding} \cite{lewenstein2004}. By making a few of them as controllers, who will be jointly modulating the entangled states shared by the remaining parties, controlled distributed dense coding can be studied. Recently Das \textit{et. al} showed that in case of many senders and two receivers the Generalized GHZ (GGHZ) states possess higher capacity of dense coding as compared to a significant fraction of pure states having the same multipartite entanglement \cite{tamoghna2014}. So GGHZ states i.e. $\sin\:\theta\:\vert 000\cdots 0\rangle_{A_{1}\:A_{2}\cdots A_{N}} + \cos\:\theta\:\vert 111\cdots 1\rangle_{A_{1}\:A_{2}\cdots A_{N}}$, where $\theta$ is the state parameter, can be utilized in CDC under controlled environment for studying how controller's joint readouts will affect the sharing of bits between the senders and the receivers. How channel capacity affects multi-partite entanglement may further be analyzed and their relationships with capacity of dense coding can also be observed. Apart from the existence of two contrasting genuine tripartite entangled states viz. GGHZ class of states and W class of states, there exist many other non-trivial types of tripartite entangled states in nature \cite{durr622000}. Such non-trivial class of tripartite entangled states have been tested with respect to controlled dense coding \cite{yynie2008,wangwu2009,sixcdc2011,cavitycdc2013}. Recently controlled dense coding has also been performed with maximal sliced states \cite{maximalslicedcdc2016}. To look for an operational criteria for controlled dense coding to classify those non-trivial tripartite entangled states which will be useful in CDC from those which cannot be used in CDC can be nice study in this direction.\\\\
Lastly, in \textit{\textbf{chapter $6$}} another interesting application of quantum information science has been explored. The concept of secret sharing though dates back to ancient era, quantum information theory promises to enhance the security of secret sharing protocols. One such secret sharing protocol has been developed here, where, BH-UQCM has played a vital role. It has been shown that a dishonest agent can be prevented by an honest one from sending secret messages to his (dishonest agent's) associates. The message en-route to the unintended recipients of the disloyal agent are cloned by the trustworthy agent using BH-UQCM to minimize the success probability.  If the loyal agent uses Wootter-Zurek quantum cloning machine, then he can stop the dishonest one in communicating the message at all.\\\\
Absolutely maximally entangled, (in short AME), states are pure multipartite states. For such states, when half or more of the parties are traced out, the maximum entropy mixed state is obtained \cite{scottame2004}. With AME states, quantum secret sharing was first studied in \cite{helwigame2012}. The protocol discussed in chapter $6$, can be built for multi-party system keeping AME states as the central point and where a number of senders and receivers will be involved into the scheme. This secret sharing scheme can also be generalized and studied for qutrit system. As the loyal agent,  by cloning the message en-route using BH-UQCM, was successful in controlling the success probability of the disloyal agent in sharing secret information to his associates, it can further be analyzed whether the scheme is successful in $3\otimes 3$ dimensional system.
\chapter{Appendix:}
\label{ch:app}
\section*{\textbf{Quantum Gates:}}
\textsf{A gate basically transforms the input state, on which it acts, to its corresponding output state according to the rules hard-wired into the truth table. Quantum gates are the building blocks of quantum computers. The physical reality is that, the quantum gate transforms the state of a quantum system into a new state. Some of the quantum gates are presented below. }\\
\subsection*{\textbf{One qubit quantum gates:}}
\begin{itemize}
\item \textbf{Identity gate:} Identity operator ($I$) leaves a qubit unchanged. It is defined as $I\:=\:\vert 0\rangle\langle 0\vert + \vert 1\rangle\langle 1\vert$.
\item  \textbf{$X$ or $NOT$ gate:} This gate transposes the components of a qubit and is defined as $X\:=\:\sigma_{x}\:=\:\vert 0\rangle\langle 1\vert + \vert 1\rangle\langle 0\vert$.
\item \textbf{$Y$  gate:} This gate multiplies the input qubit by $i$ and flips the two components of the qubit. It is defined as $Y\:=\:\sigma_{y}\:=\:-i\:\vert 0\rangle\langle 1\vert + i\: \vert 1\rangle\langle 0\vert$.
\item \textbf{$Z$  gate:} This gate changes the phase (flips the sign) of a qubit and is defined as $Z\:=\:\sigma_{z}\:=\:\vert 1\rangle\langle 0\vert - \vert 0\rangle\langle 1\vert$.
\item \textbf{$Hadamard$ gate:} The Hadamard gate $H$, when applied to a pure state, $\vert 0\rangle$ or $\vert 1\rangle$, creates a superposition state, i.e. $H\:\vert 0\rangle\:=\:\frac{1}{\sqrt{2}}\:(\:\vert 0\rangle + \vert 1\rangle\:)$ and $H\:\vert 1\rangle\:=\:\frac{1}{\sqrt{2}}\:(\:\vert 0\rangle - \vert 1\rangle\:)$.\\
\end{itemize}
\subsection*{\textbf{Two qubit quantum gates:}}
\begin{itemize}
\item \textbf{Controlled-NOT or $CNOT$:} It is the prototypical multi-qubit quantum logic gate. This gate has two input qubits, known as the \textbf{control} qubit and the \textbf{target} qubit respectively. The action of such gates is that, if the control qubit is set to $0$, then the target qubit is left alone. If the control qubit is set to $1$, then the target qubit is flipped. This is shown below.\\
\beq
\vert 00\rangle\rightarrow \vert 00\rangle,\:\:\vert 01\rangle\rightarrow \vert 01\rangle,\:\: \vert 10\rangle\rightarrow \vert 11\rangle,\:\: \vert 11\rangle\rightarrow \vert 10\rangle.
\eeq\\
\end{itemize}
\subsection*{\textbf{Three qubit quantum gates:}}
\begin{itemize}
\item \textbf{Fredkin Gate:} 
Three qubit quantum gates and their classical counterparts have three inputs $a$, $b$ and $c$ and three outputs $a^{/}$, $b^{/}$ and $c^{/}$. One or two of the inputs are referred to as control qubit(s) and are transferred directly to the output. The other input(s) are referred to as target qubit (s). The matrix representation of the gate is shown below.\\
\begin{eqnarray}
G_{Toffoli} = \left(%
\begin{array}{ccccccccc}
1& 0& 0& 0& 0& 0& 0& 0\\
0& 1& 0& 0& 0& 0& 0& 0\\
0& 0& 1& 0& 0& 0& 0& 0\\
0& 0& 0& 0& 0& 1& 0& 0\\
0& 0& 0& 0& 1& 0& 0& 0\\
0& 0& 0& 1& 0& 0& 0& 0\\
0& 0& 0& 0& 0& 0& 1& 0\\
0& 0& 0& 0& 0& 0& 0& 1
\end{array}%
\right).
\end{eqnarray}\\
\item  \textbf{Toffoli Gate:} While the Fredkin gate had only one control input, the Toffoli gate has two control inputs $a$ and $b$, and one target input $c$. The outputs are $a^{/}=a$, $b^{/}=b\: and\: \:c$. The Toffoli gate is a universal gate. The matrix representation of this gate is shown below.\\
\begin{eqnarray}
G_{Fredkin} = \left(%
\begin{array}{ccccccccc}
1& 0& 0& 0& 0& 0& 0& 0\\
0& 1& 0& 0& 0& 0& 0& 0\\
0& 0& 1& 0& 0& 0& 0& 0\\
0& 0& 0& 1& 0& 0& 0& 0\\
0& 0& 0& 0& 1& 0& 0& 0\\
0& 0& 0& 0& 0& 1& 0& 0\\
0& 0& 0& 0& 0& 0& 0& 1\\
0& 0& 0& 0& 0& 0& 1& 0
\end{array}%
\right).
\end{eqnarray}
\end{itemize}
\chapter*{List of Publications}
\begin{itemize}
\item S. Adhikari, A. S. Majumdar, S. Roy, B. Ghosh and N. Nayak, \textbf{$`$Teleportation via maximally and non-maximally entangled mixed states'}, \textit{Quantum Information and Computation}, Vol. \textbf{10}, No. 5/6, 0398-0419, (2010), Rinton Press.\\
\item S. Roy, N. Ganguly, A. Kumar, S. Adhikari and A. S. Majumdar, \textbf{$`$A cloned qutrit and its utility in information processing tasks'}, \textit{Quantum Information Processing}, Vol. \textbf{13}, No. 3, 629-638, (2014), Springer.\\
\item S. Adhikari, S. Roy, S. Chakraborty, V. Jagadish, M. K. Haris and A. Kumar, \textbf{$`$Controlled Secret Sharing Protocol using a Quantum Cloning Circuit'}, \textit{Quantum Information Processing}, Vol. \textbf{13}, No. 9, 2071-2080, (2014), Springer.\\
\item S. Roy and B. Ghosh, \textbf{$`$Study of Controlled Dense Coding with some Discrete Tripartite and Quadripartite States'}, \textit{International Journal of Quantum Information}, Vol. \textbf{13}, No. 5, 1550033-(1-20), (2015), World Scientific.\\
\end{itemize}

\printindex
\chapter*{List of Corrections}
\begin{itemize}
\item \textbf {$`$ In eq. (2.2) on page $7$ the $\lbrace\vert \psi_{i}\rangle\rbrace$ form an orthonormal set'}.\\\\
\textbf{Reply:} After eq. (2.2), as suggested by reviewer the line $``$ $\lbrace\vert \psi_{i}\rangle\rbrace$ form an orthonormal set has been added".
\item \textbf{$`$pp $7$ just before eq. (2.3), the definition of a unitary operator is required. It is given in a footnote on page $37$'}.\\\\
\textbf{Reply:} The definition of unitary operator has been added in the footnote of page $7$ which was earlier provided in page $37$.
\item \textbf{$`$ pp $8$, in eq. (2.4), the condition $\sum_{i}A_{i}=I$ is needed'}.\\\\
\textbf{Reply:} After eq. (2.4) in page $8$, the condition $\sum_{i}\vert \psi\rangle\langle \psi_{i}\vert = I$ has been added.
\item \textbf{$`$ pp $8$, after eq. (2.5), replace the sentences $``$Now...pure state. In this...$\rho=\vert \psi\rangle\langle \psi\vert$" by $``$If the vector state $\vert \psi\rangle$ of a system is normalized, the state $\rho = \vert \psi\rangle\langle \psi\vert$ is pure and $\rho^{2}=\rho$" '}.\\\\
\textbf{Reply:} The desired change has been made in page number $8$ just after the eq. (2.5).
\item \textbf{$`$ Omit the word \textit{different} and the phrase \textit{in the ensemble for $\rho$} in the penultimate sentence.}'\\\\
\textbf{Reply:} The words \textit{different} and the phrase \textit{in the ensemble for $\rho$} have been omitted.
\item \textbf{$`$ In Article $2.4$, pp $9$, second paragraph after $\rho^{AB}$, write acting on the Hilbert space $H^{AB}=H^{A}\otimes H^{B}$, the tensor product of $H^{A}$ and $H^{B}$'.}\\\\
\textbf{Reply:} In Article $2.4$, pp $9$, second paragraph after $\rho^{AB}$, the sentence \textit{acting on the Hilbert space $H^{AB}=H^{A}\otimes H^{B}$, the tensor product of $H^{A}$ and $H^{B}$} has been added , as suggested by the reviewer. 
\item $`$ \textbf{In Article $2.4$, after eq. (2.6), replace \textit{the partial trace over the system $B$ (or $A$)} by over the Hilbert space $H^{B}$ (or $H^{A}$)'.}\\\\
\textbf{Reply:} After eq. (2.6), the sentence \textit{the partial trace over the system $B$ (or $A$)} has been replaced by \textit{the partial trace over the system $H^{B}$ (or $H^{A}$)}.
\item \textbf{$`$ In Article $2.5$, pp $10$, replace \textit{orthonormal states} by  \textit{orthonormal vectors} wherever they appear'.}\\\\
\textbf{Reply:} The phrase \textit{orthonormal states} has been replaced by  \textit{orthonormal vectors} wherever they appear.
\item \textbf{$`$ In Article $2.5$, pp $10$, in line $5$ of paragraph beginning \textit{purification}, replace, \textit{The system $R$ has the same state space} by $``$ The system $R$ has a Hilbert space $H^{R}$ unitarily equivalent to $H^{A}$" '.}\\\\
\textbf{Reply:} As suggested by the reviewer, the line \textit{The system $R$ has a Hilbert space $H_{R}$ unitarily equivalent to $H_{A}$} has been added after article $2.5$ in pp $10$.
\item $`$ \textbf{In Article $2.5$, in line $7$, put, $``$ the pure state $\vert AR\rangle \in H^{A}\otimes H^{R}$ "'.}\\\\
\textbf{Reply:} $\vert AR\rangle \in H^{A}\otimes H^{R}$ has been added as suggested by the reviewer in article $2.5$ (line $7$).
\item \textbf{$`$ In Article $2.6$, pp $12$, line $5$, it is not proven that mixed states are due to the de-coherence effect of nature. This requires a proof or a reference or just omit the phrase '.}\\\\
\textbf{Reply:}  As suggested by the referee we have omitted the phrase $``$the mixed states are due to the de-coherence effect of nature". The respective paragraph on mixed state has been modified.
\item \textbf{$`$ In Article $2.9$, pp $16$, line $1$, add \textit{of a state} after \textit{maximal singlet fraction}.}\\\\
\textbf{Reply:} In pp $16$ of article $2.9$, in the definition of maximal singlet fraction \textit{maximal singlet fraction of a state} has been written.
\item \textbf{$`$ In Article $2.11.1$, pp $20$, eq. (2.27), put $E(\vert \psi\rangle_{AB})$ ... or $E(\vert \psi\rangle_{AB}\langle \psi\vert)$, because the entropy of entanglement is a function of the state $\vert \psi\rangle_{AB}\langle \psi\vert$ '.}\\\\
\textbf{Reply:} In the formula of $`$ Entropy of Entanglement' in eq. (2.27) of page $20$, the notation $E(\vert \psi\rangle_{AB})$ has been introduced as suggested by reviewer.
\item \textbf{In Article $2.16.2$, page $33$, clarification required... $4$ lines above eq. (2.46), \textit{neither $X$...nor $A$... have any features of their own left} and $2$ lines after eq. (2.46), \textit{but they did not have their own private properties any more before they were observed}. Both assertions nee explanations, they can't mean that $X$ an $A$ don't have reduced states. This is particularly puzzling when in the next line the assertion \textit{particles $X$ and $A$ are identical} is made. What can this mean if they don't have their own properties? The argument to prove teleportation, ending after eq. (2.46) is only valid when $|a|^{2}=|b|^{2}=\frac{1}{2}$, not in general.}\\\\
\textbf{Reply:}  The ambiguities, as pointed out by the reviewer, has been omitted from this section. The entire paragraph in pp $3.3$ before eq. (2.46) have written afresh. Moreover the form of eq. (2.46) has been changed. The paragraph after eq. (2.46) has also been modified.
\item \textbf{$`$ In article $3.2$. after eq. (3.1), definitions of $S$ and $M$ must be given here. They are eventually partially defined on page $48$, $5$ lines after eq. (3.2), ($M$ denotes the uniform distribution (of what $?$) on the Bloch sphere $S$)'.}\\\\
\textbf{Reply:} In article $3.2$ defining teleportation fidelity, the meaning of uniform distribution $M$ as well as that of Bloch sphere $S$ have been explicitly defined in the footnote. Also it has been mentioned clearly that this uniform distribution is defined on all the input states on the Bloch sphere.
\item \textbf{$`$ In article $3.2$, pp $45$, write $dM(\rho_{\phi})$ in place of $dM (\phi)$ for consistency.}\\\\
\textbf{Reply:} In article $3.2$ of pp $45$, in eq. (3.1), the suggested change in the notation $dM (\phi)$ has been made by replacing this with $dM(\rho_{\phi})$.
\item \textbf{pp $46$, the matrix $T$ is a function of the state $\rho$, this should be made clear to the reader.}\\\\
\textbf{Reply:} Just before eq. (3.2), as suggested by reviewer, it has been made clear how the matrix $T$ is dependent on the input states in some cases. 
\end{itemize}

\begin{thebibliography}{99}
\bibitem{epr1935} A. Einstein, B. Podolsky and N. Rosen, \textbf{Can the quantum mechanical description of reality be considered complete?}, \textit{Physical Review}, 47, 777, (1935).
\bibitem{sch1935} E. Schrodinger, \textbf{Die gegenwartige Situation in der Quantenmechanik}, \textit{Naturwissenschaften}, 23, 807, (1935).
\bibitem{feynman1977} R. P. Feynman, R. B. Leighton and M. Sands, \textbf{The Feynman Lectures in Physics, vol. 1,2 and 3}, Addison - Wesley, Reading, \textit{MA}, (1977).
\bibitem{bell1964} J. S. Bell, \textit{Physics}, 1, 195, (1964),\textbf{On the Problem of Hidden Variables in Quantum Mechanics}, \textbf{Rev. Mod. Phys.}, 38, 447, (1966).
\bibitem{aspect1981} A. Aspect, P. Grangier and G. Roger, \textbf{Experimental Tests of Realistic Local Theories via Bell's Theorem}, \textit{Phys. Rev. Lett.}, 47, 460, (1981).
\bibitem{aspect1982} A. Aspect, J. Dalibard and G. Roger, \textbf{Experimental Test of Bell's Inequalities Using Time- Varying Analyzers}, \textit{Phys. Rev. Lett.}, 49, 1804, (1982).
\bibitem{ghz1989} D. M. Greenberger, M. A. Horne and A. Zeilinger, \textbf{Going beyond Bell's theorem in Bell's theorem, Quantum Theory and Conceptions of the Universe}, Kluwer, Academic \textit{Dorthecht}, (1989).
\bibitem{bennett1993} C H Bennett, G Brassard, C Crepeau, R Jozsa, A Peres and W K Wootters, \textbf{Teleporting an unknown quantum state via dual classical and Einstein-Podolsky-Rosen channels}, \textit{Phys. Rev. Lett.} 70, 1895, (1993).
\bibitem{bouwmeester1997} D. Bouwmeester, J. W. Pan, K. Mattle, M. Eibl, H. Weinfurter and A. Zeilinger, \textbf{Experimental quantum teleportation}, \textit{Nature}, 390, 575, (1997).
\bibitem{bennett1992} C. H Bennett and S. J, Wiesner, \textbf{Communication via one- and two-particle operators on Einstein-Podolsky-Rosen states}, \textit{Phys. Rev. Lett.}, 69, 2881, (1992).
\bibitem{mattle1996} K. Mattle, H. Weinfurter, P. G. Kwiat and A. Zeilenger, \textbf{Dense Coding in Experimental Quantum Communication}, \textit{Phys. Rev. Lett.}, 76, 4656 (1996).
\bibitem{hillery1999} M. Hillery, V. Buzek and A. Berthiaume, \textbf{Quantum secret sharing}, \textit{Phys. Rev. A}, 59, 1829, (1999).
\bibitem{cleve1999} R. Cleve and H. K. Lo, \textbf{How to Share a Quantum Secret}, \textit{Phys. Rev. Lett.}, 83, 648, (1999).
\bibitem{bennett1984} C. H. Bennett and G. Brassard, \textbf{Quantum Cryptography}, \textit{Proceedings of IEEE International Conference on Computers, Systems and Signal Processing Bangalore, India. IEEE}, New York, p. 175 (1984); C. H. Bennett, G. Brassard and N. Mermin, \textbf{Quantum Cryptography without Bell's theorem}, \textit{Phys. Rev. Lett.}, 68, 557, (1992).
\bibitem{ekert1991} A. K. Ekert, \textbf{Quantum cryptography based on Bell$'$s theorem}, \textit{Phys. Rev. Lett.}, 67, 661, (1991).
\bibitem{bennett1992(2)} C. H. Bennett, F. Bessette, G. Brassard and L. Salvail, \textbf{Experimental quantum cryptography}, \textit{Jour. of Cryptology}, 5, 3, (1992).
 \bibitem{barenco1994} A. Barenco and A. K. Ekert, \textbf{Dense Coding Based on Quantum Entanglement}, \textit{Jour. of Mod. Optics}, 42,  1253, (1994).
 \bibitem{bose1999} S. Bose, M. B. Plenio and V. Vedral, \textbf{Mixed state dense coding and its relation to entanglement measures}, \textit{Jour. of Mod. Optics}, 47, 291, (1999).
 \bibitem{verstraete2003} F. Verstraete and H. Verschelde, \textbf{Optimal Teleportation with a Mixed State of Two Qubits}, \textit{Phys. Rev. Lett.}, 90, 097901, (2003).
 \bibitem{albeverio2003} S. Albeverio, S. M. Fei and W. L. Yang, \textbf{Foundations of Probability and Physics - 2}, Vaexjoe University Press, 37-56, (2003).
 \bibitem{gottesman2000} D. Gottesman, \textbf{Theory of quantum secret sharing}, \textit{Phys. Rev. A}, 61, 042311, (2000).
 \bibitem{adhikari2010} S. Adhikari, \textbf{Quantum secret sharing with two qubit bipartite mixed states} \textit{quant-ph}, 1011.2868, (2010).
 \bibitem{scarani2005} V. Scarani, S. Iblisdir and N. Gisin, \textbf{Quantum Cloning}, \textit{Rev. Mod. Phys.}, 77, 1225, (2005).
 \bibitem{heisenberg1925} W. Heisenberg, \textbf{Uber quantentheoretische Umdeutung kinematischer und mechanischer Beziehungen}, \textit{Physik}, 33, 879, (1925).
 \bibitem{wootters1982} W. K. Wootters and W. H. Zurek, \textbf{A single quantum cannot be cloned}, \textit{Nature}, 299, 802, (1982).
 \bibitem{buzek1996} V. Buzek and M. Hillery, \textbf{Quantum copying: Beyond the no-cloning theorem}, \textit{Phys. Rev. A}, 54, 3, (1996).
 \bibitem{gisin1997} N. Gisin and S. Massar, \textbf{Optimal Quantum Cloning Machines}, \textit{Phys. Rev. Lett.}, 79, 11, (1997).
 \bibitem{buzek1998} V. Buzek and M. Hillery, \textbf{Universal Optimal Cloning of Arbitrary Quantum States: From Qubits to Quantum Registers}, \textit{Phys. Rev. Lett.}, 81, 22, (1998).
 \bibitem{bru1998} D. Bru$\beta$, D. P. Vincenzo, A. Ekert, C. A. Fuchs, C. Macchiavello and J. A. Smolin, \textbf{Optimal universal and state dependent quantum cloning}, \textit{Phys. Rev. A}, 57, 4, (1998).
 \bibitem{duan1998} L. M. Duan and G. C. Guo, \textbf{Probabilistic Cloning and Identification of Linearly Independent Quantum States}, \textit{Phys. Rev. Lett.}, 80, 4999, (1998).
 \bibitem{zanardi1998} P. Zanardi, \textbf{Quantum cloning in d dimensions}, \textit{Phys. Rev. A}, 58, 5, (1998).
 \bibitem{fan2001} H. Fen, K. Matsumoto and M. Wadati, \textbf{Quantum cloning machines of a d - level system}, \textit{Phys. Rev. A}, 64, 064301, (2001).
 \bibitem{delgado2007} Y. Delgado, L. Lamata, J. Leon, D. Salgado and E. Solano, \textbf{Sequential Quantum Cloning}, \textit{Phys. Rev. Lett.}, 98, 150502, (2007).
 \bibitem{dang2008} G. F. Dang and H. Fan, \textbf{General sequential quantum cloning}, \textit{Jour. Phys. A: Mathematical and Theoretical}, 41, 155303, (2008).
 \bibitem{adhikari2007} S. Adhikari, A. K. Pati, I. Chakrabarty and B. S. Choudhury, \textbf{Hybrid Quantum Cloning Machine}, \textit{Quant. Inf. Processing}, 6, 4, (2007).
 \bibitem{adhikari2008} S. Adhikari, N. Ganguly, I. Chakrabarty and B. S. Choudhury, \textbf{Quantum Cloning, Bell's Inequality and Teleportation}, \textit{Jour. Phys. A: Mathematical and Theoretical}, 41, 415302, (2008).
 \bibitem{nielsenbook} M. A. Nielsen and I. L. Chuang, \textbf{Quantum Computation and Quantum Inforation}, \textit{Cambridge university press}, 10th anniversary edition published, ISBN:978-1-107-00217-3, (2010).
 \bibitem{mcmahonbook} D. McMahon, \textbf{Quantum Computing Explained}, \textit{A John Wiley and Sons, Inc., Publication}, ISBN 978-0-470-09699-4 , (2008).
 \bibitem{mintertreview2005} F. Mintert, A. R. R. Carvalho, M. Kus and A. Buchleitner, \textbf{Measures and dynamics of entangled states}, \textit{Physics Reports}, 415, 207-259, (2005).
 \bibitem{horodeckireview} R. Horodecki, P. Horodecki, M. Horodecki and K. Horodecki, \textbf{Quantum Entanglement}, \textit{Rev. Mod. Phys.}, 81, 865, (2009).
 \bibitem{plenioreview} M. B. Plenio and S. Virmani, \textbf{An introduction to entanglement measures}, \textit{Quant. Inf. and Comp.}, 7, 1, 2007.
 \bibitem{adhikarithesis} S. Adhikari, \textbf{Quantum Cloning and Deletion in Quantum Information Theory}, quant-ph: 0902.1622, (2009).
\bibitem{bennett1996} C. H. Bennett, D. P. DiVincenzo, J. A. Smolin and W. K. Wootters, \textbf{Mixed state entanglement and quantum error correction}, \textit{Phys. Rev. A}, 54, 3824, (1996).
 \bibitem{horodecki1998} M. Horodecki, P. Horodecki and R. Horodecki, \textbf{Mixed-State Entanglement and Distillation: Is there a Bound Entanglement in Nature ?}, \textit{Phys. Rev. A}, 80, 5239, (1998).
 \bibitem{kent1999} A. Kent, N. Linden and S. Massar, \textbf{Optimal Entanglement Enhancement for Mixed States}, \textit{Phys. Rev. Lett.}, 83, 2656, (1999).
 \bibitem{alberbook} G. Alber, T. Beth, M. Horodecki, P. Horodecki, R. Horodecki, M. Rotteler, H. Weinfurter, R. Werner and A. Zeilinger, \textbf{Quantum Information: An Introduction to Basic Theoretical Concepts and Experiments}, \textit{Springer}, Springer tracts in modern physics 173.
 \bibitem{chsh1969} J. F. Clauser, M. A. Horne, A. Shimony and R. A. Holt, \textbf{Proposed Experiment to Test Local Hidden-Variable Theories}, \textit{Phys. Rev. Lett.}, 23, 880, (1969).
 \bibitem{cirelson1980} B. S. Cirelson, \textbf{Quantum Generalizations of Bell's Inequality}, \textit{Letters in Mathematical Physics}, 4, 93,(1980). 
 \bibitem{horodecki601999} M. Horodecki, P. Horodecki and R. Horodecki, \textbf{General teleportation channel, singlet fraction, and quasidistillation}, \textit{Phys. Rev. A}, 60, 1888, (1999).
\bibitem{verstraete2002} F. Verstraete and H, Verschilde, \textbf{Fidelity of mixed states of two qubits}, \textit{Phys. Rev. A}, 66, 022307, (2002).
 \bibitem{bose2000} S. Bose and V. Vedral, \textbf{Mixedness and Teleportation}, \textit{Phys. Rev. A}, 61, 040101, (2000).
 \bibitem{wei2003} T. C. Wei, K. Nemoto, P. M. Goldbart, P. G. Kwait, W. J. Munro and F. Verstraete, \textbf{Maximal entanglement versus entropy for mixed quantum states}, \textit{Phys. Rev. A}, 67, 022110, (2003).
 \bibitem{ishizaka2000} S. Ishizaka and T. Hiroshima, \textbf{Maximally entangled mixed states under nonlocal unitary operations in two qubits}, \textit{Phys. Rev. A}, 62, 022310, (2000).
  \bibitem{werner401989} R. F. Werner, \textbf{Quantum states with Einstein-Podolsky-Rosen correlations admitting a hidden variable model}, \textit{Phys. Rev. A}, 40, 4277, (1989).
 \bibitem{wootters1998} W K Wootters, \textbf{Entanglement of Formation of an Arbitrary State of Two Qubits}, \textit{Phys. Rev. Lett.}, 80,  2245, (1998).
 \bibitem{munro2001} W. J. Munro, D. F. V. James, A. G. White and P. G. Kwiat, \textbf{Maximizing the entanglement of two mixed qubits}, \textit{Phys. Rev. A}, 64, 030302, (2001).
 \bibitem{hiroshima2000} T. Hiroshima and S. Ishizaka, \textbf{Local and nonlocal properties of Werner states}, \textit{Phys. Rev. A}, 62, 044302, (2000).
 \bibitem{coffman2000} V. Coffman, J. Kundu and W. K. Wootters, \textbf{Distributed entanglement}, \textit{Phys. Rev. A}, 61, 052306, (2000).
 \bibitem{wootters1997} S Hill and W K Wootters , \textbf{Entanglement of a Pair of Quantum Bits}, \textit{Phys. Rev. Lett.}, 78, 5022, (1997). 
\bibitem{audenaert2005} K. M. R. Audenaert  and J.Eisert, \textbf{Continuity bounds on the quantum relative entropy}, \textit{J. Math. Phys.}, 46, 102104 (2005).
\bibitem{miranowicz2004} A. Miranowicz and A. Grudka, \textbf{A comparative study of relative entropy of entanglement, concurrence and negativity}, \textit{J. Opt. B: Quantum and  Semiclassical Optics}, 6, 542, (2004).
\bibitem{plenio1998} M. B. Plenio  and V. Vedral, \textbf{Teleportation, entanglement and thermodynamics in the quantum world}, \textit{Contemp. Phys.},  39, 431 (1998).
\bibitem{horodecki1996} M. Horodecki, P. Horodecki, R. Horodecki, \textbf{Separability of mixed states: necessary and sufficient conditions}, \textit{Phys. Lett. A}, 223, 1 (1996).
\bibitem{peres1996} A. Peres, \textbf{Separability Criterion for Density Matrices}, \textit{Phys. Rev. Lett.}, 77, 1413 (1996).
\bibitem{audenaert2001} F. Verstraete, K. Audenaert, J. Dehaene and B. D.  Moor, \textbf{A comparison of the entanglement measures negativity and concurrence}, \textit{Jour. of Phys. A}, 34, 10327, (2001).
\bibitem{eisert1999} J. Eisert and M. B. Plenio, \textbf{A Comparison of Entanglement Measures} \textit{Jour. of Mod. Opt.}, 46, 145, (1999).
\bibitem{audenaertarxiv} K. Audenaert, F. Verstraete, T. D. Bie and B. D. Moor, \textbf{Negativity and Concurrence of mixed 2$\times$2 states}, \textit{QIP2001 workshop}, quantph-0012074, (2001).
\bibitem{slee2003} S. Lee, D. P. Chi, S. D. Oh and J. Kim, \textbf{Convex-roof extended negativity as an entanglement measure for bipartite quantum systems}, \textit{Phys. Rev. A}, 68, 062304, (2003).
\bibitem{guhnereview} O. Guhne, \textbf{Entanglement Detection}, \textit{Physics Reports}, 474, 1, (2009).
\bibitem{horodeckireview2009} R. Horodecki, P. Horodecki and M. Horodecki, \textbf{Quantum entanglement}, \textit{Review. Mod. Phys.}, 81, 865, (2009).
\bibitem{uhlmann} A. Uhlmann, \textbf{The transition probability in the state space of a $*$-algebra}, \textit{Rep. Math. Phys.}, 9, 273, (1976).
\bibitem{joz1994} R. Jozsa, \textbf{Fidelity for mixed quantum states}, \textit{Jour. of Mod. Opt.}, 41, 2315, (1994).
\bibitem{filip2002} R. Filip, \textbf{Overlap and entanglement-witness measurements}, \textit{Phys. Rev. A},  65, 062320, (2002).
\bibitem{plenioconjecture} V. Vedral and M. B. Plenio, \textbf{Entanglement measures and purification procedures},  \textit{Phys. Rev. A}, 57, 1619, (1998).
\bibitem{hahnbanach}  E. Kreyszig, \textbf{Introductory Functional Analysis with Applications}, \textit{Wiley Publication}, 3rd edition, ISBN: 978-81-265-1191-4, (2007).
\bibitem{werner2001} R. F. Werner, \textbf{All teleportation and dense coding schemes}, \textit{Jour. of Physics A: Mathematical and General}, 34, 35, (2001).
\bibitem{danceofphoton} A. Zeilinger, \textbf{Dance of Photons: From Einstein to Quantum Teleportation},  \textit{Farrar, Straus and Giroux}, Newyork, First edition, ISBN 978-0-374-23966-4 (hardcover), (2010).
\bibitem{physicsinf} D. Bowmeester, A. Ekert and A. Zeilinger, \textbf{The Physics of Quantum Information}, \textit{Springer-Verlag Berlin Heidelberg}, ISBN: 978-3-642-08607-6, (2000).
\bibitem{senshi2001} B. S. Shi, Y. K. Jiang and G. C. Guo, \textbf{Probabilistic teleportation of two particle entangled state}, \textit{Phys. Lett. A}, 268, 161, (2001).
\bibitem{leekim2000} J. Lee and M. S. Kim, \textbf{Entanglement Teleportation via Werner States}, \textit{Phys. Rev. Lett.}, 84, 18, (2000).
\bibitem{hao2001} J. C. Hao, C. F. Li and G. C. Guo, \textbf{Controlled dense coding using the Greenberger-Horne-Zeilinger state}, \textit{Phys. Rev. A}, 63, 054301, (2001).
\bibitem{ghzstate} D M Greenberger, M Horne and A Zeilinger, \textit{Bell's theorem, Quantum theory and Conceptions of the Universe} \textit{(U.S.A: Springer) (1989 ed.)}  M Kafatos 1 (1989).; D. M. Greenberger, M. A. Horne and A. Zeilinger, \textbf{Bell's theorem without inequalities}, \textit{Am. J. Phys.}, 58, 1131, (1990).
\bibitem{zukowski1995} M. Zukowski, A. Zeilinger and H. Weinfurther, \textbf{Entangling Photons Radiated by Independent Pulsed Sources}, \textit{Ann. N. Y. Acad. Sci.}, 755, 91, (1995).
\bibitem{pan1999} D. Bowmeester, J. W. Pan, M. Daniell. H. Weinfurther and A. Zeilinger, \textbf{Observation of three photon Greenberger-Horne-Zeilinger entanglement}, \textit{Phys. Rev. Lett.}, 82, 1345, (1999).
\bibitem{brucebook} B. Schneier, \textbf{Applied Cryptography}, \textit{Wiley, New York}, p-70, (1996); J. Gruska, \textbf{Foundations of Computing}, \textit{Thompson Computer Press, London}, p-504, (1997).
\bibitem{werner1827} R. F. Werner, \textbf{Optimal Cloning of Pure States}, \textit{Phys. Rev. A}, 58, 1827, (1998).
\bibitem{mhorodecki1996} R. Horodecki, M. Horodecki and P. Horodecki, \textbf{Teleportation, Bell's inequalities and inseparability}, \textit{Phys. Lett. A}, 222, 1-2, (1996).
\bibitem{horodecki541996} R. Horodecki and M. Horodecki, \textbf{Information-theoretic aspects of inseparability of mixed states}, \textit{Phys. Rev. A}, 54, 1838, (1996).
\bibitem{badziag2000} P. Badziag, M. Horodecki, P. Horodecki and R. Horodecki, \textbf{Local environment can enhance fidelity of quantum teleportation}, \textit{Phys. Rev. A}, 62, 012311, (2000).
\bibitem{mista2002} L. Mista Jr., R. Filip and J. Fiurasek, \textbf{Continuous variable Werner state: Separability, nonlocality, squeezing, and teleportation}, \textit{Phys. Rev. A}, 65, 062315, (2002).
\bibitem{munro482001} W. J. Munro, K. Nemoto and A. G. White, \textbf{The bell inequality: A measure of entanglement?}, \textit{Jour. of Mod. Optics}, 48, 1239, (2001).
\bibitem{wstate} A Zeilinger,  M A Horne and D Greenberger \textit{NASA Conf. Publ. No.} 3235 \textit{((Washington DC: Code NTT)} (1997)); W. D$\ddot{u}$r, G. Vidaland J. I. Cirac, \textbf{Three qubits can be entangled in two inequivalent ways}  \textit{Phys. Rev. A}, 62, 062314, (2000).
\bibitem{bruss332003} D. Bru$\beta$ and C. Macchiavello, \textbf{On the entanglement structure in quantum cloning}, \textit{Found.of  Phys.}, 33, 1617, (2003).
\bibitem{horodeckis1999} M. Horodecki and P. Horodecki, \textbf{Reduction criterion of separability and limits for a class of distillation protocols}, \textit{Phys. Rev. A}, 59, 4206, (1999). 
\bibitem{karimipour2006} V. Karimipour and L. Memarzadeh, \textbf{Equientangled bases in arbitrary dimensions}, \textit{Phys. Rev. A}, 73, 012329, (2006).
\bibitem{lewenstein2004} D. Bru\ss , G. M. D'Ariano, M. Lewenstein, C. Macchiavello, A. Sen (De), U. Sen, \textbf{Distributed quantum dense coding},  \textit{Phys. Rev. Lett.}, 93, 210501, (2004).
\bibitem{nirman1} N. Ganguly, S. Adhikari, A. S. Majumdar and J. Chatterjee, \textbf{Entanglement witness operator for quantum teleportation}, \textit{Phys. Rev. Lett.}, 107, 270501, (2011).
\bibitem{nirman2} A. Kumar, S. Adhikari, and P. Agrawal, \textbf{Generalized for of optimal teleportation witness}, \textit{Quant. Inf. Proc.}, 12, 2475, (2013). 
\bibitem{bennett761996} C. H. Bennett, G. Brassard, S. Popescu, B. Schumacher, J. A. Smolin and W. K. Wootters, \textbf{Purification of Noisy Entanglement and Faithful Teleportation via Noisy Channels}, \textit{Phys. Rev. Lett.}, 76, 722, (1996).
\bibitem{deutsch771996} D. Deutsch, A. Ekert, R. Jozsa, C. Macchiavello, S. Popescu, and A. Sanpera, \textbf{Quantum Privacy Amplification and the Security of Quantum Cryptography over Noisy Channels}, \textit{Phys. Rev. Lett.}, 77, 2818, (1996).
\bibitem{bennett541996} C. H. Bennett, D. P. Di-Vincenzo, J. A. Smolin, and W. K. Wootters, \textbf{Mixed-state entanglement and quantum error correction}, \textit{Phys. Rev. A}, 54, 3824, (1996). 
\bibitem{horodeckis781997} M. Horodecki, P. Horodecki, and R. Horodecki,  \textbf{Inseparable Two Spin - $\frac{1}{2}$ Density Matrices Can Be Distilled to a Singlet Form}, \textit{Phys. Rev. Lett.}, 78, 574, (1997).
\bibitem{karlsson1998} A. Karlsson and M. Bourennane, \textbf{Quantum teleportation using three-particle entanglement}, \textit{Phys. Rev. A}, 58, 4394, (1998).
\bibitem{yan2003} F. Yan and D. Wang ,\textbf{Probabilistic and controlled teleportation of unknown quantum states}, \textit{Phys. Lett. A}, 316, 5, (2003).
\bibitem{ting2005} G. Ting, Y. F. Li and W. Z. Xi, \textbf{Controlled quantum teleportation and secure direct communication}, \textit{Chinese Physics}, 14, 5, (2005).
\bibitem{lxhan2007} L. X. Han, D. F. Guo and Z. H. Yu, \textbf{Controlled teleportation of an arbitrary multi qudit state in a general form with d-dimensional GHZ states}, \textit{Chinese Phys. Lett.}, 24, 5, (2007).
\bibitem{man2007} Z. X. Man, Y. J. Xia and N. B. An \textbf{Genuine multiqubit entanglement and controlled teleportation}, \textit{Phys. Rev. A}, 75, 052306, (2007).
\bibitem{song2008} L. S. Song, N. Y. You, H. Z. Hui, Y. X. Zie and H. Y. Bin, \textbf{Controlled Teleportation Using Four-Particle Cluster State}, \textit{Comm. in Theor. Phys.}, 50, 3, (2008).
\bibitem{gao2008} T. Gao, F. L. Yan and Y. C. Li, \textbf{Optimal controlled teleportation}, \textit{Euro Phys. Lett.}, \textbf{84}, 50001, (2008).
\bibitem{zha2013} X. W. Zha, Z. C. Zou, J. X. Qi and H. Y. Song, \textbf{Bidirectional quantum controlled teleportation via five-qubit cluster state}, \textit{Int. Jour. of Theor. Phys.}, 52, 6, (2013).
\bibitem{peng2002} J. Zhang, C. Xie and K. Peng, \textbf{Controlled dense coding for continuous variables using three-particle entangled states}, \textit{Phys. Rev. A}, 66, 032318, (2002).
\bibitem{yynie2008} Y. B. Huang, S. S. Li and Y. Y. Nie, \textbf{Controlled Dense Coding between Multi-Parties}, \textit{Int. Jour of Theor. Phys.}, 48, 95, (2008).
\bibitem{wangwu2009} D. Y. Jiang, R. S. Wu, S. S. Li and Z. S. Wang, \textbf{Controlled dense coding with symmetric states}, \textit{Int. Jour. of Theor. Phys.}, 48, 2297, (2009).
\bibitem{sixcdc2011} X. J. Yi, J. M. Wang and G. Q. Huang, \textbf{Controlled Dense Coding with Six-Qubit Cluster State}, \textit{Int. Jour. of Theor. Phys.}, 50, 364, (2011).
\bibitem{cavitycdc2013} Y. Y. Nie, Y. H. Li, X. P. Wang and M. H. Sang, \textbf{Controlled dense coding using a five-atom cluster state in cavity QED}, \textit{Quant. Inf. Process.}, 12, 1851, (2013).
\bibitem{pathakbook} A. Pathak, \textbf{Elements of Quantum Computation and Quantum Communication}, \textit{CRC Press, Taylor and Francis Group}, ISBN (ebook):978-1-4665-1792-9, (2013).
\bibitem{cereceda2001} J. L. Cereceda, \textbf{Quantum dense coding using three qubits}, \textit{	arXiv:quant-ph/0105096}, (2001).
\bibitem{liqiu2007} L. Li and D. Qiu, \textbf{The states of W class as shared resources for perfect teleportation and super dense coding}, \textit{Jour. of. Phys. A: Math. Theor.}, 40, 10871, (2007).
\bibitem{patiparaagra2005} A. K. Pati, P. Parashar and P. Agrawal, \textbf{Probabilistic Super Dense Coding}, \textit{Phys. Rev. A}, 72, 012329, (2005).
\bibitem{fuxializh2005} C. Fu, Y. Xia, B. Liu, S. Zhang, K. Hwang and C. I. Um, \textbf{Controlled Quantum Dense Coding in a Four-particle Non-maximally Entangled State via Local Measurements}, \textit{Jour. of Korean Phys. Soc.}, 46, 1080, (2005).
\bibitem{durr622000} W. D\"{u}r, G. Vidal and J. I. Cirac, \textbf{Three Qubits can be Entangled in Two Inequivalent Ways}, 2000 \textit{Phys. Rev. A}, 62, 062314, (2000).
\bibitem{agrapati2006} P. Agrawal and A. K. Pati, \textbf{Perfect Teleportation and Superdense Coding With W-States}, \textit{Phys. Rev. A}, 74, 062320, (2006).
\bibitem{liulongtongli2002} X. S. Liu, G. L. Long, D. M. Tong and F. Li, \textbf{General scheme for superdense coding between multiparties}, \textit{Phys. Rev. A}, 65, 022304, (2002).
\bibitem{mintertsalway2012} F. Mintert, B. Salway and A. Buchleitner, \textbf{Many-body entanglement: Permutations and equivalence classes}, \textit{Phys. Rev. A}, 86, 052330, (2012).
\bibitem{abdesselam2010} B. Abdesselam, A. Chakraborty, V. K. Dobrev and S. G. Mihov, \textbf{Exotic Bialgebras from 9x9 Unitary Braid Matrices},\textit{Physics of Atomic Nuclei - PHYS ATOM NUCL-ENGL TR}, 74, 824, (2011).
\bibitem{karlssonsecret1999} A. Karlsson, M. Koashi and N. Imoto, \textbf{Quantum entanglement for secret sharing and secret splitting}, \textit{Phys. Rev. A}, 59, 162, (1999).
\bibitem{bandyopadhyay2000} S. Bandyopadhyay, \textbf{Teleportation and secret sharing with pure entangled states}, \textit{Phys. Rev. A}, 62, 012308, (2000).
\bibitem{bagherinezad2003} S. Bagherinezad and V. Karimipour, \textbf{Quantum secret sharing based on reusable Greenberger-Horne-Zeilinger states as secured carriers}, \textit{Phys. Rev. A}, 67, 044302, (2003).
\bibitem{lance2004} A. M. Lance, T. Symul, W. P. Bowen, B. C. Sanders and P. K. Lam, \textbf{Tripartite quantum state sharing}, \textit{Phys. Rev. Lett.}, 92, 177903, (2004).
\bibitem{gordon2006} G. Gordon and G. Rigolin, \textbf{Generalized quantum state sharing}, \textit{Phys. Rev. A}, 73, 062316, (2006).
\bibitem{zhengsecret2006} S. B. Zheng, \textbf{Splitting quantum information via W states}, \textit{Phys. Rev. A}, 74, 054303, (2006).
\bibitem{tittel2001} W. Tittel, H. Zbinden and N. Gisin, \textbf{Experimental demonstration of quantum secret sharing}, \textit{Phys. Rev. A}, 63, 042301, (2001).
\bibitem{schmidt2005} C. Schmidt, P. Trojek, M. Bourennane, C. Kurtsiefer, M. Zukowski and H. Weinfurter, \textbf{Experimental single qubit quantum secret sharing}, \textit{Phys. Rev. Lett.}, 95, 230505, (2005).
\bibitem{schmidt2006} C. Schmidt, P. Trojek, S. Gaertner, M. Bourennane, C. Kurtsiefer, M. Zukowski and H. Weinfurter, \textbf{Experimental quantum secret sharing}, \textit{Fortschritte der Physik}, 54, 831, (2006).
\bibitem{bogdanski2008} J. Bogdanski, N. Rafiei and M. Bourennane, \textbf{Experimental quantum secret sharing using telecommunication fibre}, \textit{Phys. Rev. A}, 78, 062307. (2008).
\bibitem{liqsecret2010} Q. Li, W. H. Chan and D. Y. Long, \textbf{Semi quantum secret sharing using entangled states}, \textit{Phys. Rev. A}, 82, 022303, (2010).
\bibitem{bertlmanwitnes2005} R. A. Bertlmann, K. Durstberger, B. C. Hiesmayr and P. Krammer, \textbf{Optimal entanglement witness for qubits and qutrits}, \textit{Phys. Rev. A}, 72, 052331, (2005).
\bibitem{sanperawitness} A. Sanpera, D. $\beta$russ and M. Lewenstein, \textbf{Schmidt-number witnesses and bound entanglement}, \textit{Phys. Rev. A}, 63, 050301, (R), (2001).
\bibitem{caves9910001} C. M. Caves and G. J. Milburn, \textbf{Qutrit Entanglement}, \textit{Optics Communication}, 179, 439, (2000). 
\bibitem{bastian2011} B. Jungnitsch, T. Moroder and O. Guhne, \textbf{Taming Multiparticle Entanglement}, \textit{Phys. Rev. Lett.}, 106, 190502, (2011).
\bibitem{zhma2011} Z. H. Ma, Z.H. Chen, J. L. Chen, C. Spengler, A. Gabriel, and M. Huber, \textbf{Measure of Genuine Multipartite Entanglement with Computable Lower Bounds}, \textit{Phys. Rev. A}, 83, 062325, (2011).
\bibitem{raf2012} S. M. H. Rafsanjani, M. Huber, C. J. Broadbent, and J. H. Eberly, \textbf{Genuinely multipartite concurrence of N-qubit X matrices}, \textit{Phy. Rev. A}, 86, 062303, (2012).
\bibitem{cerfjmo2011} N. J. Cerf, \textbf{Cloning a Qutrit}, \textit{Jour. of Mod. Opt.}, 49, 1355, (2011).
\bibitem{tamoghna2014} T. Das, R. Prabhu, A. Sen (De) and U. Sen, \textbf{Distributed quantum dense coding with two receivers in noisy environments}, \textit{Phys. Rev. A}, 92, 052330, (2015).
\bibitem{maximalslicedcdc2016} J. Liu, Z-w Mo and S-q. Sun, \textbf{Controlled dense coding using the maximal sliced states}, \textit{Int. Jour. of Theor. Phys.}, 55, 2182, (2016).
\bibitem{scottame2004} A. J. Scott, \textbf{Multipartite Entanglement, Quantum Error Correcting Codes and Entangling Power of Quantum Evolutions}, \textit{Phys. Rev. A}, 69, 2004.
\bibitem{helwigame2012} W. Helwig, W, Cui, J. I. Lattore, A. Riera and H. K. Lo, \textbf{Absolute Maximal Entanglement and Quantum Secret Sharing}, \textit{Phys. Rev. A}, 86, 052335, (2012).
\end{thebibliography}
\end{document}